\documentclass[reprint,superscriptaddress,
amsmath, amssymb,aps, prd,notitlepage,longbibliography,floatfix,
nofootinbib,onecolumn
]{revtex4-1}

\usepackage{tensor}     
\usepackage{graphicx}   
\usepackage{subfigure}
\usepackage[
colorlinks=true,        
citecolor=blue,         
linkcolor=blue,         
urlcolor=blue           
]{hyperref}             
\usepackage{bm}         
\usepackage{xcolor}     
\usepackage{lipsum}
\usepackage{color}      
\usepackage[utf8]{inputenc} 
\usepackage[section]{placeins} 
\usepackage{multirow}
\usepackage[title]{appendix}
\newcommand{\nc}{\newcommand*}

\nc{\al}{\alpha}
\nc{\s}{\sigma}
\nc{\kp}{\kappa}
\nc{\dt}{\delta}
\nc{\Dt}{\Delta}
\nc{\Ld}{\Lambda}
\nc{\p}{\partial}
\nc{\Gm}{\Gamma}
\nc{\om}{\omega}
\nc{\Om}{\Omega}
\nc{\rd}{\mathrm{d}}
\def\({\left(}
\def\){\right)}
\def\[{\left[}
\def\]{\right]}
\def\e{\begin{equation}}
\def\q{\end{equation}}
\def\m{\begin{eqnarray}}
\def\n{\end{eqnarray}}
\nc{\Eq}[1]{Eq.~\eqref{#1}}     
\nc{\Fig}[1]{Fig.~\ref{#1}}     
\nc{\Table}[1]{Table~\ref{#1}}  
\nc{\Sec}[1]{Sec.~\ref{#1}}     
\nc{\Msun}{M_\odot}             
\nc{\fpbh}{f_{\mathrm{pbh}}}    
\nc{\mpbh}{m_{\mathrm{pbh}}}    
\nc{\fpbhn}{f_{\mathrm{pbh0}}}    
\nc{\mR}{\mathcal{R}} 
\nc{\seq}{\sigma_{\mathrm{eq}}}
\nc{\ogw}{\Omega_{\mathrm{GW}}}
\nc{\gpcyr}{\mathrm{Gpc}^{-3}\,\mathrm{yr}^{-1}}
\nc{\lvc}{LIGO/Virgo} 
\nc{\SNR}{\mathrm{SNR}} 
\nc{\mmin}{{m_{\mathrm{min}}}}
\nc{\mmax}{{m_{\mathrm{max}}}}
\nc{\Mmin}{{M_{\mathrm{min}}}}
\nc{\fmin}{{f_{\mathrm{min}}}}
\nc{\VT}{\mathrm{VT}}
\nc{\rhoGW}{\rho_{\mathrm{GW}}}
\nc{\vth}{\vec{\theta}}
\nc{\vd}{\vec{d}}
\nc{\vla}{\vec{\lambda}}
\nc{\Nobs}{N_{\mathrm{obs}}}
\nc{\av}[1]{\langle #1 \rangle} 
\nc{\km}{\mathrm{km}}
\nc{\Mpc}{\mathrm{Mpc}}
\nc{\Tobs}{T_{\mathrm{obs}}}
\nc{\Ntemp}{N_{\mathrm{temp}}}
\nc{\fyr}{f_{\mathrm{yr}}}
\nc{\addref}{[\textcolor{red}{add ref}] } 
\nc{\eg}{\textit{e.g.~}}
\nc{\app}{\approx}
\nc{\hf}{\frac{1}{2}}
\nc{\discuss}{\textcolor{red}{Add discussion here!}}
\nc{\red}[1]{\textcolor{red}{#1}}
\nc{\hp}{h_+} 
\nc{\hc}{h_{\times}} 
\nc{\Oh}{\hat{\Omega}}
\nc{\vx}{\vec{x}}
\nc{\mh}{\hat{m}}
\nc{\nh}{\hat{n}}
\nc{\zh}{\hat{z}}
\nc{\ph}{\hat{p}}
\nc{\A}[1]{\mathcal{A}_{#1}}
\nc{\Ogw}[1]{\Omega_{\mathrm{#1}}}
\nc{\bn}[1]{\dt\bm{t}_{\text{#1}}}
\nc{\bC}[1]{\bm{C}_{\text{#1}}}
\nc{\NTOA}{N_{\text{TOA}}}
\nc{\Nmode}{{N_{\text{mode}}}}
\nc{\ARN}{A_{\rm{RN}}}
\nc{\gRN}{\gamma_{\rm{RN}}}
\nc{\bS}{\mathbf{\Sigma}}
\nc{\br}{\mathbf{r}}
\nc{\bN}{\mathbf{R}}
\nc{\Agw}{A_\mathrm{GWB}}
\nc{\UCP}{\mathrm{UCP}}
\nc{\TT}{\mathrm{TT}}
\nc{\ST}{\mathrm{ST}}
\nc{\SL}{\mathrm{SL}}
\nc{\VL}{\mathrm{VL}}
\nc{\mH}{\mathcal{H}}
\nc{\BFST}{$107 \pm 7$}

\begin{document}
	
\title{Primordial black holes generated by the non-minimal spectator field}


\author{De-Shuang Meng}
\email{mengdeshuang@itp.ac.cn}
\affiliation{CAS Key Laboratory of Theoretical Physics,
	Institute of Theoretical Physics, Chinese Academy of Sciences,
	Beijing 100190, China}
\affiliation{School of Physical Sciences,
	University of Chinese Academy of Sciences,
	No. 19A Yuquan Road, Beijing 100049, China}
\author{Chen Yuan}
\email{Corresponding author: yuanchen@itp.ac.cn}
\affiliation{CAS Key Laboratory of Theoretical Physics,
Institute of Theoretical Physics, Chinese Academy of Sciences,
Beijing 100190, China}
\affiliation{School of Physical Sciences,
University of Chinese Academy of Sciences,
No. 19A Yuquan Road, Beijing 100049, China}

\author{Qing-Guo Huang}
\email{Corresponding author: huangqg@itp.ac.cn}
\affiliation{CAS Key Laboratory of Theoretical Physics,
Institute of Theoretical Physics, Chinese Academy of Sciences,
Beijing 100190, China}
\affiliation{School of Physical Sciences,
University of Chinese Academy of Sciences,
No. 19A Yuquan Road, Beijing 100049, China}
\affiliation{School of Fundamental Physics and Mathematical Sciences
Hangzhou Institute for Advanced Study, UCAS, Hangzhou 310024, China}

\date{\today}

\begin{abstract}

We improve and generalize the non-minimal curvaton model originally proposed in arXiv:2112.12680 to a model in which a spectator field non-minimally couples to an inflaton field and the power spectrum of the perturbation of spectator field at small scales is dramatically enhanced by the sharp feature in the form of non-minimal coupling. At or after the end of inflation, the perturbation of the spectator field is converted into curvature perturbation and leads to the formation of primordial black holes (PBHs). Furthermore, for example, we consider three phenomenological models for generating PBHs with mass function peaked at $\sim10^{-12}M_\odot$ and representing all the cold dark matter in our Universe and find that the scalar induced gravitational waves generated by the curvature perturbation can be detected by the future space-borne gravitational-wave detectors such as Taiji, TianQin and LISA.

\end{abstract}

\maketitle

\maketitle

\section{Introduction}
Primordial black holes (PBHs) formed from the collapse of large density perturbations in the very early universe \cite{Zeldovich:1967lct,Hawking:1971ei,Carr:1974nx,Carr:1975qj} can contribute to the cold dark matter (CDM) and provide a possible explanation \cite{Bird:2016dcv,Sasaki:2016jop,Clesse:2016vqa,Wang:2016ana,Ali-Haimoud:2017rtz,Chen:2018czv,Chen:2018rzo,Kavanagh:2018ggo,Raidal:2018bbj,Liu:2018ess,Chen:2019irf,Yuan:2019udt,Liu:2019rnx,DeLuca:2020sae,Vattis:2020iuz,Wang:2021iwp,Chen:2021nxo} to the gravitational-wave (GW) events from the mergers of binary black holes detected by LIGO-Virgo collaboration \cite{LIGOScientific:2016dsl,LIGOScientific:2021djp}.

Currently, various independent observations \cite{Carr:2009jm,Graham:2015apa,Niikura:2017zjd,EROS-2:2006ryy,Niikura:2019kqi,Wang:2016ana,Chen:2019irf,Brandt:2016aco,Chen:2019xse,Montero-Camacho:2019jte,Laha:2019ssq,Dasgupta:2019cae,Laha:2020ivk,Saha:2021pqf,Ray:2021mxu} have placed upper limits at the percent level on the fraction of PBHs in CDM, leaving only two mass windows $\sim[10^{-16},10^{-14}]\Msun$ and $\sim[10^{-13},10^{-12}]\Msun$ where PBHs may still constitute all the CDM. See some  recent reviews in  \cite{Carr:2020gox,Escriva:2022duf}. In order to produce a sizable amount of PBHs to explain most of the CDM, the amplitude of the power spectrum $\mathcal{P}_\zeta$ of curvature perturbation should be significantly enhanced to  $\mathcal{O}(10^{-2})$ on small scales from $\mathcal{O}(10^{-9})$  on the cosmic microwave background (CMB) scales  \cite{Planck:2018vyg}. Such an enhancement of the curvature power spectrum on small scales can be realized in many scenarios, including single-field inflation models \cite{Yokoyama:1998pt,Kinney:2005vj,Choudhury:2013woa,Garcia-Bellido:2016dkw,Cheng:2016qzb,Garcia-Bellido:2017mdw,Cheng:2018yyr,Dalianis:2018frf,Tada:2019amh,Xu:2019bdp,Mishra:2019pzq,Bhaumik:2019tvl,Liu:2020oqe,Atal:2019erb,Fu:2020lob,Vennin:2020kng,Ragavendra:2020sop,Gao:2021dfi,Cai:2022erk,Karam:2022nym,Di:2017ndc,Cai:2018tuh,Chen:2020uhe,Cai:2019jah,Cai:2019bmk,Cotner:2016cvr,Cotner:2017tir,Cotner:2018vug,Cotner:2019ykd,Escriva:2022yaf,Pi:2022zxs,Garcia-Bellido:2017mdw,Germani:2017bcs,Byrnes:2018txb,Passaglia:2018ixg,Fu:2019ttf,Fu:2019vqc,Liu:2020oqe,Fu:2020lob,Inomata:2021tpx,Tasinato:2020vdk,Ragavendra:2020sop,Cole:2022xqc,Karam:2022nym,Fu:2022ypp,Peng:2021zon,Zhai:2022mpi,Kannike:2017bxn,Gao:2018pvq,Cheong:2019vzl,Cheong:2020rao,Fu:2019ttf,Dalianis:2019vit,Fu:2019vqc,Martin:2019nuw,Martin:2020fgl,Lin:2020goi,Yi:2020cut,Gao:2020tsa,Gao:2021vxb,Wu:2021zta,Teimoori:2021thk,Kawai:2021edk,Zhang:2021rqs,Yi:2022anu,Gu:2022pbo,Cook:2022zol,Hidalgo:2022yed,Animali:2022otk,Fu:2022ypp,Papanikolaou:2022did,Braglia:2022phb,Kawaguchi:2022nku,Ashoorioon:2019xqc,Fu:2022ssq,Ahmed:2021ucx} and multi-field models \cite{Garcia-Bellido:1996mdl,Kawasaki:1997ju,Yokoyama:1995ex,Frampton:2010sw,Giovannini:2010tk,Clesse:2015wea,Inomata:2017okj,Inomata:2018cht,Espinosa:2017sgp,Kawasaki:2019hvt,Palma:2020ejf,Fumagalli:2020adf,Braglia:2020eai,Anguelova:2020nzl,Romano:2020gtn,Gundhi:2020zvb,Gundhi:2020kzm,Cai:2021wzd,Ishikawa:2021xya,Spanos:2021hpk,Hooshangi:2022lao,Chen:2019zza,Fu:2022ssq,Kohri:2012yw,Kawasaki:2012wr,Pi:2017gih,Liu:2021rgq,Pi:2021dft,Hooshangi:2022lao,Kawai:2022emp,Ashoorioon:2020hln,Ashoorioon:2022raz}, etc. 
Usually the enhancement of the power spectrum of curvature perturbation at small scales may lead to a  non-Gaussian distribution for the curvature perturbation. According to the explicit calculation of the one-loop correction to the power spectrum of curvature perturbation with local-type non-Guassianity, we conclude that the enhanced curvature perturbation for the formation of PBHs should be nearly Gaussian \cite{Meng:2022ixx};  otherwise, the power spectrum will be dominated by the one-loop correction and then the perturbed description of curvature perturbation breaks down. 
Even though we only focus on the local-type non-Gaussianity, our conclusion is expected to be qualitatively reliable for the non-local-type non-Gaussianity as well. Along this line of thought, the single-field inflation models for the formation of PBHs might have been ruled out \cite{Kristiano:2022maq,Inomata:2022yte}, or the scenario is not reliable at least. See more recent discussions in \cite{Riotto:2023hoz,Choudhury:2023vuj}.



Even though the minimally coupled multi-field inflation has been widely explored, the non-minimally coupled multi-field inflation is also attracted much attention in literature, e.g.  \cite{Lalak:2007vi,vandeBruck:2014ata,Braglia:2020fms,Cai:2021wzd}. Recently, a curvaton model with a sharp dip in the non-minimal coupling $f(\phi)$ is originally proposed in \cite{Pi:2021dft} where the perturbation of such a curvaton field is supposed to be enhanced by the inverse of $f(\phi)$ and peaked around the mode $k=k_*$ stretching outside the horizon at the dip.
In this paper, we improve and generalize the non-minimal curvaton model to more general non-minimal spectator field model. The main differences of our model from  \cite{Pi:2021dft} are: \\ 
\noindent 1) Considering that the non-minimal coupling dramatically changes the perturbation equation of the spectator field, we numerically solve the perturbation equation and find that the shape, peak location and magnification of spectator field perturbation are quite different from those given in \cite{Pi:2021dft}. \\
\noindent 2) The perturbation of the non-minimal spectator field can be amplified by not only the sharp dip proposed in \cite{Pi:2021dft}, but also some other sharp features, e.g. an oscillating feature.  \\
\noindent 3) The non-Gaussianity of curvature perturbation can significantly alter both the abundance of PBHs and the scalar induced gravitational waves (SIGWs) which are two important observables associated with PBHs. Different from \cite{Pi:2021dft} in which the non-Gaussianity can be large, we take into account the requirement of perturbativity condition \cite{Meng:2022ixx} and only focus on the nearly Gaussian curvature perturbation; otherwise, the power spectrum calculated at the tree level in our paper should not be reliable any more. 



Even though the energy density of the spectator field is subdominant during inflation, the spectator perturbation is converted into curvature perturbation at or after the end of inflation and then leads to  producing a large amount of PBHs. 
This paper is organized as follows. In Sec.~\ref{powerspectrum}, we derive the equations of motion for the background and perturbations in the non-minimal spectator model and numerically calculate the power spectrum of spectator perturbation with three phenomenological sharp features in coupling. In Sec.~\ref{PBHandSIGW}, we evaluate the PBHs mass function and calculate the corresponding SIGWs. Finally, we give a brief summary and discussion in Sec.~\ref{SD}.

\section{Power spectrum of the non-minimal spectator}
\label{powerspectrum}

The action for an inflaton field $\phi$ and a non-minimal spectator field $\chi$ takes the following form:
\e
S[\phi,\chi]=\int \mathrm{d}^4 x\sqrt{-g} \[-\frac{1}{2}(\partial \phi)^{2}-V(\phi)-\frac{1}{2} f(\phi)^{2}(\partial \chi)^{2}-\frac{1}{2} m^{2} \chi^{2}\],
\label{actionpc}
\q
where $f(\phi)$ denotes the non-minimal couple between $\phi$ and $\chi$. In the conformal coordinate system, the equations of motion in the background for these two fields are given by 
\m
\phi''+2{\cal H} \phi'+a^2V_{,\phi}&=&f f_{,\phi} \chi'^2, \\
\chi''+2\({\cal H}+{f'\over f}\)\chi'+{m^2a^2\over f^2}\chi&=&0,
\n 
where a prime denotes the derivative with respect to the conformal time $\tau$, ${\cal H}\equiv a'/ a$ is the comoving Hubble parameter during inflation, $f_{,\phi}\equiv df(\phi)/d\phi$ and $V_{,\phi}\equiv dV(\phi)/d\phi$. The inflaton field, $\phi$, slowly rolls down its potential $V(\phi)$ during inflation. Here  $m/f$ can be taken as the effective mass of spectator field $\chi$. 

From Eq.~(\ref{actionpc}), the action for the perturbations of these two fields $\delta \phi$ and $\delta \chi$ reads
\m
S^{(2)}= \int\mathrm{d}\tau\mathrm{d}^3x {1\over2}a^2\Big\{ 
&&\delta\phi'^2-(\nabla\delta\phi)^2+\[(f_{,\phi}^2+ff_{,\phi\phi})\chi'^2-a^2 V_{,\phi\phi}\]\delta\phi^2+4ff_{,\phi}\chi'\delta\phi\delta\chi'\nonumber\\
&&+f^2\[
\delta\chi'^2-(\nabla\delta\chi)^2-{m^2 a^2\over f^2} \delta\chi^2\]
\Big\}. 
\label{actionpert}
\n
In momentum space, the equations of motion for both $\delta\phi$ and $\delta\chi$ can be written as
\m\label{sigma}
\delta\phi_k''+2{\cal H} \delta\phi_k'+\[k^2+a^2V_{,\phi\phi}-(f_{,\phi}^2+ff_{,\phi\phi})\chi'^2\]\delta\phi_k-2ff_{,\phi}\chi'\delta\chi_k'&=0,\\
\delta\chi_k''+2\({\cal H}+{f'\over f}\)\delta\chi_k'+\(k^2+{m^2a^2\over f^2}\)\delta\chi_k+{2\over a^2f^2}{d\over d\tau}\(a^2ff_{,\phi}\chi'\delta\phi_k\)&=0.
\n 
Notice that $\delta\chi$ is not a canonical variable and we introduce a canonical variable $\varphi_k$ which is related to $\delta\chi_k$ by $\delta \chi_k=\varphi_k/(af)$. The equation of motion for $\varphi_k$ reads
\m\label{sigma}
\varphi_k''+\[k^2-\frac{(af)''}{af}+\frac{m^{2}a^2}{f^{2}}\]\varphi_k+\frac{2}{af}{d\over d\tau}\(a^2ff_{,\phi} \chi'\delta \phi_k\)=0.
\n
In this paper, the effective mass of spectator field is supposed to be much less than the Hubble parameter $H$ during inflation. In this sense, all the terms with $\chi'$ can be neglected and Eq.~(\ref{sigma}) can be simplified as
\m\label{sigmam}
\varphi_k''+\[k^2-\frac{(af)''}{af}\]\varphi_k=0.
\n
During inflation, the Hubble parameter $H={\cal H}/a$ is roughly a constant and $a=-1/(H\tau)$. In order to significantly enhance the power spectrum of $\delta\chi$ at small scales compared to the CMB scales, $f(\phi)$ is supposed to have a narrow feature around $\phi=\phi_*$ and $f(\phi)=1$ out of the feature. Such a sharp feature of $f(\phi)$ will dramatically affect the behavior of $\varphi_k$ around $\phi=\phi_*$. In the sub-horizon limit $(-k\tau\gg 1)$, we choose the Bunch-Davies adiabatic vacuum as the boundary condition, namely 
\e
\varphi_k (k\tau\rightarrow -\infty)\rightarrow {1\over \sqrt{2k}} e^{-ik\tau},
\q
and then fully solve Eq.~(\ref{sigmam}). The power spectrum of $\delta\chi$ is given by 
\m
{\cal P}_{\delta\chi}=\({H\over 2\pi}\)^2\times 2k^3\left| {\varphi_k(\tau)\over -f(\tau)/\tau}\right|^2_{k\tau\rightarrow 0^-}, 
\n
where ${\cal P}_{\delta\chi}$ is defined by 
\m
\lim _{k \tau \to 0^{-}}\left\langle\delta \chi_{\boldsymbol{k}}(\tau) \delta \chi_{\boldsymbol{k}^{\prime}}(\tau)\right\rangle=(2 \pi)^{3} \delta^{(3)}\left(\boldsymbol{k}+\boldsymbol{k}^{\prime}\right) \frac{2 \pi^{2}}{k^{3}} \mathcal{P}_{\delta \chi}(k).
\n
Note that the power spectrum of $\delta\chi$ is evaluated at the end of inflation. 

\begin{figure}
	\centering
	\includegraphics[width=0.7\columnwidth]{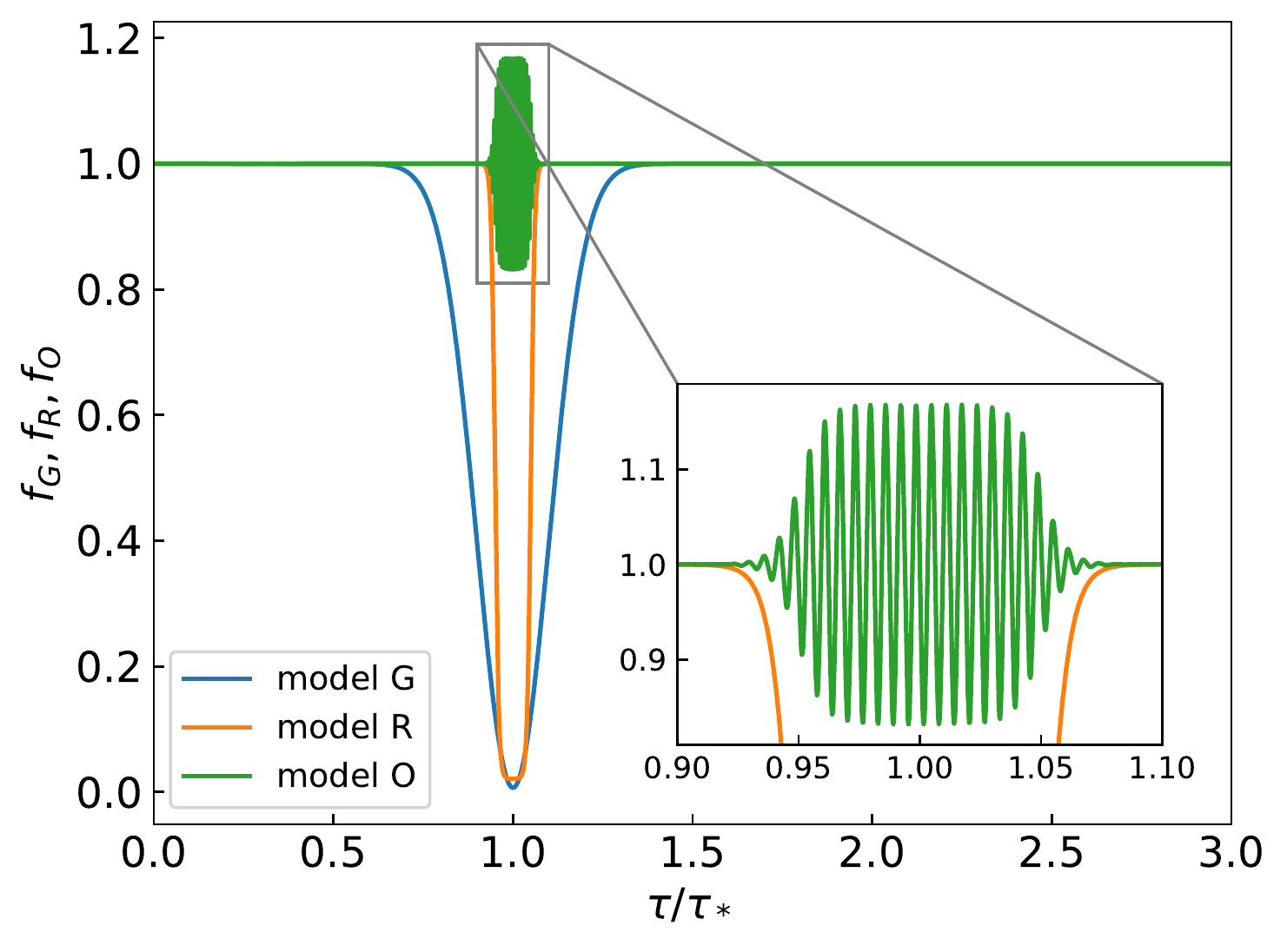}
	\caption{The non-minimal coupling $f(\tau/\tau_*)$ in Eqs.~(\ref{modelG}), (\ref{modelR}) and (\ref{modelO}). Here we set $\Delta=0.1$, $\Lambda=0.01$, $\xi=0.001$ and the values of $A_\text{G}$, $A_\text{R}$ and $A_\text{O}$ are chosen for PBHs making up all of the CDM.}
	\label{fplot}
\end{figure}

\begin{figure}
	\centering
	\includegraphics[width=0.7\columnwidth]{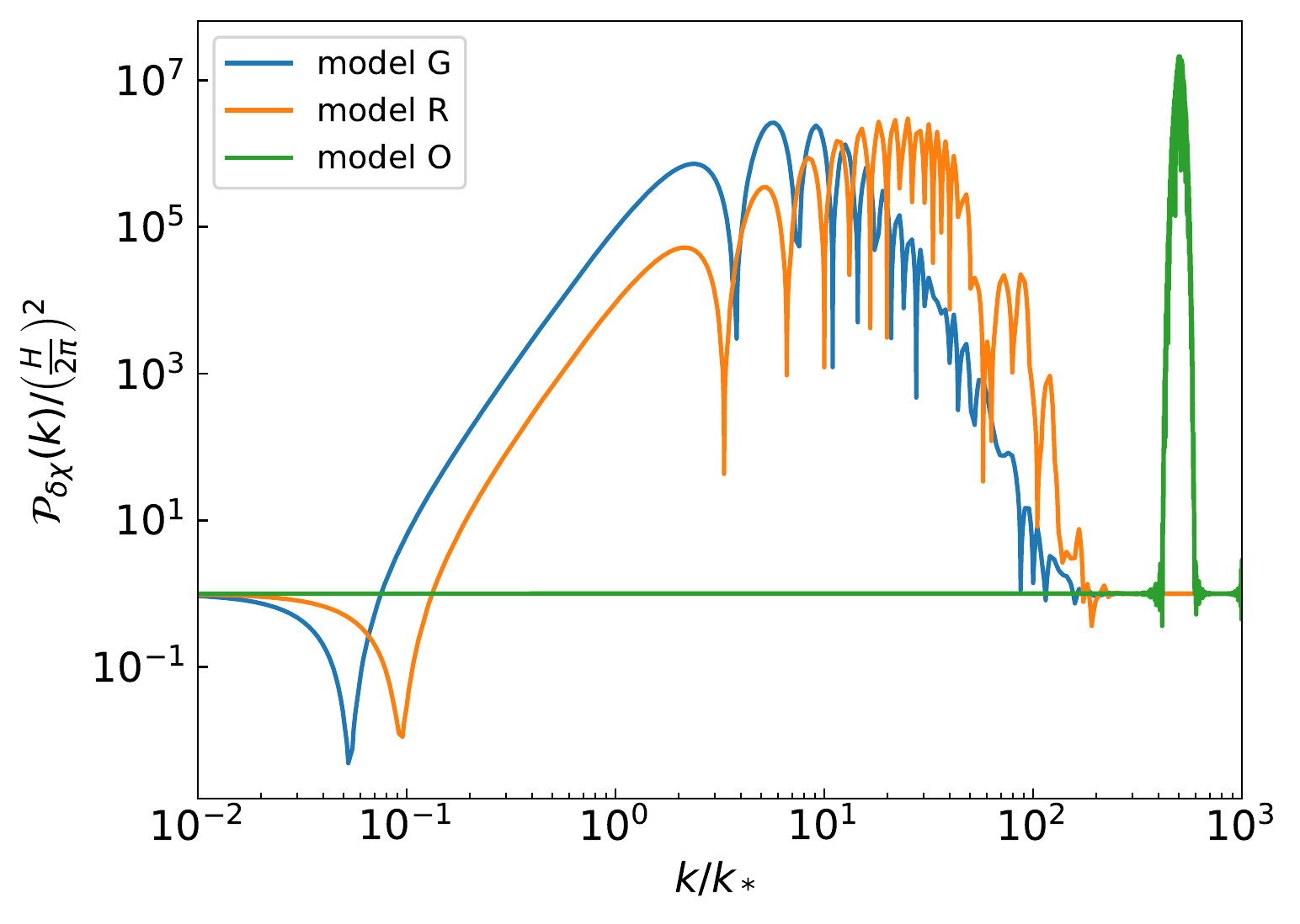}
	\caption{The power spectra of $\delta \chi$ for the three models corresponding to the features shown in  \Fig{fplot}.}
	\label{Pchi}
\end{figure}

To illustrate the effects of $f(\phi)$ around $\phi_*$, we consider three phenomenological forms (denoted by model G, model R and model O) of $f(\phi)$ as follows 
\m
f_\text{G}(\phi)=1-A_\text{G} \exp\[{-{(\phi-\phi_*)^2\over 2\Delta_\phi^2}}\],
\n
\m
f_\text{R}(\phi)=1-{A_\text{R}\over 2}\[\text{Tanh}{\phi-(\phi_*-\Delta_\phi/2)\over \Lambda_\phi}-\text{Tanh}{\phi-(\phi_*+\Delta_\phi/2)\over \Lambda_\phi}\], 
\n
\m
f_\text{O}(\phi)=1-{A_\text{O}\over 2}\[\text{Tanh}{\phi-(\phi_*-\Delta_\phi/2)\over \Lambda_\phi}-\text{Tanh}{\phi-(\phi_*+\Delta_\phi/2)\over \Lambda_\phi}\]\sin {\phi-\phi_*\over \xi_\phi}, 
\n
where $A_\text{G}$, $A_\text{R}$ and $A_\text{O}$ denote the sizes of the features for these three models. The evolution of $\phi$ around $\phi_*$ is approximately given by $\phi(\tau)\simeq \phi_*+\phi_*'(\tau-\tau_*)$, where $\phi_*'$ is the velocity of $\phi$ at the conformal time $\tau_*$ when $\phi=\phi_*$, and then 
\m\label{modelG}
f_\text{G}(x)&=&1-A_\text{G} \exp\[{-{(x-1)^2\over 2\Delta^2}}\], 
\n
\m\label{modelR}
f_\text{R}(x)&=&1-{A_\text{R}\over 2}\[\text{Tanh}{x-(1-\Delta/2) \over \Lambda}-\text{Tanh}{x-(1+\Delta/2)\over \Lambda}\],
\n
\m\label{modelO}
f_\text{O}(x)&=&1-{A_\text{O}\over 2}\[\text{Tanh}{x-(1-\Delta/2) \over \Lambda}-\text{Tanh}{x-(1+\Delta/2)\over \Lambda}\]\sin {x-1\over \xi},
\n
where $x\equiv \tau/\tau_*$ is the dimensionless conformal time,  $\Delta\equiv \Delta_\phi/(\phi_*' \tau_*)$ is a constant characterizing the width of the feature in $f(\phi)$,  $\Lambda\equiv \Lambda_\phi/(\phi_*' \tau_*)$ and $\xi\equiv \xi_\phi/(\phi_*'\tau_*)$. Without loss of generality, we assume  $\phi_*'>0$. In this paper, we take  $\Delta=0.1$, $\Lambda=0.01$, $\xi=0.001$ and $A_\text{G}$, $A_\text{R}$ and $A_\text{O}$ are chosen for  PBHs consisting all of the CDM whose mass function is peaking at $10^{-12}\Msun$. The coupling $f(x)$ for these three models are shown in \Fig{fplot} and our numerical results for the power spectrum of  $\delta\chi$ are illustrated in \Fig{Pchi}, where $k_*=a(\tau_*)H$ is the perturbation mode stretching outside the horizon at the time of $\tau_*$. 
From \Fig{Pchi}, the power spectra of $\delta \chi$ in the model G and R with a dip reach the maximum values at  $\mathcal{O}(10)k_*$, not $k_*$, and the maximum magnifications are much larger than $1/f^2(\phi_*)$. For model O, our numerical results indicate that the peak of the power spectrum of $\delta\chi$ is roughly located at $k=k_*/(2\xi)$ and the maximum magnification is also sensitive to the value of $\xi$. 
Actually the growth of the perturbation of non-minimal spectator field is due to the parametric resonance for the  model O. 
For simplicity, we  focus on the  perturbation modes relevant to the oscillating feature, and the non-minimal coupling is roughly given by 
\begin{equation}
f_\text{O}(x)\simeq 1-A_\text{O} \sin {x-1\over \xi},    
\end{equation}
where $A_\text{O}\ll 1$. The equation of motion for $\varphi_{k}$ reads
\begin{equation}
    {\mathrm{d} ^2\varphi_{k}\over\mathrm{d}x^2} +\left[\left(\frac{k}{k_*} \right)^{2}-\frac{1}{a f}{\mathrm{d} ^2(af)\over\mathrm{d}x^2}\right] \varphi_{k}=0,
\label{varphimodelO}
\end{equation}
where $k_*=-1/\tau_*$ and 
\begin{equation}
    \frac{1}{a f}{\mathrm{d}^2(af)\over\mathrm{d}x^2}=\frac{2}{x^{2}}+\frac{2 A_\text{O}\cos\left(\frac{x-1}{\xi}\right)}{x\xi}+\frac{A_\text{O} \sin\left(\frac{x-1}{\xi}\right)}{\xi^{2}}+\mathcal{O}(A_\text{O}^{2}).
\end{equation}
For the modes deep inside the horizon, the first two terms on the right hand side of the above equation are negligible. Introducing a new coordinate $y=x/(2\xi)$, we can re-write Eq.~(\ref{varphimodelO}) in the following form 
\begin{equation}
  {\mathrm{d} ^2\varphi_{k}\over\mathrm{d}y^2}+\left[B_k-2q\cos\left(2y\right)\right]\varphi_{k}=0,
\end{equation}
where $B_k=\left(2\xi k/k_* \right)^2$,  $q={A_\text{O}/2}$. Here we can neglect the phase in the cosine function as long as the oscillation period is much shorter than the time scale of the feature in the non-minimal coupling. The resonance bands are located in narrow ranges around $B_k\sim n$ where $n$ is a positive integer and the first one corresponding to $k\simeq k_*/(2\xi)$ is mostly enhanced. It is consistent with our numerical results.


\section{Formation of PBHs and scalar induced gravitational waves}
\label{PBHandSIGW}

The energy density of the spectator field is subdominant and therefore the perturbation of the spectator field does not contribute to the curvature perturbation during inflation. However, its perturbation can be converted into curvature perturbation at or after the end of inflation, such as in the model with nontrivial reheating field space surface \cite{Sasaki:2008uc,Huang:2009vk}, modulated reheating model \cite{Suyama:2007bg,Dvali:2003em,Kofman:2003nx}, curvaton mechanism \cite{Mollerach:1989hu,Linde:1996gt,Enqvist:2001zp,Lyth:2001nq,Moroi:2001ct,Sasaki:2006kq,Enqvist:2005pg,Huang:2008bg,Huang:2008zj,Chingangbam:2009xi,Chingangbam:2010xn,Kawasaki:2011pd} and so on, and may generate the local-type non-Gaussianity of curvature perturbation. Here we need to stress that the perturbativity condition \cite{Kristiano:2021urj,Meng:2022ixx} requires that such a curvature perturbation should be nearly Gaussian if the PBHs consist of most of the CDM in our Universe.

In this paper, we focus on the model in which the curvature perturbation is mainly produced by the spectator field. 
In the model with a nontrivial reheating surface in field space, the power spectrum of curvature perturbation generated by the spectator field is 
\e
{\cal P}_\zeta=\cot^2\theta \({{\cal H}\over \phi'}\)^2 {\cal P}_{\delta \chi},
\q
where $\theta$ is the angle between the reheating surface and the inflaton trajectory. In this scenario, the local-type non-Gaussianity is small as long as the reheating surface is a straight line \cite{Huang:2009vk}. 
In the modulated reheating model, the decay rate of the inflaton field is related to the expectation value of $\chi$ by $\Gamma_\phi=\Gamma_\phi (\chi)$. The power spectrum of curvature perturbation is 
\e
{\cal P}_\zeta=\alpha^2 \({\Gamma_{,\chi}\over \Gamma}\)^2 {\cal P}_{\delta \chi},
\q
where $\alpha$ is a parameter depending on the ratio of $\Gamma$ to the Hubble parameter at the end of inflation. The local-type non-Gaussianity can be also small, for example, if the decay rate $\Gamma_\phi(\chi)$ linearly depends on $\chi$.
In the curvaton scenario, for the curvaton field with quadratic potential, the curvaton linearly evolves after the end of inflation and then the poewer spectrum of curvature perturbation is 
\e
{\cal P}_\zeta={4\over 9} {f_D^2\over \chi^2} {\cal P}_{\delta \chi},
\q
where 
\e
f_D=\left. {3\rho_\chi\over 3\rho_\chi+4\rho_R} \right|_{t_\text{decay}},
\q
$\rho_\chi$ and $\rho_R$ are the energy density of curvaton field and radiation at time of curvaton decay. In the curvaton model with quadratic potential, the non-Gaussianity parameter $f_\text{NL}$ is, \cite{Sasaki:2006kq}, 
\e
f_\text{NL}={5\over 4f_D}-{5\over 3}-{5f_D\over 6}. 
\q
The smallness of non-Gaussianity due to the perturbativity condition \cite{Meng:2022ixx} yields $f_D\simeq 1$, implying that the curvaton field becomes dominant when it decays.

\begin{figure}
	\centering
	\includegraphics[width=0.7\columnwidth]{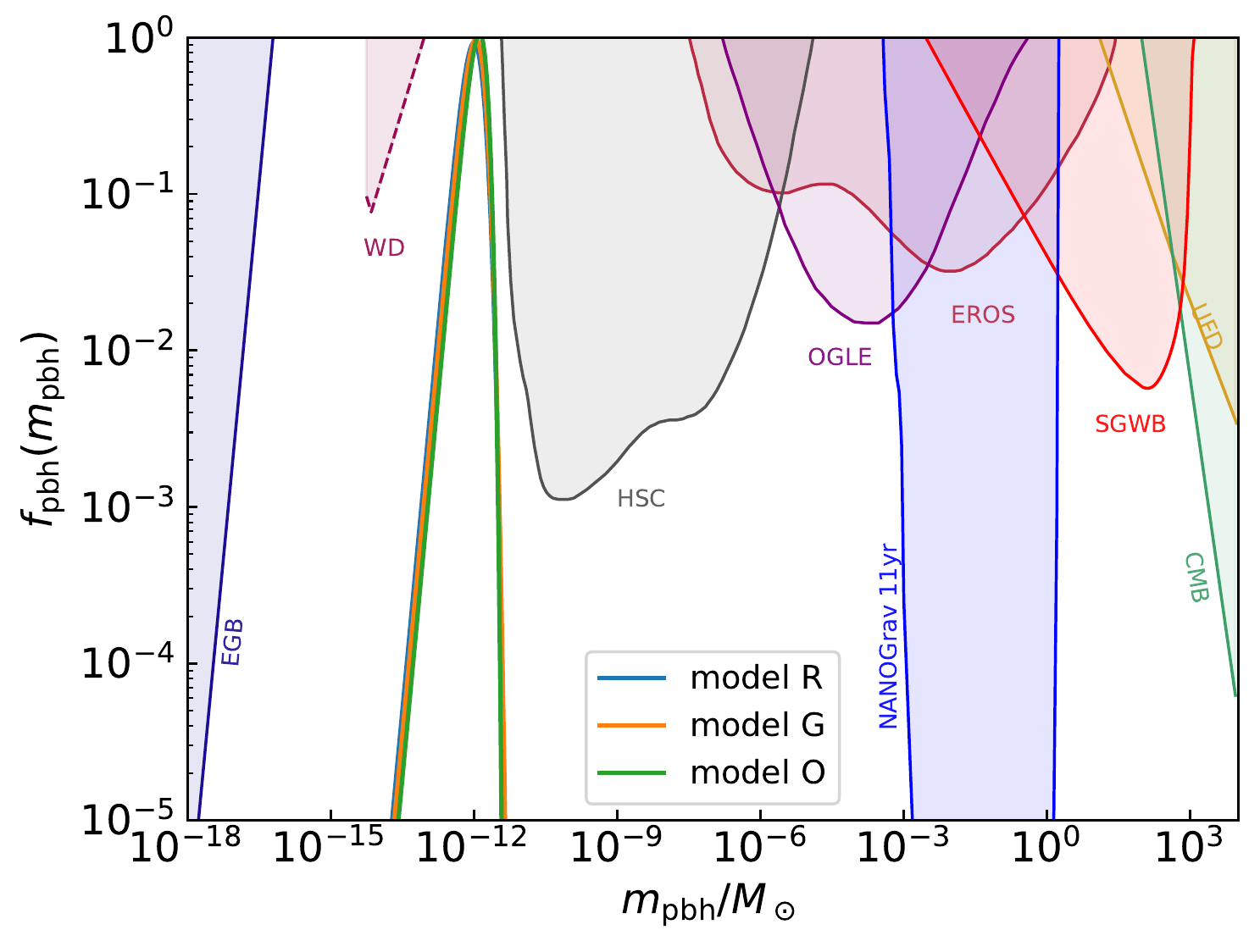}
	\caption{The fraction of PBHs, $\fpbh(\mpbh)$, for the three models corresponding to the features shown in  \Fig{fplot}. Here the PBHs make up all of the CDM. We also plot an overview of the current observational constraints on $\fpbh$, including extra-galactic Gamma-ray background (EGB) \cite{Carr:2009jm}, white dwarfs (WD) \cite{Graham:2015apa} (note that this constraint might be relaxed, see \cite{Montero-Camacho:2019jte}), Subaru/HSC \cite{Niikura:2017zjd}, EROS/MACHO \cite{EROS-2:2006ryy}, OGLE \cite{Niikura:2019kqi}, stochastic gravitational wave background (SGWB) from binary PBHs \cite{Wang:2016ana,Chen:2019irf}, ultra-faint dwarf galaxies (UFD) \cite{Brandt:2016aco} and NANOGrav \cite{Chen:2019xse}.}
	\label{fpbhplot}
\end{figure}

The mass function of the PBHs at the formation time, $\beta(\mpbh)$, can be estimated using the Press-Schechter formalism \cite{Press:1973iz}, namely by integrating the probability distribution function (PDF) of the density contrast $P(\delta)$ over the region $\delta>\delta_c$,
\m\label{betam}
\beta(\mpbh)=\int_{\delta_{c}}^{\infty} d \delta \frac{\mpbh}{M_{H}} P(\delta),
\n
where $\delta_c\sim0.41$ is the critical value to form a single PBH \cite{Harada:2013epa}. 
The horizon mass $M_H$ is related to the comoving wavelength by
\e
M_{H} \simeq 17\left(\frac{g}{10.75}\right)^{-1 / 6}\left(\frac{k}{10^{6} \mathrm{Mpc}^{-1}}\right)^{-2} M_{\odot},
\q
and $g$ is the degress of freedom of relativistic particles at the formation time. The PDF, $P(\delta)$, takes the form of 
\m
P(\delta)=\frac{1}{\sqrt{2 \pi \sigma_{k}^{2}}} \exp \left(-\frac{\delta^{2}}{2 \sigma_{k}^{2}}\right),
\n
with $\sigma_{k}^{2}(R)$ being smoothed variance of the density contrast on a comoving scale $R_H$:
\m
\sigma_{k}^{2}=\left(\frac{4}{9}\right)^{2} \int_{0}^{\infty} \frac{d q}{q} W^{2}\left(q, R_{H}\right)\left(\frac{q}{k}\right)^{4} T^{2}\left(q, R_{H}\right) \mathcal{P}_{\zeta}(q).
\n
Here, $W(k,R_H)$ is the window function which we adopt a top-hat window function in real space, namely
\e
W\left(k, R_{H}\right)=3 \frac{\sin \left(k R_{H}\right)-\left(k R_{H}\right) \cos \left(k R_{H}\right)}{\left(k R_{H}\right)^{3}}.
\q
$T(q, R_{H})$ denotes the transfer function during radiation dominated era which takes the form
\begin{equation}
T\left(k, R_H\right)=3 \frac{\sin \left(k R_H / \sqrt{3}\right)-\left(k R_H / \sqrt{3}\right) \cos \left(k R_H / \sqrt{3}\right)}{\left(k R_H / \sqrt{3}\right)^3}
\end{equation}
The mass of PBH in Eq.~(\ref{betam}) is realted to $\delta$ by $\mpbh=M_{H} \kappa\left(\delta_{m}-\delta_{c}\right)^{\gamma}$ \cite{Choptuik:1992jv,Evans:1994pj,Niemeyer:1997mt}, where $\kappa=3.3$, $\gamma=0.36$ \cite{Koike:1995jm} and $\delta_m=\delta-3/8\delta^2$ accounts for nonlinear effects \cite{Young:2019yug,DeLuca:2019qsy,Kawasaki:2019mbl}.
The relation between the fraction of PBHs in the CDM at present, $\fpbh(\mpbh)$, and $\beta(\mpbh)$ can be written as
\m
f_{\mathrm{pbh}}(\mpbh)=\frac{1}{\Omega_{\mathrm{CDM}}}\left(\frac{M_{\mathrm{eq}}}{\mpbh}\right)^{1 / 2} \beta(\mpbh),
\n
where we use the convention such that $\fpbh\equiv{\Omega_{\mathrm{PBH}}/\Omega_{\mathrm{CDM}}}=\int \fpbh (\mpbh)\mathrm{d}\ln \mpbh$ and $M_{\mathrm{eq}}=2.8\times 10^{17}\Msun$.  The numeric results of $\fpbh(\mpbh)$ are shown in Fig.~\ref{fpbhplot}. Although various independent constraints on $\fpbh$ have excluded PBHs in the mass range $\sim [10^{-18},10^3]\Msun$ to a percent level, PBHs are still able to represent all the CDM within $\sim[10^{-16},10^{-14}]\Msun$ and $\sim[10^{-13},10^{-12}]\Msun$. We choose $k_*=7\times10^{11}\mathrm{Mpc^{-1}}$, $k_*=2\times10^{11}\mathrm{Mpc^{-1}}$ and $k_*=9\times10^{9}\mathrm{Mpc^{-1}}$ in model G, model R and model O respectively, and then all these models can generate sufficient PBHs peaked at $\sim 10^{-12}\Msun$ that can represent all the CDM. It can be seen that the $\fpbh(\mpbh)$ for these three models are all compatible with current observational constraints.

\begin{figure}
	\centering
	\includegraphics[width=0.7\columnwidth]{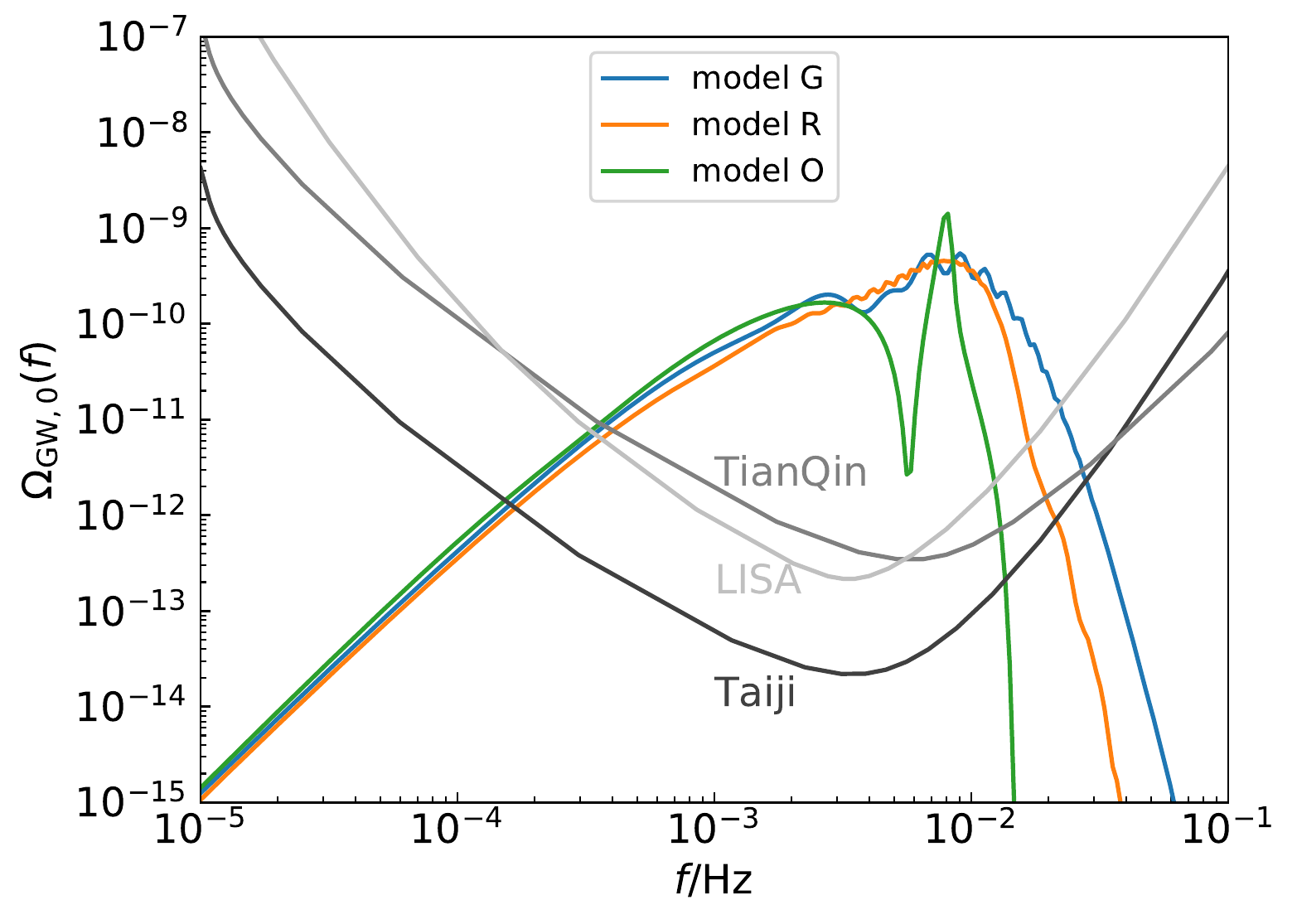}
	\caption{The energy spectrum of SIGWs for the three models corresponding to the features shown in  \Fig{fplot}. Here black, dark grey and light grey curves correspond to the power-law integrated sensitivity curves for Taiji \cite{Hu:17}, TianQin \cite{TianQin:2015yph} and LISA \cite{Audley:2017drz} respectively, assuming a four year detection.}
	\label{SIGW}
\end{figure}

In addition, the linear scalar perturbations would source second-order tensor perturbations during the radiation dominated era, also dubbed as SIGWs
\cite{tomita1967non,Matarrese:1992rp,Matarrese:1993zf,Matarrese:1997ay,Noh:2004bc,Carbone:2004iv,Nakamura:2004rm}. SIGWs were inevitably generated during the formation of PBHs and can be a powerful tool to search or constrain PBHs \cite{Ananda:2006af,Baumann:2007zm,Saito:2008jc,Arroja:2009sh,Assadullahi:2009jc,Bugaev:2009kq,Bugaev:2009zh,Saito:2009jt,Bugaev:2010bb,Alabidi:2013lya,Nakama:2016enz,Nakama:2016gzw,Inomata:2016rbd,Orlofsky:2016vbd,Garcia-Bellido:2017aan,Sasaki:2018dmp,Espinosa:2018eve,Kohri:2018awv,Cai:2018dig,Bartolo:2018evs,Bartolo:2018rku,Unal:2018yaa,Byrnes:2018txb,Inomata:2018epa,Clesse:2018ogk,Cai:2019amo,Inomata:2019zqy,Inomata:2019ivs,Cai:2019elf,Yuan:2019udt,Cai:2019cdl,Lu:2019sti,Yuan:2019wwo,Tomikawa:2019tvi,DeLuca:2019ufz,Yuan:2019fwv,Inomata:2020tkl,Inomata:2020yqv,Inomata:2020lmk,Yuan:2020iwf,Papanikolaou:2020qtd,Zhang:2020ptw,Kapadia:2020pnr,Zhang:2020uek,Domenech:2020ssp,Dalianis:2020gup,Atal:2021jyo,Chen:2021nxo,Franciolini:2021nvv,Witkowski:2021raz,Balaji:2022dbi,Cang:2022oia,Gehrman:2022imk,Braglia:2020taf,Papanikolaou:2022chm}. For review of SIGW, see \cite{Yuan:2021qgz,Domenech:2021ztg}.
The superposition of SIGWs all over the sky will form a stochastic gravitational wave background whose energy spectrum is defined as the energy of GWs per logarithm frequency normalized by the critical energy. The energy spectrum of SIGWs at equality time can be evaluated semi-analytically as \cite{Kohri:2018awv}
\m
\Omega_{\mathrm{GW}}(k)=\frac{1}{6} \int_{0}^{\infty} \mathrm{d} u \int_{|1-u|}^{1+u} \mathrm{~d} v \frac{v^{2}}{u^{2}}\left[1-\left(\frac{1+v^{2}-u^{2}}{2 v}\right)^{2}\right]^{2} \mathcal{P}_{\zeta}(u k)  \mathcal{P}_{\zeta}(v k) \overline{I^{2}(u, v, x \rightarrow \infty)},
\n
where the oscillating average of the square of the kernel function in the sub-horizon limit is given by \cite{Kohri:2018awv}
\m
\overline{I^{2}}=\frac{9\left(u^{2}+v^{2}-3\)^{2}}{32 u^{6} v^{6}}\Bigg\{\[-4 u v+\left(u^{2}+v^{2}-3\right) \ln \left|\frac{3-(u+v)^{2}}{3-(u-v)^{2}}\right|\]^{2} +\pi^{2}\left(u^{2}+v^{2}-3\right)^{2} \Theta(u+v-\sqrt{3})\Bigg\}.
\n
The energy spectrum of SGIWs at present, $\Omega_{\mathrm{GW,0}}(k)$, is evaluated by $\Omega_{\mathrm{GW,0}}(k)=\Omega_r\Omega_{\mathrm{GW}}(k)$, where $\Omega_r$ is the density parameter of radiation by today. The results of $\Omega_{\mathrm{GW,0}}(f)$ are shown in Fig.~\ref{SIGW}.
Here, the wavelength $k$ is realted to the frequency by $k=2\pi f$ in the $c=1$ unit. It can be seen that the SIGWs accompanying the formation of $\sim 10^{-12}\Msun$ PBHs generated by model G, model R and model O can be detected by the future space-borne GW detectors such as Taiji \cite{Hu:17}, TianQin \cite{TianQin:2015yph} and LISA \cite{Audley:2017drz}.

\section{Summary and Discussion}
\label{SD}

In this paper, we improve and generalize the non-minimal curvaton model with a sharp dip \cite{Pi:2021dft} to a more general non-minimal spectator model for the formation of PBHs. Since the coupling between the inflaton and spectator fields is supposed to have a sharp feature controlled by the expectation value of the inflaton field, the power spectrum of the perturbation of the spectator field for some perturbation modes stretching outside the horizon when the inflaton field rolls in the region of the feature are significantly amplified. At or after the end of inflation, the perturbation of the spectator field is converted into the curvature perturbation and the enhanced curvature perturbation at small scales due to the feature in the coupling between spectator and inflaton fields leads to the formation of PBHs. Our model can produce a sizeable amount of PBHs peaked at $\sim 10^{-12}\Msun$ for consisting of all the CDM and the SIGWs can be detected by future space-borne GW detectors such as Taiji, TianQin and LISA.

\vspace{5mm}
{\it Acknowledgments. } 
This work is supported by the National Key Research and Development Program of China Grant No.2020YFC2201502, grants from NSFC (grant No. 11975019, 11991052, 12047503), Key Research Program of Frontier Sciences, CAS, Grant NO. ZDBS-LY-7009, CAS Project for Young Scientists in Basic Research YSBR-006, the Key Research Program of the Chinese Academy of Sciences (Grant NO. XDPB15). 
We acknowledge the use of HPC Cluster of ITP-CAS and  \texttt{GWSC.jl} package (https://github.com/bingining/gwsc.jl) for plotting the power-law integrated sensitivity curves of Taiji, TianQin and LISA.

\bibliography{./ref}

\begin{thebibliography}{236}%
\makeatletter
\providecommand \@ifxundefined [1]{%
 \@ifx{#1\undefined}
}%
\providecommand \@ifnum [1]{%
 \ifnum #1\expandafter \@firstoftwo
 \else \expandafter \@secondoftwo
 \fi
}%
\providecommand \@ifx [1]{%
 \ifx #1\expandafter \@firstoftwo
 \else \expandafter \@secondoftwo
 \fi
}%
\providecommand \natexlab [1]{#1}%
\providecommand \enquote  [1]{``#1''}%
\providecommand \bibnamefont  [1]{#1}%
\providecommand \bibfnamefont [1]{#1}%
\providecommand \citenamefont [1]{#1}%
\providecommand \href@noop [0]{\@secondoftwo}%
\providecommand \href [0]{\begingroup \@sanitize@url \@href}%
\providecommand \@href[1]{\@@startlink{#1}\@@href}%
\providecommand \@@href[1]{\endgroup#1\@@endlink}%
\providecommand \@sanitize@url [0]{\catcode `\\12\catcode `\$12\catcode
  `\&12\catcode `\#12\catcode `\^12\catcode `\_12\catcode `\%12\relax}%
\providecommand \@@startlink[1]{}%
\providecommand \@@endlink[0]{}%
\providecommand \url  [0]{\begingroup\@sanitize@url \@url }%
\providecommand \@url [1]{\endgroup\@href {#1}{\urlprefix }}%
\providecommand \urlprefix  [0]{URL }%
\providecommand \Eprint [0]{\href }%
\providecommand \doibase [0]{http://dx.doi.org/}%
\providecommand \selectlanguage [0]{\@gobble}%
\providecommand \bibinfo  [0]{\@secondoftwo}%
\providecommand \bibfield  [0]{\@secondoftwo}%
\providecommand \translation [1]{[#1]}%
\providecommand \BibitemOpen [0]{}%
\providecommand \bibitemStop [0]{}%
\providecommand \bibitemNoStop [0]{.\EOS\space}%
\providecommand \EOS [0]{\spacefactor3000\relax}%
\providecommand \BibitemShut  [1]{\csname bibitem#1\endcsname}%
\let\auto@bib@innerbib\@empty
\bibitem [{\citenamefont {Zel'dovich}\ and\ \citenamefont
  {Novikov}(1967)}]{Zeldovich:1967lct}%
  \BibitemOpen
  \bibfield  {author} {\bibinfo {author} {\bibfnamefont {Ya.~B.}\ \bibnamefont
  {Zel'dovich}}\ and\ \bibinfo {author} {\bibfnamefont {I.~D.}\ \bibnamefont
  {Novikov}},\ }\bibfield  {title} {\enquote {\bibinfo {title} {{The Hypothesis
  of Cores Retarded during Expansion and the Hot Cosmological Model}},}\
  }\href@noop {} {\bibfield  {journal} {\bibinfo  {journal} {Soviet Astron. AJ
  (Engl. Transl. ),}\ }\textbf {\bibinfo {volume} {10}},\ \bibinfo {pages}
  {602} (\bibinfo {year} {1967})}\BibitemShut {NoStop}%
\bibitem [{\citenamefont {Hawking}(1971)}]{Hawking:1971ei}%
  \BibitemOpen
  \bibfield  {author} {\bibinfo {author} {\bibfnamefont {Stephen}\ \bibnamefont
  {Hawking}},\ }\bibfield  {title} {\enquote {\bibinfo {title}
  {{Gravitationally collapsed objects of very low mass}},}\ }\href@noop {}
  {\bibfield  {journal} {\bibinfo  {journal} {Mon. Not. Roy. Astron. Soc.}\
  }\textbf {\bibinfo {volume} {152}},\ \bibinfo {pages} {75} (\bibinfo {year}
  {1971})}\BibitemShut {NoStop}%
\bibitem [{\citenamefont {Carr}\ and\ \citenamefont
  {Hawking}(1974)}]{Carr:1974nx}%
  \BibitemOpen
  \bibfield  {author} {\bibinfo {author} {\bibfnamefont {Bernard~J.}\
  \bibnamefont {Carr}}\ and\ \bibinfo {author} {\bibfnamefont {S.~W.}\
  \bibnamefont {Hawking}},\ }\bibfield  {title} {\enquote {\bibinfo {title}
  {{Black holes in the early Universe}},}\ }\href {\doibase
  10.1093/mnras/168.2.399} {\bibfield  {journal} {\bibinfo  {journal} {Mon.
  Not. Roy. Astron. Soc.}\ }\textbf {\bibinfo {volume} {168}},\ \bibinfo
  {pages} {399--415} (\bibinfo {year} {1974})}\BibitemShut {NoStop}%
\bibitem [{\citenamefont {Carr}(1975)}]{Carr:1975qj}%
  \BibitemOpen
  \bibfield  {author} {\bibinfo {author} {\bibfnamefont {Bernard~J.}\
  \bibnamefont {Carr}},\ }\bibfield  {title} {\enquote {\bibinfo {title} {{The
  Primordial black hole mass spectrum}},}\ }\href {\doibase 10.1086/153853}
  {\bibfield  {journal} {\bibinfo  {journal} {Astrophys. J.}\ }\textbf
  {\bibinfo {volume} {201}},\ \bibinfo {pages} {1--19} (\bibinfo {year}
  {1975})}\BibitemShut {NoStop}%
\bibitem [{\citenamefont {Bird}\ \emph {et~al.}(2016)\citenamefont {Bird},
  \citenamefont {Cholis}, \citenamefont {Muñoz}, \citenamefont {Ali-Haïmoud},
  \citenamefont {Kamionkowski}, \citenamefont {Kovetz}, \citenamefont
  {Raccanelli},\ and\ \citenamefont {Riess}}]{Bird:2016dcv}%
  \BibitemOpen
  \bibfield  {author} {\bibinfo {author} {\bibfnamefont {Simeon}\ \bibnamefont
  {Bird}}, \bibinfo {author} {\bibfnamefont {Ilias}\ \bibnamefont {Cholis}},
  \bibinfo {author} {\bibfnamefont {Julian~B.}\ \bibnamefont {Muñoz}},
  \bibinfo {author} {\bibfnamefont {Yacine}\ \bibnamefont {Ali-Haïmoud}},
  \bibinfo {author} {\bibfnamefont {Marc}\ \bibnamefont {Kamionkowski}},
  \bibinfo {author} {\bibfnamefont {Ely~D.}\ \bibnamefont {Kovetz}}, \bibinfo
  {author} {\bibfnamefont {Alvise}\ \bibnamefont {Raccanelli}}, \ and\ \bibinfo
  {author} {\bibfnamefont {Adam~G.}\ \bibnamefont {Riess}},\ }\bibfield
  {title} {\enquote {\bibinfo {title} {{Did LIGO detect dark matter?}}}\ }\href
  {\doibase 10.1103/PhysRevLett.116.201301} {\bibfield  {journal} {\bibinfo
  {journal} {Phys. Rev. Lett.}\ }\textbf {\bibinfo {volume} {116}},\ \bibinfo
  {pages} {201301} (\bibinfo {year} {2016})},\ \Eprint
  {http://arxiv.org/abs/1603.00464} {arXiv:1603.00464 [astro-ph.CO]}
  \BibitemShut {NoStop}%
\bibitem [{\citenamefont {Sasaki}\ \emph {et~al.}(2016)\citenamefont {Sasaki},
  \citenamefont {Suyama}, \citenamefont {Tanaka},\ and\ \citenamefont
  {Yokoyama}}]{Sasaki:2016jop}%
  \BibitemOpen
  \bibfield  {author} {\bibinfo {author} {\bibfnamefont {Misao}\ \bibnamefont
  {Sasaki}}, \bibinfo {author} {\bibfnamefont {Teruaki}\ \bibnamefont
  {Suyama}}, \bibinfo {author} {\bibfnamefont {Takahiro}\ \bibnamefont
  {Tanaka}}, \ and\ \bibinfo {author} {\bibfnamefont {Shuichiro}\ \bibnamefont
  {Yokoyama}},\ }\bibfield  {title} {\enquote {\bibinfo {title} {{Primordial
  Black Hole Scenario for the Gravitational-Wave Event GW150914}},}\ }\href
  {\doibase 10.1103/PhysRevLett.117.061101} {\bibfield  {journal} {\bibinfo
  {journal} {Phys. Rev. Lett.}\ }\textbf {\bibinfo {volume} {117}},\ \bibinfo
  {pages} {061101} (\bibinfo {year} {2016})},\ \bibinfo {note} {[Erratum:
  Phys.Rev.Lett. 121, 059901 (2018)]},\ \Eprint
  {http://arxiv.org/abs/1603.08338} {arXiv:1603.08338 [astro-ph.CO]}
  \BibitemShut {NoStop}%
\bibitem [{\citenamefont {Clesse}\ and\ \citenamefont
  {García-Bellido}(2017)}]{Clesse:2016vqa}%
  \BibitemOpen
  \bibfield  {author} {\bibinfo {author} {\bibfnamefont {Sebastien}\
  \bibnamefont {Clesse}}\ and\ \bibinfo {author} {\bibfnamefont {Juan}\
  \bibnamefont {García-Bellido}},\ }\bibfield  {title} {\enquote {\bibinfo
  {title} {{The clustering of massive Primordial Black Holes as Dark Matter:
  measuring their mass distribution with Advanced LIGO}},}\ }\href {\doibase
  10.1016/j.dark.2016.10.002} {\bibfield  {journal} {\bibinfo  {journal} {Phys.
  Dark Univ.}\ }\textbf {\bibinfo {volume} {15}},\ \bibinfo {pages} {142--147}
  (\bibinfo {year} {2017})},\ \Eprint {http://arxiv.org/abs/1603.05234}
  {arXiv:1603.05234 [astro-ph.CO]} \BibitemShut {NoStop}%
\bibitem [{\citenamefont {Wang}\ \emph {et~al.}(2018)\citenamefont {Wang},
  \citenamefont {Wang}, \citenamefont {Huang},\ and\ \citenamefont
  {Li}}]{Wang:2016ana}%
  \BibitemOpen
  \bibfield  {author} {\bibinfo {author} {\bibfnamefont {Sai}\ \bibnamefont
  {Wang}}, \bibinfo {author} {\bibfnamefont {Yi-Fan}\ \bibnamefont {Wang}},
  \bibinfo {author} {\bibfnamefont {Qing-Guo}\ \bibnamefont {Huang}}, \ and\
  \bibinfo {author} {\bibfnamefont {Tjonnie G.~F.}\ \bibnamefont {Li}},\
  }\bibfield  {title} {\enquote {\bibinfo {title} {{Constraints on the
  Primordial Black Hole Abundance from the First Advanced LIGO Observation Run
  Using the Stochastic Gravitational-Wave Background}},}\ }\href {\doibase
  10.1103/PhysRevLett.120.191102} {\bibfield  {journal} {\bibinfo  {journal}
  {Phys. Rev. Lett.}\ }\textbf {\bibinfo {volume} {120}},\ \bibinfo {pages}
  {191102} (\bibinfo {year} {2018})},\ \Eprint
  {http://arxiv.org/abs/1610.08725} {arXiv:1610.08725 [astro-ph.CO]}
  \BibitemShut {NoStop}%
\bibitem [{\citenamefont {Ali-Haïmoud}\ \emph {et~al.}(2017)\citenamefont
  {Ali-Haïmoud}, \citenamefont {Kovetz},\ and\ \citenamefont
  {Kamionkowski}}]{Ali-Haimoud:2017rtz}%
  \BibitemOpen
  \bibfield  {author} {\bibinfo {author} {\bibfnamefont {Yacine}\ \bibnamefont
  {Ali-Haïmoud}}, \bibinfo {author} {\bibfnamefont {Ely~D.}\ \bibnamefont
  {Kovetz}}, \ and\ \bibinfo {author} {\bibfnamefont {Marc}\ \bibnamefont
  {Kamionkowski}},\ }\bibfield  {title} {\enquote {\bibinfo {title} {{Merger
  rate of primordial black-hole binaries}},}\ }\href {\doibase
  10.1103/PhysRevD.96.123523} {\bibfield  {journal} {\bibinfo  {journal} {Phys.
  Rev.}\ }\textbf {\bibinfo {volume} {D96}},\ \bibinfo {pages} {123523}
  (\bibinfo {year} {2017})},\ \Eprint {http://arxiv.org/abs/1709.06576}
  {arXiv:1709.06576 [astro-ph.CO]} \BibitemShut {NoStop}%
\bibitem [{\citenamefont {Chen}\ and\ \citenamefont
  {Huang}(2018)}]{Chen:2018czv}%
  \BibitemOpen
  \bibfield  {author} {\bibinfo {author} {\bibfnamefont {Zu-Cheng}\
  \bibnamefont {Chen}}\ and\ \bibinfo {author} {\bibfnamefont {Qing-Guo}\
  \bibnamefont {Huang}},\ }\bibfield  {title} {\enquote {\bibinfo {title}
  {{Merger Rate Distribution of Primordial-Black-Hole Binaries}},}\ }\href
  {\doibase 10.3847/1538-4357/aad6e2} {\bibfield  {journal} {\bibinfo
  {journal} {Astrophys. J.}\ }\textbf {\bibinfo {volume} {864}},\ \bibinfo
  {pages} {61} (\bibinfo {year} {2018})},\ \Eprint
  {http://arxiv.org/abs/1801.10327} {arXiv:1801.10327 [astro-ph.CO]}
  \BibitemShut {NoStop}%
\bibitem [{\citenamefont {Chen}\ \emph {et~al.}(2019)\citenamefont {Chen},
  \citenamefont {Huang},\ and\ \citenamefont {Huang}}]{Chen:2018rzo}%
  \BibitemOpen
  \bibfield  {author} {\bibinfo {author} {\bibfnamefont {Zu-Cheng}\
  \bibnamefont {Chen}}, \bibinfo {author} {\bibfnamefont {Fan}\ \bibnamefont
  {Huang}}, \ and\ \bibinfo {author} {\bibfnamefont {Qing-Guo}\ \bibnamefont
  {Huang}},\ }\bibfield  {title} {\enquote {\bibinfo {title} {{Stochastic
  Gravitational-wave Background from Binary Black Holes and Binary Neutron
  Stars and Implications for LISA}},}\ }\href {\doibase
  10.3847/1538-4357/aaf581} {\bibfield  {journal} {\bibinfo  {journal}
  {Astrophys. J.}\ }\textbf {\bibinfo {volume} {871}},\ \bibinfo {pages} {97}
  (\bibinfo {year} {2019})},\ \Eprint {http://arxiv.org/abs/1809.10360}
  {arXiv:1809.10360 [gr-qc]} \BibitemShut {NoStop}%
\bibitem [{\citenamefont {Kavanagh}\ \emph {et~al.}(2018)\citenamefont
  {Kavanagh}, \citenamefont {Gaggero},\ and\ \citenamefont
  {Bertone}}]{Kavanagh:2018ggo}%
  \BibitemOpen
  \bibfield  {author} {\bibinfo {author} {\bibfnamefont {Bradley~J.}\
  \bibnamefont {Kavanagh}}, \bibinfo {author} {\bibfnamefont {Daniele}\
  \bibnamefont {Gaggero}}, \ and\ \bibinfo {author} {\bibfnamefont
  {Gianfranco}\ \bibnamefont {Bertone}},\ }\bibfield  {title} {\enquote
  {\bibinfo {title} {{Merger rate of a subdominant population of primordial
  black holes}},}\ }\href {\doibase 10.1103/PhysRevD.98.023536} {\bibfield
  {journal} {\bibinfo  {journal} {Phys. Rev.}\ }\textbf {\bibinfo {volume}
  {D98}},\ \bibinfo {pages} {023536} (\bibinfo {year} {2018})},\ \Eprint
  {http://arxiv.org/abs/1805.09034} {arXiv:1805.09034 [astro-ph.CO]}
  \BibitemShut {NoStop}%
\bibitem [{\citenamefont {Raidal}\ \emph {et~al.}(2019)\citenamefont {Raidal},
  \citenamefont {Spethmann}, \citenamefont {Vaskonen},\ and\ \citenamefont
  {Veermäe}}]{Raidal:2018bbj}%
  \BibitemOpen
  \bibfield  {author} {\bibinfo {author} {\bibfnamefont {Martti}\ \bibnamefont
  {Raidal}}, \bibinfo {author} {\bibfnamefont {Christian}\ \bibnamefont
  {Spethmann}}, \bibinfo {author} {\bibfnamefont {Ville}\ \bibnamefont
  {Vaskonen}}, \ and\ \bibinfo {author} {\bibfnamefont {Hardi}\ \bibnamefont
  {Veermäe}},\ }\bibfield  {title} {\enquote {\bibinfo {title} {{Formation and
  Evolution of Primordial Black Hole Binaries in the Early Universe}},}\ }\href
  {\doibase 10.1088/1475-7516/2019/02/018} {\bibfield  {journal} {\bibinfo
  {journal} {JCAP}\ }\textbf {\bibinfo {volume} {1902}},\ \bibinfo {pages}
  {018} (\bibinfo {year} {2019})},\ \Eprint {http://arxiv.org/abs/1812.01930}
  {arXiv:1812.01930 [astro-ph.CO]} \BibitemShut {NoStop}%
\bibitem [{\citenamefont {Liu}\ \emph {et~al.}(2019{\natexlab{a}})\citenamefont
  {Liu}, \citenamefont {Guo},\ and\ \citenamefont {Cai}}]{Liu:2018ess}%
  \BibitemOpen
  \bibfield  {author} {\bibinfo {author} {\bibfnamefont {Lang}\ \bibnamefont
  {Liu}}, \bibinfo {author} {\bibfnamefont {Zong-Kuan}\ \bibnamefont {Guo}}, \
  and\ \bibinfo {author} {\bibfnamefont {Rong-Gen}\ \bibnamefont {Cai}},\
  }\bibfield  {title} {\enquote {\bibinfo {title} {{Effects of the surrounding
  primordial black holes on the merger rate of primordial black hole
  binaries}},}\ }\href {\doibase 10.1103/PhysRevD.99.063523} {\bibfield
  {journal} {\bibinfo  {journal} {Phys. Rev.}\ }\textbf {\bibinfo {volume}
  {D99}},\ \bibinfo {pages} {063523} (\bibinfo {year} {2019}{\natexlab{a}})},\
  \Eprint {http://arxiv.org/abs/1812.05376} {arXiv:1812.05376 [astro-ph.CO]}
  \BibitemShut {NoStop}%
\bibitem [{\citenamefont {Chen}\ and\ \citenamefont
  {Huang}(2020)}]{Chen:2019irf}%
  \BibitemOpen
  \bibfield  {author} {\bibinfo {author} {\bibfnamefont {Zu-Cheng}\
  \bibnamefont {Chen}}\ and\ \bibinfo {author} {\bibfnamefont {Qing-Guo}\
  \bibnamefont {Huang}},\ }\bibfield  {title} {\enquote {\bibinfo {title}
  {{Distinguishing Primordial Black Holes from Astrophysical Black Holes by
  Einstein Telescope and Cosmic Explorer}},}\ }\href {\doibase
  10.1088/1475-7516/2020/08/039} {\bibfield  {journal} {\bibinfo  {journal}
  {JCAP}\ }\textbf {\bibinfo {volume} {08}},\ \bibinfo {pages} {039} (\bibinfo
  {year} {2020})},\ \Eprint {http://arxiv.org/abs/1904.02396} {arXiv:1904.02396
  [astro-ph.CO]} \BibitemShut {NoStop}%
\bibitem [{\citenamefont {Yuan}\ \emph
  {et~al.}(2019{\natexlab{a}})\citenamefont {Yuan}, \citenamefont {Chen},\ and\
  \citenamefont {Huang}}]{Yuan:2019udt}%
  \BibitemOpen
  \bibfield  {author} {\bibinfo {author} {\bibfnamefont {Chen}\ \bibnamefont
  {Yuan}}, \bibinfo {author} {\bibfnamefont {Zu-Cheng}\ \bibnamefont {Chen}}, \
  and\ \bibinfo {author} {\bibfnamefont {Qing-Guo}\ \bibnamefont {Huang}},\
  }\bibfield  {title} {\enquote {\bibinfo {title} {{Probing
  Primordial-Black-Hole Dark Matter with Scalar Induced Gravitational
  Waves}},}\ }\href {\doibase 10.1103/PhysRevD.100.081301} {\bibfield
  {journal} {\bibinfo  {journal} {Phys. Rev.}\ }\textbf {\bibinfo {volume}
  {D100}},\ \bibinfo {pages} {081301} (\bibinfo {year} {2019}{\natexlab{a}})},\
  \Eprint {http://arxiv.org/abs/1906.11549} {arXiv:1906.11549 [astro-ph.CO]}
  \BibitemShut {NoStop}%
\bibitem [{\citenamefont {Liu}\ \emph {et~al.}(2019{\natexlab{b}})\citenamefont
  {Liu}, \citenamefont {Guo},\ and\ \citenamefont {Cai}}]{Liu:2019rnx}%
  \BibitemOpen
  \bibfield  {author} {\bibinfo {author} {\bibfnamefont {Lang}\ \bibnamefont
  {Liu}}, \bibinfo {author} {\bibfnamefont {Zong-Kuan}\ \bibnamefont {Guo}}, \
  and\ \bibinfo {author} {\bibfnamefont {Rong-Gen}\ \bibnamefont {Cai}},\
  }\bibfield  {title} {\enquote {\bibinfo {title} {{Effects of the merger
  history on the merger rate density of primordial black hole binaries}},}\
  }\href@noop {} {\  (\bibinfo {year} {2019}{\natexlab{b}})},\ \Eprint
  {http://arxiv.org/abs/1901.07672} {arXiv:1901.07672 [astro-ph.CO]}
  \BibitemShut {NoStop}%
\bibitem [{\citenamefont {De~Luca}\ \emph {et~al.}(2021)\citenamefont
  {De~Luca}, \citenamefont {Desjacques}, \citenamefont {Franciolini},
  \citenamefont {Pani},\ and\ \citenamefont {Riotto}}]{DeLuca:2020sae}%
  \BibitemOpen
  \bibfield  {author} {\bibinfo {author} {\bibfnamefont {V.}~\bibnamefont
  {De~Luca}}, \bibinfo {author} {\bibfnamefont {V.}~\bibnamefont {Desjacques}},
  \bibinfo {author} {\bibfnamefont {G.}~\bibnamefont {Franciolini}}, \bibinfo
  {author} {\bibfnamefont {P.}~\bibnamefont {Pani}}, \ and\ \bibinfo {author}
  {\bibfnamefont {A.}~\bibnamefont {Riotto}},\ }\bibfield  {title} {\enquote
  {\bibinfo {title} {{GW190521 Mass Gap Event and the Primordial Black Hole
  Scenario}},}\ }\href {\doibase 10.1103/PhysRevLett.126.051101} {\bibfield
  {journal} {\bibinfo  {journal} {Phys. Rev. Lett.}\ }\textbf {\bibinfo
  {volume} {126}},\ \bibinfo {pages} {051101} (\bibinfo {year} {2021})},\
  \Eprint {http://arxiv.org/abs/2009.01728} {arXiv:2009.01728 [astro-ph.CO]}
  \BibitemShut {NoStop}%
\bibitem [{\citenamefont {Vattis}\ \emph {et~al.}(2020)\citenamefont {Vattis},
  \citenamefont {Goldstein},\ and\ \citenamefont
  {Koushiappas}}]{Vattis:2020iuz}%
  \BibitemOpen
  \bibfield  {author} {\bibinfo {author} {\bibfnamefont {Kyriakos}\
  \bibnamefont {Vattis}}, \bibinfo {author} {\bibfnamefont {Isabelle~S.}\
  \bibnamefont {Goldstein}}, \ and\ \bibinfo {author} {\bibfnamefont
  {Savvas~M.}\ \bibnamefont {Koushiappas}},\ }\bibfield  {title} {\enquote
  {\bibinfo {title} {{Could the 2.6 $M_\odot$ object in GW190814 be a
  primordial black hole?}}}\ }\href {\doibase 10.1103/PhysRevD.102.061301}
  {\bibfield  {journal} {\bibinfo  {journal} {Phys. Rev. D}\ }\textbf {\bibinfo
  {volume} {102}},\ \bibinfo {pages} {061301} (\bibinfo {year} {2020})},\
  \Eprint {http://arxiv.org/abs/2006.15675} {arXiv:2006.15675 [astro-ph.HE]}
  \BibitemShut {NoStop}%
\bibitem [{\citenamefont {Wang}\ and\ \citenamefont
  {Zhao}(2022)}]{Wang:2021iwp}%
  \BibitemOpen
  \bibfield  {author} {\bibinfo {author} {\bibfnamefont {Sai}\ \bibnamefont
  {Wang}}\ and\ \bibinfo {author} {\bibfnamefont {Zhi-Chao}\ \bibnamefont
  {Zhao}},\ }\bibfield  {title} {\enquote {\bibinfo {title} {{GW200105 and
  GW200115 are compatible with a scenario of primordial black hole binary
  coalescences}},}\ }\href {\doibase 10.1140/epjc/s10052-021-09981-1}
  {\bibfield  {journal} {\bibinfo  {journal} {Eur. Phys. J. C}\ }\textbf
  {\bibinfo {volume} {82}},\ \bibinfo {pages} {9} (\bibinfo {year} {2022})},\
  \Eprint {http://arxiv.org/abs/2107.00450} {arXiv:2107.00450 [astro-ph.CO]}
  \BibitemShut {NoStop}%
\bibitem [{\citenamefont {Chen}\ \emph {et~al.}(2022)\citenamefont {Chen},
  \citenamefont {Yuan},\ and\ \citenamefont {Huang}}]{Chen:2021nxo}%
  \BibitemOpen
  \bibfield  {author} {\bibinfo {author} {\bibfnamefont {Zu-Cheng}\
  \bibnamefont {Chen}}, \bibinfo {author} {\bibfnamefont {Chen}\ \bibnamefont
  {Yuan}}, \ and\ \bibinfo {author} {\bibfnamefont {Qing-Guo}\ \bibnamefont
  {Huang}},\ }\bibfield  {title} {\enquote {\bibinfo {title} {{Confronting the
  primordial black hole scenario with the gravitational-wave events detected by
  LIGO-Virgo}},}\ }\href {\doibase 10.1016/j.physletb.2022.137040} {\bibfield
  {journal} {\bibinfo  {journal} {Phys. Lett. B}\ }\textbf {\bibinfo {volume}
  {829}},\ \bibinfo {pages} {137040} (\bibinfo {year} {2022})},\ \Eprint
  {http://arxiv.org/abs/2108.11740} {arXiv:2108.11740 [astro-ph.CO]}
  \BibitemShut {NoStop}%
\bibitem [{\citenamefont {Abbott}\ \emph {et~al.}(2016)\citenamefont {Abbott}
  \emph {et~al.}}]{LIGOScientific:2016dsl}%
  \BibitemOpen
  \bibfield  {author} {\bibinfo {author} {\bibfnamefont {B.~P.}\ \bibnamefont
  {Abbott}} \emph {et~al.} (\bibinfo {collaboration} {LIGO Scientific,
  Virgo}),\ }\bibfield  {title} {\enquote {\bibinfo {title} {{Binary Black Hole
  Mergers in the first Advanced LIGO Observing Run}},}\ }\href {\doibase
  10.1103/PhysRevX.6.041015} {\bibfield  {journal} {\bibinfo  {journal} {Phys.
  Rev. X}\ }\textbf {\bibinfo {volume} {6}},\ \bibinfo {pages} {041015}
  (\bibinfo {year} {2016})},\ \bibinfo {note} {[Erratum: Phys.Rev.X 8, 039903
  (2018)]},\ \Eprint {http://arxiv.org/abs/1606.04856} {arXiv:1606.04856
  [gr-qc]} \BibitemShut {NoStop}%
\bibitem [{\citenamefont {Abbott}\ \emph {et~al.}(2021)\citenamefont {Abbott}
  \emph {et~al.}}]{LIGOScientific:2021djp}%
  \BibitemOpen
  \bibfield  {author} {\bibinfo {author} {\bibfnamefont {R.}~\bibnamefont
  {Abbott}} \emph {et~al.} (\bibinfo {collaboration} {LIGO Scientific, VIRGO,
  KAGRA}),\ }\bibfield  {title} {\enquote {\bibinfo {title} {{GWTC-3: Compact
  Binary Coalescences Observed by LIGO and Virgo During the Second Part of the
  Third Observing Run}},}\ }\href@noop {} {\  (\bibinfo {year} {2021})},\
  \Eprint {http://arxiv.org/abs/2111.03606} {arXiv:2111.03606 [gr-qc]}
  \BibitemShut {NoStop}%
\bibitem [{\citenamefont {Carr}\ \emph {et~al.}(2010)\citenamefont {Carr},
  \citenamefont {Kohri}, \citenamefont {Sendouda},\ and\ \citenamefont
  {Yokoyama}}]{Carr:2009jm}%
  \BibitemOpen
  \bibfield  {author} {\bibinfo {author} {\bibfnamefont {B.~J.}\ \bibnamefont
  {Carr}}, \bibinfo {author} {\bibfnamefont {Kazunori}\ \bibnamefont {Kohri}},
  \bibinfo {author} {\bibfnamefont {Yuuiti}\ \bibnamefont {Sendouda}}, \ and\
  \bibinfo {author} {\bibfnamefont {Jun'ichi}\ \bibnamefont {Yokoyama}},\
  }\bibfield  {title} {\enquote {\bibinfo {title} {{New cosmological
  constraints on primordial black holes}},}\ }\href {\doibase
  10.1103/PhysRevD.81.104019} {\bibfield  {journal} {\bibinfo  {journal} {Phys.
  Rev. D}\ }\textbf {\bibinfo {volume} {81}},\ \bibinfo {pages} {104019}
  (\bibinfo {year} {2010})},\ \Eprint {http://arxiv.org/abs/0912.5297}
  {arXiv:0912.5297 [astro-ph.CO]} \BibitemShut {NoStop}%
\bibitem [{\citenamefont {Graham}\ \emph {et~al.}(2015)\citenamefont {Graham},
  \citenamefont {Rajendran},\ and\ \citenamefont {Varela}}]{Graham:2015apa}%
  \BibitemOpen
  \bibfield  {author} {\bibinfo {author} {\bibfnamefont {Peter~W.}\
  \bibnamefont {Graham}}, \bibinfo {author} {\bibfnamefont {Surjeet}\
  \bibnamefont {Rajendran}}, \ and\ \bibinfo {author} {\bibfnamefont {Jaime}\
  \bibnamefont {Varela}},\ }\bibfield  {title} {\enquote {\bibinfo {title}
  {{Dark Matter Triggers of Supernovae}},}\ }\href {\doibase
  10.1103/PhysRevD.92.063007} {\bibfield  {journal} {\bibinfo  {journal} {Phys.
  Rev. D}\ }\textbf {\bibinfo {volume} {92}},\ \bibinfo {pages} {063007}
  (\bibinfo {year} {2015})},\ \Eprint {http://arxiv.org/abs/1505.04444}
  {arXiv:1505.04444 [hep-ph]} \BibitemShut {NoStop}%
\bibitem [{\citenamefont {Niikura}\ \emph
  {et~al.}(2019{\natexlab{a}})\citenamefont {Niikura} \emph
  {et~al.}}]{Niikura:2017zjd}%
  \BibitemOpen
  \bibfield  {author} {\bibinfo {author} {\bibfnamefont {Hiroko}\ \bibnamefont
  {Niikura}} \emph {et~al.},\ }\bibfield  {title} {\enquote {\bibinfo {title}
  {{Microlensing constraints on primordial black holes with Subaru/HSC
  Andromeda observations}},}\ }\href {\doibase 10.1038/s41550-019-0723-1}
  {\bibfield  {journal} {\bibinfo  {journal} {Nature Astron.}\ }\textbf
  {\bibinfo {volume} {3}},\ \bibinfo {pages} {524--534} (\bibinfo {year}
  {2019}{\natexlab{a}})},\ \Eprint {http://arxiv.org/abs/1701.02151}
  {arXiv:1701.02151 [astro-ph.CO]} \BibitemShut {NoStop}%
\bibitem [{\citenamefont {Tisserand}\ \emph {et~al.}(2007)\citenamefont
  {Tisserand} \emph {et~al.}}]{EROS-2:2006ryy}%
  \BibitemOpen
  \bibfield  {author} {\bibinfo {author} {\bibfnamefont {P.}~\bibnamefont
  {Tisserand}} \emph {et~al.} (\bibinfo {collaboration} {EROS-2}),\ }\bibfield
  {title} {\enquote {\bibinfo {title} {{Limits on the Macho Content of the
  Galactic Halo from the EROS-2 Survey of the Magellanic Clouds}},}\ }\href
  {\doibase 10.1051/0004-6361:20066017} {\bibfield  {journal} {\bibinfo
  {journal} {Astron. Astrophys.}\ }\textbf {\bibinfo {volume} {469}},\ \bibinfo
  {pages} {387--404} (\bibinfo {year} {2007})},\ \Eprint
  {http://arxiv.org/abs/astro-ph/0607207} {arXiv:astro-ph/0607207} \BibitemShut
  {NoStop}%
\bibitem [{\citenamefont {Niikura}\ \emph
  {et~al.}(2019{\natexlab{b}})\citenamefont {Niikura}, \citenamefont {Takada},
  \citenamefont {Yokoyama}, \citenamefont {Sumi},\ and\ \citenamefont
  {Masaki}}]{Niikura:2019kqi}%
  \BibitemOpen
  \bibfield  {author} {\bibinfo {author} {\bibfnamefont {Hiroko}\ \bibnamefont
  {Niikura}}, \bibinfo {author} {\bibfnamefont {Masahiro}\ \bibnamefont
  {Takada}}, \bibinfo {author} {\bibfnamefont {Shuichiro}\ \bibnamefont
  {Yokoyama}}, \bibinfo {author} {\bibfnamefont {Takahiro}\ \bibnamefont
  {Sumi}}, \ and\ \bibinfo {author} {\bibfnamefont {Shogo}\ \bibnamefont
  {Masaki}},\ }\bibfield  {title} {\enquote {\bibinfo {title} {{Constraints on
  Earth-mass primordial black holes from OGLE 5-year microlensing events}},}\
  }\href {\doibase 10.1103/PhysRevD.99.083503} {\bibfield  {journal} {\bibinfo
  {journal} {Phys. Rev. D}\ }\textbf {\bibinfo {volume} {99}},\ \bibinfo
  {pages} {083503} (\bibinfo {year} {2019}{\natexlab{b}})},\ \Eprint
  {http://arxiv.org/abs/1901.07120} {arXiv:1901.07120 [astro-ph.CO]}
  \BibitemShut {NoStop}%
\bibitem [{\citenamefont {Brandt}(2016)}]{Brandt:2016aco}%
  \BibitemOpen
  \bibfield  {author} {\bibinfo {author} {\bibfnamefont {Timothy~D.}\
  \bibnamefont {Brandt}},\ }\bibfield  {title} {\enquote {\bibinfo {title}
  {{Constraints on MACHO Dark Matter from Compact Stellar Systems in
  Ultra-Faint Dwarf Galaxies}},}\ }\href {\doibase 10.3847/2041-8205/824/2/L31}
  {\bibfield  {journal} {\bibinfo  {journal} {Astrophys. J. Lett.}\ }\textbf
  {\bibinfo {volume} {824}},\ \bibinfo {pages} {L31} (\bibinfo {year}
  {2016})},\ \Eprint {http://arxiv.org/abs/1605.03665} {arXiv:1605.03665
  [astro-ph.GA]} \BibitemShut {NoStop}%
\bibitem [{\citenamefont {Chen}\ \emph
  {et~al.}(2020{\natexlab{a}})\citenamefont {Chen}, \citenamefont {Yuan},\ and\
  \citenamefont {Huang}}]{Chen:2019xse}%
  \BibitemOpen
  \bibfield  {author} {\bibinfo {author} {\bibfnamefont {Zu-Cheng}\
  \bibnamefont {Chen}}, \bibinfo {author} {\bibfnamefont {Chen}\ \bibnamefont
  {Yuan}}, \ and\ \bibinfo {author} {\bibfnamefont {Qing-Guo}\ \bibnamefont
  {Huang}},\ }\bibfield  {title} {\enquote {\bibinfo {title} {{Pulsar Timing
  Array Constraints on Primordial Black Holes with NANOGrav 11-Year
  Dataset}},}\ }\href {\doibase 10.1103/PhysRevLett.124.251101} {\bibfield
  {journal} {\bibinfo  {journal} {Phys. Rev. Lett.}\ }\textbf {\bibinfo
  {volume} {124}},\ \bibinfo {pages} {251101} (\bibinfo {year}
  {2020}{\natexlab{a}})},\ \Eprint {http://arxiv.org/abs/1910.12239}
  {arXiv:1910.12239 [astro-ph.CO]} \BibitemShut {NoStop}%
\bibitem [{\citenamefont {Montero-Camacho}\ \emph {et~al.}(2019)\citenamefont
  {Montero-Camacho}, \citenamefont {Fang}, \citenamefont {Vasquez},
  \citenamefont {Silva},\ and\ \citenamefont
  {Hirata}}]{Montero-Camacho:2019jte}%
  \BibitemOpen
  \bibfield  {author} {\bibinfo {author} {\bibfnamefont {Paulo}\ \bibnamefont
  {Montero-Camacho}}, \bibinfo {author} {\bibfnamefont {Xiao}\ \bibnamefont
  {Fang}}, \bibinfo {author} {\bibfnamefont {Gabriel}\ \bibnamefont {Vasquez}},
  \bibinfo {author} {\bibfnamefont {Makana}\ \bibnamefont {Silva}}, \ and\
  \bibinfo {author} {\bibfnamefont {Christopher~M.}\ \bibnamefont {Hirata}},\
  }\bibfield  {title} {\enquote {\bibinfo {title} {{Revisiting constraints on
  asteroid-mass primordial black holes as dark matter candidates}},}\ }\href
  {\doibase 10.1088/1475-7516/2019/08/031} {\bibfield  {journal} {\bibinfo
  {journal} {JCAP}\ }\textbf {\bibinfo {volume} {08}},\ \bibinfo {pages} {031}
  (\bibinfo {year} {2019})},\ \Eprint {http://arxiv.org/abs/1906.05950}
  {arXiv:1906.05950 [astro-ph.CO]} \BibitemShut {NoStop}%
\bibitem [{\citenamefont {Laha}(2019)}]{Laha:2019ssq}%
  \BibitemOpen
  \bibfield  {author} {\bibinfo {author} {\bibfnamefont {Ranjan}\ \bibnamefont
  {Laha}},\ }\bibfield  {title} {\enquote {\bibinfo {title} {{Primordial Black
  Holes as a Dark Matter Candidate Are Severely Constrained by the Galactic
  Center 511 keV $\gamma$ -Ray Line}},}\ }\href {\doibase
  10.1103/PhysRevLett.123.251101} {\bibfield  {journal} {\bibinfo  {journal}
  {Phys. Rev. Lett.}\ }\textbf {\bibinfo {volume} {123}},\ \bibinfo {pages}
  {251101} (\bibinfo {year} {2019})},\ \Eprint
  {http://arxiv.org/abs/1906.09994} {arXiv:1906.09994 [astro-ph.HE]}
  \BibitemShut {NoStop}%
\bibitem [{\citenamefont {Dasgupta}\ \emph {et~al.}(2020)\citenamefont
  {Dasgupta}, \citenamefont {Laha},\ and\ \citenamefont
  {Ray}}]{Dasgupta:2019cae}%
  \BibitemOpen
  \bibfield  {author} {\bibinfo {author} {\bibfnamefont {Basudeb}\ \bibnamefont
  {Dasgupta}}, \bibinfo {author} {\bibfnamefont {Ranjan}\ \bibnamefont {Laha}},
  \ and\ \bibinfo {author} {\bibfnamefont {Anupam}\ \bibnamefont {Ray}},\
  }\bibfield  {title} {\enquote {\bibinfo {title} {{Neutrino and positron
  constraints on spinning primordial black hole dark matter}},}\ }\href
  {\doibase 10.1103/PhysRevLett.125.101101} {\bibfield  {journal} {\bibinfo
  {journal} {Phys. Rev. Lett.}\ }\textbf {\bibinfo {volume} {125}},\ \bibinfo
  {pages} {101101} (\bibinfo {year} {2020})},\ \Eprint
  {http://arxiv.org/abs/1912.01014} {arXiv:1912.01014 [hep-ph]} \BibitemShut
  {NoStop}%
\bibitem [{\citenamefont {Laha}\ \emph {et~al.}(2020)\citenamefont {Laha},
  \citenamefont {Mu\~noz},\ and\ \citenamefont {Slatyer}}]{Laha:2020ivk}%
  \BibitemOpen
  \bibfield  {author} {\bibinfo {author} {\bibfnamefont {Ranjan}\ \bibnamefont
  {Laha}}, \bibinfo {author} {\bibfnamefont {Julian~B.}\ \bibnamefont
  {Mu\~noz}}, \ and\ \bibinfo {author} {\bibfnamefont {Tracy~R.}\ \bibnamefont
  {Slatyer}},\ }\bibfield  {title} {\enquote {\bibinfo {title} {{INTEGRAL
  constraints on primordial black holes and particle dark matter}},}\ }\href
  {\doibase 10.1103/PhysRevD.101.123514} {\bibfield  {journal} {\bibinfo
  {journal} {Phys. Rev. D}\ }\textbf {\bibinfo {volume} {101}},\ \bibinfo
  {pages} {123514} (\bibinfo {year} {2020})},\ \Eprint
  {http://arxiv.org/abs/2004.00627} {arXiv:2004.00627 [astro-ph.CO]}
  \BibitemShut {NoStop}%
\bibitem [{\citenamefont {Saha}\ and\ \citenamefont
  {Laha}(2022)}]{Saha:2021pqf}%
  \BibitemOpen
  \bibfield  {author} {\bibinfo {author} {\bibfnamefont {Akash~Kumar}\
  \bibnamefont {Saha}}\ and\ \bibinfo {author} {\bibfnamefont {Ranjan}\
  \bibnamefont {Laha}},\ }\bibfield  {title} {\enquote {\bibinfo {title}
  {{Sensitivities on nonspinning and spinning primordial black hole dark matter
  with global 21-cm troughs}},}\ }\href {\doibase 10.1103/PhysRevD.105.103026}
  {\bibfield  {journal} {\bibinfo  {journal} {Phys. Rev. D}\ }\textbf {\bibinfo
  {volume} {105}},\ \bibinfo {pages} {103026} (\bibinfo {year} {2022})},\
  \Eprint {http://arxiv.org/abs/2112.10794} {arXiv:2112.10794 [astro-ph.CO]}
  \BibitemShut {NoStop}%
\bibitem [{\citenamefont {Ray}\ \emph {et~al.}(2021)\citenamefont {Ray},
  \citenamefont {Laha}, \citenamefont {Mu\~noz},\ and\ \citenamefont
  {Caputo}}]{Ray:2021mxu}%
  \BibitemOpen
  \bibfield  {author} {\bibinfo {author} {\bibfnamefont {Anupam}\ \bibnamefont
  {Ray}}, \bibinfo {author} {\bibfnamefont {Ranjan}\ \bibnamefont {Laha}},
  \bibinfo {author} {\bibfnamefont {Julian~B.}\ \bibnamefont {Mu\~noz}}, \ and\
  \bibinfo {author} {\bibfnamefont {Regina}\ \bibnamefont {Caputo}},\
  }\bibfield  {title} {\enquote {\bibinfo {title} {{Near future MeV telescopes
  can discover asteroid-mass primordial black hole dark matter}},}\ }\href
  {\doibase 10.1103/PhysRevD.104.023516} {\bibfield  {journal} {\bibinfo
  {journal} {Phys. Rev. D}\ }\textbf {\bibinfo {volume} {104}},\ \bibinfo
  {pages} {023516} (\bibinfo {year} {2021})},\ \Eprint
  {http://arxiv.org/abs/2102.06714} {arXiv:2102.06714 [astro-ph.CO]}
  \BibitemShut {NoStop}%
\bibitem [{\citenamefont {Carr}\ \emph {et~al.}(2020)\citenamefont {Carr},
  \citenamefont {Kohri}, \citenamefont {Sendouda},\ and\ \citenamefont
  {Yokoyama}}]{Carr:2020gox}%
  \BibitemOpen
  \bibfield  {author} {\bibinfo {author} {\bibfnamefont {Bernard}\ \bibnamefont
  {Carr}}, \bibinfo {author} {\bibfnamefont {Kazunori}\ \bibnamefont {Kohri}},
  \bibinfo {author} {\bibfnamefont {Yuuiti}\ \bibnamefont {Sendouda}}, \ and\
  \bibinfo {author} {\bibfnamefont {Jun'ichi}\ \bibnamefont {Yokoyama}},\
  }\bibfield  {title} {\enquote {\bibinfo {title} {{Constraints on Primordial
  Black Holes}},}\ }\href@noop {} {\  (\bibinfo {year} {2020})},\ \Eprint
  {http://arxiv.org/abs/2002.12778} {arXiv:2002.12778 [astro-ph.CO]}
  \BibitemShut {NoStop}%
\bibitem [{\citenamefont {Escriv\`a}\ \emph {et~al.}(2022)\citenamefont
  {Escriv\`a}, \citenamefont {Kuhnel},\ and\ \citenamefont
  {Tada}}]{Escriva:2022duf}%
  \BibitemOpen
  \bibfield  {author} {\bibinfo {author} {\bibfnamefont {Albert}\ \bibnamefont
  {Escriv\`a}}, \bibinfo {author} {\bibfnamefont {Florian}\ \bibnamefont
  {Kuhnel}}, \ and\ \bibinfo {author} {\bibfnamefont {Yuichiro}\ \bibnamefont
  {Tada}},\ }\bibfield  {title} {\enquote {\bibinfo {title} {{Primordial Black
  Holes}},}\ }\href@noop {} {\  (\bibinfo {year} {2022})},\ \Eprint
  {http://arxiv.org/abs/2211.05767} {arXiv:2211.05767 [astro-ph.CO]}
  \BibitemShut {NoStop}%
\bibitem [{\citenamefont {Aghanim}\ \emph {et~al.}(2020)\citenamefont {Aghanim}
  \emph {et~al.}}]{Planck:2018vyg}%
  \BibitemOpen
  \bibfield  {author} {\bibinfo {author} {\bibfnamefont {N.}~\bibnamefont
  {Aghanim}} \emph {et~al.} (\bibinfo {collaboration} {Planck}),\ }\bibfield
  {title} {\enquote {\bibinfo {title} {{Planck 2018 results. VI. Cosmological
  parameters}},}\ }\href {\doibase 10.1051/0004-6361/201833910} {\bibfield
  {journal} {\bibinfo  {journal} {Astron. Astrophys.}\ }\textbf {\bibinfo
  {volume} {641}},\ \bibinfo {pages} {A6} (\bibinfo {year} {2020})},\ \bibinfo
  {note} {[Erratum: Astron.Astrophys. 652, C4 (2021)]},\ \Eprint
  {http://arxiv.org/abs/1807.06209} {arXiv:1807.06209 [astro-ph.CO]}
  \BibitemShut {NoStop}%
\bibitem [{\citenamefont {Yokoyama}(1998)}]{Yokoyama:1998pt}%
  \BibitemOpen
  \bibfield  {author} {\bibinfo {author} {\bibfnamefont {Jun'ichi}\
  \bibnamefont {Yokoyama}},\ }\bibfield  {title} {\enquote {\bibinfo {title}
  {{Chaotic new inflation and formation of primordial black holes}},}\ }\href
  {\doibase 10.1103/PhysRevD.58.083510} {\bibfield  {journal} {\bibinfo
  {journal} {Phys. Rev. D}\ }\textbf {\bibinfo {volume} {58}},\ \bibinfo
  {pages} {083510} (\bibinfo {year} {1998})},\ \Eprint
  {http://arxiv.org/abs/astro-ph/9802357} {arXiv:astro-ph/9802357} \BibitemShut
  {NoStop}%
\bibitem [{\citenamefont {Kinney}(2005)}]{Kinney:2005vj}%
  \BibitemOpen
  \bibfield  {author} {\bibinfo {author} {\bibfnamefont {William~H.}\
  \bibnamefont {Kinney}},\ }\bibfield  {title} {\enquote {\bibinfo {title}
  {{Horizon crossing and inflation with large eta}},}\ }\href {\doibase
  10.1103/PhysRevD.72.023515} {\bibfield  {journal} {\bibinfo  {journal} {Phys.
  Rev. D}\ }\textbf {\bibinfo {volume} {72}},\ \bibinfo {pages} {023515}
  (\bibinfo {year} {2005})},\ \Eprint {http://arxiv.org/abs/gr-qc/0503017}
  {arXiv:gr-qc/0503017} \BibitemShut {NoStop}%
\bibitem [{\citenamefont {Choudhury}\ and\ \citenamefont
  {Mazumdar}(2014)}]{Choudhury:2013woa}%
  \BibitemOpen
  \bibfield  {author} {\bibinfo {author} {\bibfnamefont {Sayantan}\
  \bibnamefont {Choudhury}}\ and\ \bibinfo {author} {\bibfnamefont {Anupam}\
  \bibnamefont {Mazumdar}},\ }\bibfield  {title} {\enquote {\bibinfo {title}
  {{Primordial blackholes and gravitational waves for an inflection-point model
  of inflation}},}\ }\href {\doibase 10.1016/j.physletb.2014.04.050} {\bibfield
   {journal} {\bibinfo  {journal} {Phys. Lett. B}\ }\textbf {\bibinfo {volume}
  {733}},\ \bibinfo {pages} {270--275} (\bibinfo {year} {2014})},\ \Eprint
  {http://arxiv.org/abs/1307.5119} {arXiv:1307.5119 [astro-ph.CO]} \BibitemShut
  {NoStop}%
\bibitem [{\citenamefont {Garcia-Bellido}\ \emph {et~al.}(2016)\citenamefont
  {Garcia-Bellido}, \citenamefont {Peloso},\ and\ \citenamefont
  {Unal}}]{Garcia-Bellido:2016dkw}%
  \BibitemOpen
  \bibfield  {author} {\bibinfo {author} {\bibfnamefont {Juan}\ \bibnamefont
  {Garcia-Bellido}}, \bibinfo {author} {\bibfnamefont {Marco}\ \bibnamefont
  {Peloso}}, \ and\ \bibinfo {author} {\bibfnamefont {Caner}\ \bibnamefont
  {Unal}},\ }\bibfield  {title} {\enquote {\bibinfo {title} {{Gravitational
  waves at interferometer scales and primordial black holes in axion
  inflation}},}\ }\href {\doibase 10.1088/1475-7516/2016/12/031} {\bibfield
  {journal} {\bibinfo  {journal} {JCAP}\ }\textbf {\bibinfo {volume} {12}},\
  \bibinfo {pages} {031} (\bibinfo {year} {2016})},\ \Eprint
  {http://arxiv.org/abs/1610.03763} {arXiv:1610.03763 [astro-ph.CO]}
  \BibitemShut {NoStop}%
\bibitem [{\citenamefont {Cheng}\ \emph {et~al.}(2017)\citenamefont {Cheng},
  \citenamefont {Lee},\ and\ \citenamefont {Ng}}]{Cheng:2016qzb}%
  \BibitemOpen
  \bibfield  {author} {\bibinfo {author} {\bibfnamefont {Shu-Lin}\ \bibnamefont
  {Cheng}}, \bibinfo {author} {\bibfnamefont {Wolung}\ \bibnamefont {Lee}}, \
  and\ \bibinfo {author} {\bibfnamefont {Kin-Wang}\ \bibnamefont {Ng}},\
  }\bibfield  {title} {\enquote {\bibinfo {title} {{Production of high
  stellar-mass primordial black holes in trapped inflation}},}\ }\href
  {\doibase 10.1007/JHEP02(2017)008} {\bibfield  {journal} {\bibinfo  {journal}
  {JHEP}\ }\textbf {\bibinfo {volume} {02}},\ \bibinfo {pages} {008} (\bibinfo
  {year} {2017})},\ \Eprint {http://arxiv.org/abs/1606.00206} {arXiv:1606.00206
  [astro-ph.CO]} \BibitemShut {NoStop}%
\bibitem [{\citenamefont {Garcia-Bellido}\ and\ \citenamefont
  {Ruiz~Morales}(2017)}]{Garcia-Bellido:2017mdw}%
  \BibitemOpen
  \bibfield  {author} {\bibinfo {author} {\bibfnamefont {Juan}\ \bibnamefont
  {Garcia-Bellido}}\ and\ \bibinfo {author} {\bibfnamefont {Ester}\
  \bibnamefont {Ruiz~Morales}},\ }\bibfield  {title} {\enquote {\bibinfo
  {title} {{Primordial black holes from single field models of inflation}},}\
  }\href {\doibase 10.1016/j.dark.2017.09.007} {\bibfield  {journal} {\bibinfo
  {journal} {Phys. Dark Univ.}\ }\textbf {\bibinfo {volume} {18}},\ \bibinfo
  {pages} {47--54} (\bibinfo {year} {2017})},\ \Eprint
  {http://arxiv.org/abs/1702.03901} {arXiv:1702.03901 [astro-ph.CO]}
  \BibitemShut {NoStop}%
\bibitem [{\citenamefont {Cheng}\ \emph {et~al.}(2018)\citenamefont {Cheng},
  \citenamefont {Lee},\ and\ \citenamefont {Ng}}]{Cheng:2018yyr}%
  \BibitemOpen
  \bibfield  {author} {\bibinfo {author} {\bibfnamefont {Shu-Lin}\ \bibnamefont
  {Cheng}}, \bibinfo {author} {\bibfnamefont {Wolung}\ \bibnamefont {Lee}}, \
  and\ \bibinfo {author} {\bibfnamefont {Kin-Wang}\ \bibnamefont {Ng}},\
  }\bibfield  {title} {\enquote {\bibinfo {title} {{Primordial black holes and
  associated gravitational waves in axion monodromy inflation}},}\ }\href
  {\doibase 10.1088/1475-7516/2018/07/001} {\bibfield  {journal} {\bibinfo
  {journal} {JCAP}\ }\textbf {\bibinfo {volume} {07}},\ \bibinfo {pages} {001}
  (\bibinfo {year} {2018})},\ \Eprint {http://arxiv.org/abs/1801.09050}
  {arXiv:1801.09050 [astro-ph.CO]} \BibitemShut {NoStop}%
\bibitem [{\citenamefont {Dalianis}\ \emph {et~al.}(2019)\citenamefont
  {Dalianis}, \citenamefont {Kehagias},\ and\ \citenamefont
  {Tringas}}]{Dalianis:2018frf}%
  \BibitemOpen
  \bibfield  {author} {\bibinfo {author} {\bibfnamefont {Ioannis}\ \bibnamefont
  {Dalianis}}, \bibinfo {author} {\bibfnamefont {Alex}\ \bibnamefont
  {Kehagias}}, \ and\ \bibinfo {author} {\bibfnamefont {George}\ \bibnamefont
  {Tringas}},\ }\bibfield  {title} {\enquote {\bibinfo {title} {{Primordial
  black holes from \ensuremath{\alpha}-attractors}},}\ }\href {\doibase
  10.1088/1475-7516/2019/01/037} {\bibfield  {journal} {\bibinfo  {journal}
  {JCAP}\ }\textbf {\bibinfo {volume} {01}},\ \bibinfo {pages} {037} (\bibinfo
  {year} {2019})},\ \Eprint {http://arxiv.org/abs/1805.09483} {arXiv:1805.09483
  [astro-ph.CO]} \BibitemShut {NoStop}%
\bibitem [{\citenamefont {Tada}\ and\ \citenamefont
  {Yokoyama}(2019)}]{Tada:2019amh}%
  \BibitemOpen
  \bibfield  {author} {\bibinfo {author} {\bibfnamefont {Yuichiro}\
  \bibnamefont {Tada}}\ and\ \bibinfo {author} {\bibfnamefont {Shuichiro}\
  \bibnamefont {Yokoyama}},\ }\bibfield  {title} {\enquote {\bibinfo {title}
  {{Primordial black hole tower: Dark matter, earth-mass, and LIGO black
  holes}},}\ }\href {\doibase 10.1103/PhysRevD.100.023537} {\bibfield
  {journal} {\bibinfo  {journal} {Phys. Rev. D}\ }\textbf {\bibinfo {volume}
  {100}},\ \bibinfo {pages} {023537} (\bibinfo {year} {2019})},\ \Eprint
  {http://arxiv.org/abs/1904.10298} {arXiv:1904.10298 [astro-ph.CO]}
  \BibitemShut {NoStop}%
\bibitem [{\citenamefont {Xu}\ \emph {et~al.}(2020)\citenamefont {Xu},
  \citenamefont {Liu}, \citenamefont {Gao},\ and\ \citenamefont
  {Guo}}]{Xu:2019bdp}%
  \BibitemOpen
  \bibfield  {author} {\bibinfo {author} {\bibfnamefont {Wu-Tao}\ \bibnamefont
  {Xu}}, \bibinfo {author} {\bibfnamefont {Jing}\ \bibnamefont {Liu}}, \bibinfo
  {author} {\bibfnamefont {Tie-Jun}\ \bibnamefont {Gao}}, \ and\ \bibinfo
  {author} {\bibfnamefont {Zong-Kuan}\ \bibnamefont {Guo}},\ }\bibfield
  {title} {\enquote {\bibinfo {title} {{Gravitational waves from
  double-inflection-point inflation}},}\ }\href {\doibase
  10.1103/PhysRevD.101.023505} {\bibfield  {journal} {\bibinfo  {journal}
  {Phys. Rev. D}\ }\textbf {\bibinfo {volume} {101}},\ \bibinfo {pages}
  {023505} (\bibinfo {year} {2020})},\ \Eprint
  {http://arxiv.org/abs/1907.05213} {arXiv:1907.05213 [astro-ph.CO]}
  \BibitemShut {NoStop}%
\bibitem [{\citenamefont {Mishra}\ and\ \citenamefont
  {Sahni}(2020)}]{Mishra:2019pzq}%
  \BibitemOpen
  \bibfield  {author} {\bibinfo {author} {\bibfnamefont {Swagat~S.}\
  \bibnamefont {Mishra}}\ and\ \bibinfo {author} {\bibfnamefont {Varun}\
  \bibnamefont {Sahni}},\ }\bibfield  {title} {\enquote {\bibinfo {title}
  {{Primordial Black Holes from a tiny bump/dip in the Inflaton potential}},}\
  }\href {\doibase 10.1088/1475-7516/2020/04/007} {\bibfield  {journal}
  {\bibinfo  {journal} {JCAP}\ }\textbf {\bibinfo {volume} {04}},\ \bibinfo
  {pages} {007} (\bibinfo {year} {2020})},\ \Eprint
  {http://arxiv.org/abs/1911.00057} {arXiv:1911.00057 [gr-qc]} \BibitemShut
  {NoStop}%
\bibitem [{\citenamefont {Bhaumik}\ and\ \citenamefont
  {Jain}(2020)}]{Bhaumik:2019tvl}%
  \BibitemOpen
  \bibfield  {author} {\bibinfo {author} {\bibfnamefont {Nilanjandev}\
  \bibnamefont {Bhaumik}}\ and\ \bibinfo {author} {\bibfnamefont
  {Rajeev~Kumar}\ \bibnamefont {Jain}},\ }\bibfield  {title} {\enquote
  {\bibinfo {title} {{Primordial black holes dark matter from inflection point
  models of inflation and the effects of reheating}},}\ }\href {\doibase
  10.1088/1475-7516/2020/01/037} {\bibfield  {journal} {\bibinfo  {journal}
  {JCAP}\ }\textbf {\bibinfo {volume} {01}},\ \bibinfo {pages} {037} (\bibinfo
  {year} {2020})},\ \Eprint {http://arxiv.org/abs/1907.04125} {arXiv:1907.04125
  [astro-ph.CO]} \BibitemShut {NoStop}%
\bibitem [{\citenamefont {Liu}\ \emph {et~al.}(2020)\citenamefont {Liu},
  \citenamefont {Guo},\ and\ \citenamefont {Cai}}]{Liu:2020oqe}%
  \BibitemOpen
  \bibfield  {author} {\bibinfo {author} {\bibfnamefont {Jing}\ \bibnamefont
  {Liu}}, \bibinfo {author} {\bibfnamefont {Zong-Kuan}\ \bibnamefont {Guo}}, \
  and\ \bibinfo {author} {\bibfnamefont {Rong-Gen}\ \bibnamefont {Cai}},\
  }\bibfield  {title} {\enquote {\bibinfo {title} {{Analytical approximation of
  the scalar spectrum in the ultraslow-roll inflationary models}},}\ }\href
  {\doibase 10.1103/PhysRevD.101.083535} {\bibfield  {journal} {\bibinfo
  {journal} {Phys. Rev. D}\ }\textbf {\bibinfo {volume} {101}},\ \bibinfo
  {pages} {083535} (\bibinfo {year} {2020})},\ \Eprint
  {http://arxiv.org/abs/2003.02075} {arXiv:2003.02075 [astro-ph.CO]}
  \BibitemShut {NoStop}%
\bibitem [{\citenamefont {Atal}\ \emph {et~al.}(2020)\citenamefont {Atal},
  \citenamefont {Cid}, \citenamefont {Escriv\`a},\ and\ \citenamefont
  {Garriga}}]{Atal:2019erb}%
  \BibitemOpen
  \bibfield  {author} {\bibinfo {author} {\bibfnamefont {Vicente}\ \bibnamefont
  {Atal}}, \bibinfo {author} {\bibfnamefont {Judith}\ \bibnamefont {Cid}},
  \bibinfo {author} {\bibfnamefont {Albert}\ \bibnamefont {Escriv\`a}}, \ and\
  \bibinfo {author} {\bibfnamefont {Jaume}\ \bibnamefont {Garriga}},\
  }\bibfield  {title} {\enquote {\bibinfo {title} {{PBH in single field
  inflation: the effect of shape dispersion and non-Gaussianities}},}\ }\href
  {\doibase 10.1088/1475-7516/2020/05/022} {\bibfield  {journal} {\bibinfo
  {journal} {JCAP}\ }\textbf {\bibinfo {volume} {05}},\ \bibinfo {pages} {022}
  (\bibinfo {year} {2020})},\ \Eprint {http://arxiv.org/abs/1908.11357}
  {arXiv:1908.11357 [astro-ph.CO]} \BibitemShut {NoStop}%
\bibitem [{\citenamefont {Fu}\ \emph {et~al.}(2020)\citenamefont {Fu},
  \citenamefont {Wu},\ and\ \citenamefont {Yu}}]{Fu:2020lob}%
  \BibitemOpen
  \bibfield  {author} {\bibinfo {author} {\bibfnamefont {Chengjie}\
  \bibnamefont {Fu}}, \bibinfo {author} {\bibfnamefont {Puxun}\ \bibnamefont
  {Wu}}, \ and\ \bibinfo {author} {\bibfnamefont {Hongwei}\ \bibnamefont
  {Yu}},\ }\bibfield  {title} {\enquote {\bibinfo {title} {{Primordial black
  holes and oscillating gravitational waves in slow-roll and slow-climb
  inflation with an intermediate noninflationary phase}},}\ }\href {\doibase
  10.1103/PhysRevD.102.043527} {\bibfield  {journal} {\bibinfo  {journal}
  {Phys. Rev. D}\ }\textbf {\bibinfo {volume} {102}},\ \bibinfo {pages}
  {043527} (\bibinfo {year} {2020})},\ \Eprint
  {http://arxiv.org/abs/2006.03768} {arXiv:2006.03768 [astro-ph.CO]}
  \BibitemShut {NoStop}%
\bibitem [{\citenamefont {Vennin}(2020)}]{Vennin:2020kng}%
  \BibitemOpen
  \bibfield  {author} {\bibinfo {author} {\bibfnamefont {Vincent}\ \bibnamefont
  {Vennin}},\ }\emph {\bibinfo {title} {{Stochastic inflation and primordial
  black holes}}},\ \href@noop {} {\bibinfo {type} {Other thesis}} (\bibinfo
  {year} {2020}),\ \Eprint {http://arxiv.org/abs/2009.08715} {arXiv:2009.08715
  [astro-ph.CO]} \BibitemShut {NoStop}%
\bibitem [{\citenamefont {Ragavendra}\ \emph {et~al.}(2020)\citenamefont
  {Ragavendra}, \citenamefont {Saha}, \citenamefont {Sriramkumar},\ and\
  \citenamefont {Silk}}]{Ragavendra:2020sop}%
  \BibitemOpen
  \bibfield  {author} {\bibinfo {author} {\bibfnamefont {H.~V.}\ \bibnamefont
  {Ragavendra}}, \bibinfo {author} {\bibfnamefont {Pankaj}\ \bibnamefont
  {Saha}}, \bibinfo {author} {\bibfnamefont {L.}~\bibnamefont {Sriramkumar}}, \
  and\ \bibinfo {author} {\bibfnamefont {Joseph}\ \bibnamefont {Silk}},\
  }\bibfield  {title} {\enquote {\bibinfo {title} {{PBHs and secondary GWs from
  ultra slow roll and punctuated inflation}},}\ }\href@noop {} {\  (\bibinfo
  {year} {2020})},\ \Eprint {http://arxiv.org/abs/2008.12202} {arXiv:2008.12202
  [astro-ph.CO]} \BibitemShut {NoStop}%
\bibitem [{\citenamefont {Gao}\ and\ \citenamefont {Yang}(2021)}]{Gao:2021dfi}%
  \BibitemOpen
  \bibfield  {author} {\bibinfo {author} {\bibfnamefont {Tie-Jun}\ \bibnamefont
  {Gao}}\ and\ \bibinfo {author} {\bibfnamefont {Xiu-Yi}\ \bibnamefont
  {Yang}},\ }\bibfield  {title} {\enquote {\bibinfo {title} {{Double peaks of
  gravitational wave spectrum induced from inflection point inflation}},}\
  }\href {\doibase 10.1140/epjc/s10052-021-09269-4} {\bibfield  {journal}
  {\bibinfo  {journal} {Eur. Phys. J. C}\ }\textbf {\bibinfo {volume} {81}},\
  \bibinfo {pages} {494} (\bibinfo {year} {2021})},\ \Eprint
  {http://arxiv.org/abs/2101.07616} {arXiv:2101.07616 [astro-ph.CO]}
  \BibitemShut {NoStop}%
\bibitem [{\citenamefont {Cai}\ \emph {et~al.}(2022)\citenamefont {Cai},
  \citenamefont {Ma}, \citenamefont {Sasaki}, \citenamefont {Wang},\ and\
  \citenamefont {Zhou}}]{Cai:2022erk}%
  \BibitemOpen
  \bibfield  {author} {\bibinfo {author} {\bibfnamefont {Yi-Fu}\ \bibnamefont
  {Cai}}, \bibinfo {author} {\bibfnamefont {Xiao-Han}\ \bibnamefont {Ma}},
  \bibinfo {author} {\bibfnamefont {Misao}\ \bibnamefont {Sasaki}}, \bibinfo
  {author} {\bibfnamefont {Dong-Gang}\ \bibnamefont {Wang}}, \ and\ \bibinfo
  {author} {\bibfnamefont {Zihan}\ \bibnamefont {Zhou}},\ }\bibfield  {title}
  {\enquote {\bibinfo {title} {{Highly non-Gaussian tails and primordial black
  holes from single-field inflation}},}\ }\href@noop {} {\  (\bibinfo {year}
  {2022})},\ \Eprint {http://arxiv.org/abs/2207.11910} {arXiv:2207.11910
  [astro-ph.CO]} \BibitemShut {NoStop}%
\bibitem [{\citenamefont {Karam}\ \emph {et~al.}(2022)\citenamefont {Karam},
  \citenamefont {Koivunen}, \citenamefont {Tomberg}, \citenamefont {Vaskonen},\
  and\ \citenamefont {Veerm\"ae}}]{Karam:2022nym}%
  \BibitemOpen
  \bibfield  {author} {\bibinfo {author} {\bibfnamefont {Alexandros}\
  \bibnamefont {Karam}}, \bibinfo {author} {\bibfnamefont {Niko}\ \bibnamefont
  {Koivunen}}, \bibinfo {author} {\bibfnamefont {Eemeli}\ \bibnamefont
  {Tomberg}}, \bibinfo {author} {\bibfnamefont {Ville}\ \bibnamefont
  {Vaskonen}}, \ and\ \bibinfo {author} {\bibfnamefont {Hardi}\ \bibnamefont
  {Veerm\"ae}},\ }\bibfield  {title} {\enquote {\bibinfo {title} {{Anatomy of
  single-field inflationary models for primordial black holes}},}\ }\href@noop
  {} {\  (\bibinfo {year} {2022})},\ \Eprint {http://arxiv.org/abs/2205.13540}
  {arXiv:2205.13540 [astro-ph.CO]} \BibitemShut {NoStop}%
\bibitem [{\citenamefont {Di}\ and\ \citenamefont {Gong}(2018)}]{Di:2017ndc}%
  \BibitemOpen
  \bibfield  {author} {\bibinfo {author} {\bibfnamefont {Haoran}\ \bibnamefont
  {Di}}\ and\ \bibinfo {author} {\bibfnamefont {Yungui}\ \bibnamefont {Gong}},\
  }\bibfield  {title} {\enquote {\bibinfo {title} {{Primordial black holes and
  second order gravitational waves from ultra-slow-roll inflation}},}\ }\href
  {\doibase 10.1088/1475-7516/2018/07/007} {\bibfield  {journal} {\bibinfo
  {journal} {JCAP}\ }\textbf {\bibinfo {volume} {07}},\ \bibinfo {pages} {007}
  (\bibinfo {year} {2018})},\ \Eprint {http://arxiv.org/abs/1707.09578}
  {arXiv:1707.09578 [astro-ph.CO]} \BibitemShut {NoStop}%
\bibitem [{\citenamefont {Cai}\ \emph {et~al.}(2018)\citenamefont {Cai},
  \citenamefont {Tong}, \citenamefont {Wang},\ and\ \citenamefont
  {Yan}}]{Cai:2018tuh}%
  \BibitemOpen
  \bibfield  {author} {\bibinfo {author} {\bibfnamefont {Yi-Fu}\ \bibnamefont
  {Cai}}, \bibinfo {author} {\bibfnamefont {Xi}~\bibnamefont {Tong}}, \bibinfo
  {author} {\bibfnamefont {Dong-Gang}\ \bibnamefont {Wang}}, \ and\ \bibinfo
  {author} {\bibfnamefont {Sheng-Feng}\ \bibnamefont {Yan}},\ }\bibfield
  {title} {\enquote {\bibinfo {title} {{Primordial Black Holes from Sound Speed
  Resonance during Inflation}},}\ }\href {\doibase
  10.1103/PhysRevLett.121.081306} {\bibfield  {journal} {\bibinfo  {journal}
  {Phys. Rev. Lett.}\ }\textbf {\bibinfo {volume} {121}},\ \bibinfo {pages}
  {081306} (\bibinfo {year} {2018})},\ \Eprint
  {http://arxiv.org/abs/1805.03639} {arXiv:1805.03639 [astro-ph.CO]}
  \BibitemShut {NoStop}%
\bibitem [{\citenamefont {Chen}\ \emph
  {et~al.}(2020{\natexlab{b}})\citenamefont {Chen}, \citenamefont {Ma},\ and\
  \citenamefont {Cai}}]{Chen:2020uhe}%
  \BibitemOpen
  \bibfield  {author} {\bibinfo {author} {\bibfnamefont {Chao}\ \bibnamefont
  {Chen}}, \bibinfo {author} {\bibfnamefont {Xiao-Han}\ \bibnamefont {Ma}}, \
  and\ \bibinfo {author} {\bibfnamefont {Yi-Fu}\ \bibnamefont {Cai}},\
  }\bibfield  {title} {\enquote {\bibinfo {title} {{Dirac-Born-Infeld
  realization of sound speed resonance mechanism for primordial black
  holes}},}\ }\href {\doibase 10.1103/PhysRevD.102.063526} {\bibfield
  {journal} {\bibinfo  {journal} {Phys. Rev. D}\ }\textbf {\bibinfo {volume}
  {102}},\ \bibinfo {pages} {063526} (\bibinfo {year} {2020}{\natexlab{b}})},\
  \Eprint {http://arxiv.org/abs/2003.03821} {arXiv:2003.03821 [astro-ph.CO]}
  \BibitemShut {NoStop}%
\bibitem [{\citenamefont {Cai}\ \emph {et~al.}(2019{\natexlab{a}})\citenamefont
  {Cai}, \citenamefont {Chen}, \citenamefont {Tong}, \citenamefont {Wang},\
  and\ \citenamefont {Yan}}]{Cai:2019jah}%
  \BibitemOpen
  \bibfield  {author} {\bibinfo {author} {\bibfnamefont {Yi-Fu}\ \bibnamefont
  {Cai}}, \bibinfo {author} {\bibfnamefont {Chao}\ \bibnamefont {Chen}},
  \bibinfo {author} {\bibfnamefont {Xi}~\bibnamefont {Tong}}, \bibinfo {author}
  {\bibfnamefont {Dong-Gang}\ \bibnamefont {Wang}}, \ and\ \bibinfo {author}
  {\bibfnamefont {Sheng-Feng}\ \bibnamefont {Yan}},\ }\bibfield  {title}
  {\enquote {\bibinfo {title} {{When Primordial Black Holes from Sound Speed
  Resonance Meet a Stochastic Background of Gravitational Waves}},}\ }\href
  {\doibase 10.1103/PhysRevD.100.043518} {\bibfield  {journal} {\bibinfo
  {journal} {Phys. Rev. D}\ }\textbf {\bibinfo {volume} {100}},\ \bibinfo
  {pages} {043518} (\bibinfo {year} {2019}{\natexlab{a}})},\ \Eprint
  {http://arxiv.org/abs/1902.08187} {arXiv:1902.08187 [astro-ph.CO]}
  \BibitemShut {NoStop}%
\bibitem [{\citenamefont {Cai}\ \emph {et~al.}(2020)\citenamefont {Cai},
  \citenamefont {Guo}, \citenamefont {Liu}, \citenamefont {Liu},\ and\
  \citenamefont {Yang}}]{Cai:2019bmk}%
  \BibitemOpen
  \bibfield  {author} {\bibinfo {author} {\bibfnamefont {Rong-Gen}\
  \bibnamefont {Cai}}, \bibinfo {author} {\bibfnamefont {Zong-Kuan}\
  \bibnamefont {Guo}}, \bibinfo {author} {\bibfnamefont {Jing}\ \bibnamefont
  {Liu}}, \bibinfo {author} {\bibfnamefont {Lang}\ \bibnamefont {Liu}}, \ and\
  \bibinfo {author} {\bibfnamefont {Xing-Yu}\ \bibnamefont {Yang}},\ }\bibfield
   {title} {\enquote {\bibinfo {title} {{Primordial black holes and
  gravitational waves from parametric amplification of curvature
  perturbations}},}\ }\href {\doibase 10.1088/1475-7516/2020/06/013} {\bibfield
   {journal} {\bibinfo  {journal} {JCAP}\ }\textbf {\bibinfo {volume} {06}},\
  \bibinfo {pages} {013} (\bibinfo {year} {2020})},\ \Eprint
  {http://arxiv.org/abs/1912.10437} {arXiv:1912.10437 [astro-ph.CO]}
  \BibitemShut {NoStop}%
\bibitem [{\citenamefont {Cotner}\ and\ \citenamefont
  {Kusenko}(2017{\natexlab{a}})}]{Cotner:2016cvr}%
  \BibitemOpen
  \bibfield  {author} {\bibinfo {author} {\bibfnamefont {Eric}\ \bibnamefont
  {Cotner}}\ and\ \bibinfo {author} {\bibfnamefont {Alexander}\ \bibnamefont
  {Kusenko}},\ }\bibfield  {title} {\enquote {\bibinfo {title} {{Primordial
  black holes from supersymmetry in the early universe}},}\ }\href {\doibase
  10.1103/PhysRevLett.119.031103} {\bibfield  {journal} {\bibinfo  {journal}
  {Phys. Rev. Lett.}\ }\textbf {\bibinfo {volume} {119}},\ \bibinfo {pages}
  {031103} (\bibinfo {year} {2017}{\natexlab{a}})},\ \Eprint
  {http://arxiv.org/abs/1612.02529} {arXiv:1612.02529 [astro-ph.CO]}
  \BibitemShut {NoStop}%
\bibitem [{\citenamefont {Cotner}\ and\ \citenamefont
  {Kusenko}(2017{\natexlab{b}})}]{Cotner:2017tir}%
  \BibitemOpen
  \bibfield  {author} {\bibinfo {author} {\bibfnamefont {Eric}\ \bibnamefont
  {Cotner}}\ and\ \bibinfo {author} {\bibfnamefont {Alexander}\ \bibnamefont
  {Kusenko}},\ }\bibfield  {title} {\enquote {\bibinfo {title} {{Primordial
  black holes from scalar field evolution in the early universe}},}\ }\href
  {\doibase 10.1103/PhysRevD.96.103002} {\bibfield  {journal} {\bibinfo
  {journal} {Phys. Rev. D}\ }\textbf {\bibinfo {volume} {96}},\ \bibinfo
  {pages} {103002} (\bibinfo {year} {2017}{\natexlab{b}})},\ \Eprint
  {http://arxiv.org/abs/1706.09003} {arXiv:1706.09003 [astro-ph.CO]}
  \BibitemShut {NoStop}%
\bibitem [{\citenamefont {Cotner}\ \emph {et~al.}(2018)\citenamefont {Cotner},
  \citenamefont {Kusenko},\ and\ \citenamefont {Takhistov}}]{Cotner:2018vug}%
  \BibitemOpen
  \bibfield  {author} {\bibinfo {author} {\bibfnamefont {Eric}\ \bibnamefont
  {Cotner}}, \bibinfo {author} {\bibfnamefont {Alexander}\ \bibnamefont
  {Kusenko}}, \ and\ \bibinfo {author} {\bibfnamefont {Volodymyr}\ \bibnamefont
  {Takhistov}},\ }\bibfield  {title} {\enquote {\bibinfo {title} {{Primordial
  Black Holes from Inflaton Fragmentation into Oscillons}},}\ }\href {\doibase
  10.1103/PhysRevD.98.083513} {\bibfield  {journal} {\bibinfo  {journal} {Phys.
  Rev. D}\ }\textbf {\bibinfo {volume} {98}},\ \bibinfo {pages} {083513}
  (\bibinfo {year} {2018})},\ \Eprint {http://arxiv.org/abs/1801.03321}
  {arXiv:1801.03321 [astro-ph.CO]} \BibitemShut {NoStop}%
\bibitem [{\citenamefont {Cotner}\ \emph {et~al.}(2019)\citenamefont {Cotner},
  \citenamefont {Kusenko}, \citenamefont {Sasaki},\ and\ \citenamefont
  {Takhistov}}]{Cotner:2019ykd}%
  \BibitemOpen
  \bibfield  {author} {\bibinfo {author} {\bibfnamefont {Eric}\ \bibnamefont
  {Cotner}}, \bibinfo {author} {\bibfnamefont {Alexander}\ \bibnamefont
  {Kusenko}}, \bibinfo {author} {\bibfnamefont {Misao}\ \bibnamefont {Sasaki}},
  \ and\ \bibinfo {author} {\bibfnamefont {Volodymyr}\ \bibnamefont
  {Takhistov}},\ }\bibfield  {title} {\enquote {\bibinfo {title} {{Analytic
  Description of Primordial Black Hole Formation from Scalar Field
  Fragmentation}},}\ }\href {\doibase 10.1088/1475-7516/2019/10/077} {\bibfield
   {journal} {\bibinfo  {journal} {JCAP}\ }\textbf {\bibinfo {volume} {10}},\
  \bibinfo {pages} {077} (\bibinfo {year} {2019})},\ \Eprint
  {http://arxiv.org/abs/1907.10613} {arXiv:1907.10613 [astro-ph.CO]}
  \BibitemShut {NoStop}%
\bibitem [{\citenamefont {Escriv\`a}\ and\ \citenamefont
  {Subils}(2022)}]{Escriva:2022yaf}%
  \BibitemOpen
  \bibfield  {author} {\bibinfo {author} {\bibfnamefont {Albert}\ \bibnamefont
  {Escriv\`a}}\ and\ \bibinfo {author} {\bibfnamefont {Javier~G.}\ \bibnamefont
  {Subils}},\ }\bibfield  {title} {\enquote {\bibinfo {title} {{Primordial
  Black Hole Formation during a Strongly Coupled Crossover}},}\ }\href@noop {}
  {\  (\bibinfo {year} {2022})},\ \Eprint {http://arxiv.org/abs/2211.15674}
  {arXiv:2211.15674 [astro-ph.CO]} \BibitemShut {NoStop}%
\bibitem [{\citenamefont {Pi}\ and\ \citenamefont {Wang}(2022)}]{Pi:2022zxs}%
  \BibitemOpen
  \bibfield  {author} {\bibinfo {author} {\bibfnamefont {Shi}\ \bibnamefont
  {Pi}}\ and\ \bibinfo {author} {\bibfnamefont {Jianing}\ \bibnamefont
  {Wang}},\ }\bibfield  {title} {\enquote {\bibinfo {title} {{Primordial Black
  Hole Formation in Starobinsky's Linear Potential Model}},}\ }\href@noop {} {\
   (\bibinfo {year} {2022})},\ \Eprint {http://arxiv.org/abs/2209.14183}
  {arXiv:2209.14183 [astro-ph.CO]} \BibitemShut {NoStop}%
\bibitem [{\citenamefont {Germani}\ and\ \citenamefont
  {Prokopec}(2017)}]{Germani:2017bcs}%
  \BibitemOpen
  \bibfield  {author} {\bibinfo {author} {\bibfnamefont {Cristiano}\
  \bibnamefont {Germani}}\ and\ \bibinfo {author} {\bibfnamefont {Tomislav}\
  \bibnamefont {Prokopec}},\ }\bibfield  {title} {\enquote {\bibinfo {title}
  {{On primordial black holes from an inflection point}},}\ }\href {\doibase
  10.1016/j.dark.2017.09.001} {\bibfield  {journal} {\bibinfo  {journal} {Phys.
  Dark Univ.}\ }\textbf {\bibinfo {volume} {18}},\ \bibinfo {pages} {6--10}
  (\bibinfo {year} {2017})},\ \Eprint {http://arxiv.org/abs/1706.04226}
  {arXiv:1706.04226 [astro-ph.CO]} \BibitemShut {NoStop}%
\bibitem [{\citenamefont {Byrnes}\ \emph {et~al.}(2019)\citenamefont {Byrnes},
  \citenamefont {Cole},\ and\ \citenamefont {Patil}}]{Byrnes:2018txb}%
  \BibitemOpen
  \bibfield  {author} {\bibinfo {author} {\bibfnamefont {Christian~T.}\
  \bibnamefont {Byrnes}}, \bibinfo {author} {\bibfnamefont {Philippa~S.}\
  \bibnamefont {Cole}}, \ and\ \bibinfo {author} {\bibfnamefont {Subodh~P.}\
  \bibnamefont {Patil}},\ }\bibfield  {title} {\enquote {\bibinfo {title}
  {{Steepest growth of the power spectrum and primordial black holes}},}\
  }\href {\doibase 10.1088/1475-7516/2019/06/028} {\bibfield  {journal}
  {\bibinfo  {journal} {JCAP}\ }\textbf {\bibinfo {volume} {06}},\ \bibinfo
  {pages} {028} (\bibinfo {year} {2019})},\ \Eprint
  {http://arxiv.org/abs/1811.11158} {arXiv:1811.11158 [astro-ph.CO]}
  \BibitemShut {NoStop}%
\bibitem [{\citenamefont {Passaglia}\ \emph {et~al.}(2019)\citenamefont
  {Passaglia}, \citenamefont {Hu},\ and\ \citenamefont
  {Motohashi}}]{Passaglia:2018ixg}%
  \BibitemOpen
  \bibfield  {author} {\bibinfo {author} {\bibfnamefont {Samuel}\ \bibnamefont
  {Passaglia}}, \bibinfo {author} {\bibfnamefont {Wayne}\ \bibnamefont {Hu}}, \
  and\ \bibinfo {author} {\bibfnamefont {Hayato}\ \bibnamefont {Motohashi}},\
  }\bibfield  {title} {\enquote {\bibinfo {title} {{Primordial black holes and
  local non-Gaussianity in canonical inflation}},}\ }\href {\doibase
  10.1103/PhysRevD.99.043536} {\bibfield  {journal} {\bibinfo  {journal} {Phys.
  Rev. D}\ }\textbf {\bibinfo {volume} {99}},\ \bibinfo {pages} {043536}
  (\bibinfo {year} {2019})},\ \Eprint {http://arxiv.org/abs/1812.08243}
  {arXiv:1812.08243 [astro-ph.CO]} \BibitemShut {NoStop}%
\bibitem [{\citenamefont {Fu}\ \emph {et~al.}(2019{\natexlab{a}})\citenamefont
  {Fu}, \citenamefont {Wu},\ and\ \citenamefont {Yu}}]{Fu:2019ttf}%
  \BibitemOpen
  \bibfield  {author} {\bibinfo {author} {\bibfnamefont {Chengjie}\
  \bibnamefont {Fu}}, \bibinfo {author} {\bibfnamefont {Puxun}\ \bibnamefont
  {Wu}}, \ and\ \bibinfo {author} {\bibfnamefont {Hongwei}\ \bibnamefont
  {Yu}},\ }\bibfield  {title} {\enquote {\bibinfo {title} {{Primordial Black
  Holes from Inflation with Nonminimal Derivative Coupling}},}\ }\href
  {\doibase 10.1103/PhysRevD.100.063532} {\bibfield  {journal} {\bibinfo
  {journal} {Phys. Rev. D}\ }\textbf {\bibinfo {volume} {100}},\ \bibinfo
  {pages} {063532} (\bibinfo {year} {2019}{\natexlab{a}})},\ \Eprint
  {http://arxiv.org/abs/1907.05042} {arXiv:1907.05042 [astro-ph.CO]}
  \BibitemShut {NoStop}%
\bibitem [{\citenamefont {Fu}\ \emph {et~al.}(2019{\natexlab{b}})\citenamefont
  {Fu}, \citenamefont {Wu},\ and\ \citenamefont {Yu}}]{Fu:2019vqc}%
  \BibitemOpen
  \bibfield  {author} {\bibinfo {author} {\bibfnamefont {Chengjie}\
  \bibnamefont {Fu}}, \bibinfo {author} {\bibfnamefont {Puxun}\ \bibnamefont
  {Wu}}, \ and\ \bibinfo {author} {\bibfnamefont {Hongwei}\ \bibnamefont
  {Yu}},\ }\bibfield  {title} {\enquote {\bibinfo {title} {{Scalar induced
  gravitational waves in inflation with gravitationally enhanced friction}},}\
  }\href@noop {} {\  (\bibinfo {year} {2019}{\natexlab{b}})},\ \Eprint
  {http://arxiv.org/abs/1912.05927} {arXiv:1912.05927 [astro-ph.CO]}
  \BibitemShut {NoStop}%
\bibitem [{\citenamefont {Inomata}\ \emph
  {et~al.}(2022{\natexlab{a}})\citenamefont {Inomata}, \citenamefont
  {McDonough},\ and\ \citenamefont {Hu}}]{Inomata:2021tpx}%
  \BibitemOpen
  \bibfield  {author} {\bibinfo {author} {\bibfnamefont {Keisuke}\ \bibnamefont
  {Inomata}}, \bibinfo {author} {\bibfnamefont {Evan}\ \bibnamefont
  {McDonough}}, \ and\ \bibinfo {author} {\bibfnamefont {Wayne}\ \bibnamefont
  {Hu}},\ }\bibfield  {title} {\enquote {\bibinfo {title} {{Amplification of
  primordial perturbations from the rise or fall of the inflaton}},}\ }\href
  {\doibase 10.1088/1475-7516/2022/02/031} {\bibfield  {journal} {\bibinfo
  {journal} {JCAP}\ }\textbf {\bibinfo {volume} {02}},\ \bibinfo {pages} {031}
  (\bibinfo {year} {2022}{\natexlab{a}})},\ \Eprint
  {http://arxiv.org/abs/2110.14641} {arXiv:2110.14641 [astro-ph.CO]}
  \BibitemShut {NoStop}%
\bibitem [{\citenamefont {Tasinato}(2021)}]{Tasinato:2020vdk}%
  \BibitemOpen
  \bibfield  {author} {\bibinfo {author} {\bibfnamefont {Gianmassimo}\
  \bibnamefont {Tasinato}},\ }\bibfield  {title} {\enquote {\bibinfo {title}
  {{An analytic approach to non-slow-roll inflation}},}\ }\href {\doibase
  10.1103/PhysRevD.103.023535} {\bibfield  {journal} {\bibinfo  {journal}
  {Phys. Rev. D}\ }\textbf {\bibinfo {volume} {103}},\ \bibinfo {pages}
  {023535} (\bibinfo {year} {2021})},\ \Eprint
  {http://arxiv.org/abs/2012.02518} {arXiv:2012.02518 [hep-th]} \BibitemShut
  {NoStop}%
\bibitem [{\citenamefont {Cole}\ \emph {et~al.}(2022)\citenamefont {Cole},
  \citenamefont {Gow}, \citenamefont {Byrnes},\ and\ \citenamefont
  {Patil}}]{Cole:2022xqc}%
  \BibitemOpen
  \bibfield  {author} {\bibinfo {author} {\bibfnamefont {Philippa~S.}\
  \bibnamefont {Cole}}, \bibinfo {author} {\bibfnamefont {Andrew~D.}\
  \bibnamefont {Gow}}, \bibinfo {author} {\bibfnamefont {Christian~T.}\
  \bibnamefont {Byrnes}}, \ and\ \bibinfo {author} {\bibfnamefont {Subodh~P.}\
  \bibnamefont {Patil}},\ }\bibfield  {title} {\enquote {\bibinfo {title}
  {{Steepest growth re-examined: repercussions for primordial black hole
  formation}},}\ }\href@noop {} {\  (\bibinfo {year} {2022})},\ \Eprint
  {http://arxiv.org/abs/2204.07573} {arXiv:2204.07573 [astro-ph.CO]}
  \BibitemShut {NoStop}%
\bibitem [{\citenamefont {Fu}\ and\ \citenamefont {Wang}(2022)}]{Fu:2022ypp}%
  \BibitemOpen
  \bibfield  {author} {\bibinfo {author} {\bibfnamefont {Chengjie}\
  \bibnamefont {Fu}}\ and\ \bibinfo {author} {\bibfnamefont {Shao-Jiang}\
  \bibnamefont {Wang}},\ }\bibfield  {title} {\enquote {\bibinfo {title}
  {{Primordial black holes and induced gravitational waves from double-pole
  inflation}},}\ }\href@noop {} {\  (\bibinfo {year} {2022})},\ \Eprint
  {http://arxiv.org/abs/2211.03523} {arXiv:2211.03523 [astro-ph.CO]}
  \BibitemShut {NoStop}%
\bibitem [{\citenamefont {Peng}\ \emph {et~al.}(2021)\citenamefont {Peng},
  \citenamefont {Fu}, \citenamefont {Liu}, \citenamefont {Guo},\ and\
  \citenamefont {Cai}}]{Peng:2021zon}%
  \BibitemOpen
  \bibfield  {author} {\bibinfo {author} {\bibfnamefont {Zhi-Zhang}\
  \bibnamefont {Peng}}, \bibinfo {author} {\bibfnamefont {Chengjie}\
  \bibnamefont {Fu}}, \bibinfo {author} {\bibfnamefont {Jing}\ \bibnamefont
  {Liu}}, \bibinfo {author} {\bibfnamefont {Zong-Kuan}\ \bibnamefont {Guo}}, \
  and\ \bibinfo {author} {\bibfnamefont {Rong-Gen}\ \bibnamefont {Cai}},\
  }\bibfield  {title} {\enquote {\bibinfo {title} {{Gravitational waves from
  resonant amplification of curvature perturbations during inflation}},}\
  }\href {\doibase 10.1088/1475-7516/2021/10/050} {\bibfield  {journal}
  {\bibinfo  {journal} {JCAP}\ }\textbf {\bibinfo {volume} {10}},\ \bibinfo
  {pages} {050} (\bibinfo {year} {2021})},\ \Eprint
  {http://arxiv.org/abs/2106.11816} {arXiv:2106.11816 [astro-ph.CO]}
  \BibitemShut {NoStop}%
\bibitem [{\citenamefont {Zhai}\ \emph {et~al.}(2022)\citenamefont {Zhai},
  \citenamefont {Yu},\ and\ \citenamefont {Wu}}]{Zhai:2022mpi}%
  \BibitemOpen
  \bibfield  {author} {\bibinfo {author} {\bibfnamefont {Rongrong}\
  \bibnamefont {Zhai}}, \bibinfo {author} {\bibfnamefont {Hongwei}\
  \bibnamefont {Yu}}, \ and\ \bibinfo {author} {\bibfnamefont {Puxun}\
  \bibnamefont {Wu}},\ }\bibfield  {title} {\enquote {\bibinfo {title} {{Growth
  of power spectrum due to decrease of sound speed during inflation}},}\ }\href
  {\doibase 10.1103/PhysRevD.106.023517} {\bibfield  {journal} {\bibinfo
  {journal} {Phys. Rev. D}\ }\textbf {\bibinfo {volume} {106}},\ \bibinfo
  {pages} {023517} (\bibinfo {year} {2022})},\ \Eprint
  {http://arxiv.org/abs/2207.12745} {arXiv:2207.12745 [gr-qc]} \BibitemShut
  {NoStop}%
\bibitem [{\citenamefont {Kannike}\ \emph {et~al.}(2017)\citenamefont
  {Kannike}, \citenamefont {Marzola}, \citenamefont {Raidal},\ and\
  \citenamefont {Veerm\"ae}}]{Kannike:2017bxn}%
  \BibitemOpen
  \bibfield  {author} {\bibinfo {author} {\bibfnamefont {Kristjan}\
  \bibnamefont {Kannike}}, \bibinfo {author} {\bibfnamefont {Luca}\
  \bibnamefont {Marzola}}, \bibinfo {author} {\bibfnamefont {Martti}\
  \bibnamefont {Raidal}}, \ and\ \bibinfo {author} {\bibfnamefont {Hardi}\
  \bibnamefont {Veerm\"ae}},\ }\bibfield  {title} {\enquote {\bibinfo {title}
  {{Single Field Double Inflation and Primordial Black Holes}},}\ }\href
  {\doibase 10.1088/1475-7516/2017/09/020} {\bibfield  {journal} {\bibinfo
  {journal} {JCAP}\ }\textbf {\bibinfo {volume} {09}},\ \bibinfo {pages} {020}
  (\bibinfo {year} {2017})},\ \Eprint {http://arxiv.org/abs/1705.06225}
  {arXiv:1705.06225 [astro-ph.CO]} \BibitemShut {NoStop}%
\bibitem [{\citenamefont {Gao}\ and\ \citenamefont {Guo}(2018)}]{Gao:2018pvq}%
  \BibitemOpen
  \bibfield  {author} {\bibinfo {author} {\bibfnamefont {Tie-Jun}\ \bibnamefont
  {Gao}}\ and\ \bibinfo {author} {\bibfnamefont {Zong-Kuan}\ \bibnamefont
  {Guo}},\ }\bibfield  {title} {\enquote {\bibinfo {title} {{Primordial Black
  Hole Production in Inflationary Models of Supergravity with a Single Chiral
  Superfield}},}\ }\href {\doibase 10.1103/PhysRevD.98.063526} {\bibfield
  {journal} {\bibinfo  {journal} {Phys. Rev. D}\ }\textbf {\bibinfo {volume}
  {98}},\ \bibinfo {pages} {063526} (\bibinfo {year} {2018})},\ \Eprint
  {http://arxiv.org/abs/1806.09320} {arXiv:1806.09320 [hep-ph]} \BibitemShut
  {NoStop}%
\bibitem [{\citenamefont {Cheong}\ \emph {et~al.}(2021)\citenamefont {Cheong},
  \citenamefont {Lee},\ and\ \citenamefont {Park}}]{Cheong:2019vzl}%
  \BibitemOpen
  \bibfield  {author} {\bibinfo {author} {\bibfnamefont {Dhong~Yeon}\
  \bibnamefont {Cheong}}, \bibinfo {author} {\bibfnamefont {Sung~Mook}\
  \bibnamefont {Lee}}, \ and\ \bibinfo {author} {\bibfnamefont {Seong~Chan}\
  \bibnamefont {Park}},\ }\bibfield  {title} {\enquote {\bibinfo {title}
  {{Primordial black holes in Higgs-$R^2$ inflation as the whole of dark
  matter}},}\ }\href {\doibase 10.1088/1475-7516/2021/01/032} {\bibfield
  {journal} {\bibinfo  {journal} {JCAP}\ }\textbf {\bibinfo {volume} {01}},\
  \bibinfo {pages} {032} (\bibinfo {year} {2021})},\ \Eprint
  {http://arxiv.org/abs/1912.12032} {arXiv:1912.12032 [hep-ph]} \BibitemShut
  {NoStop}%
\bibitem [{\citenamefont {Cheong}\ \emph {et~al.}(2020)\citenamefont {Cheong},
  \citenamefont {Lee},\ and\ \citenamefont {Park}}]{Cheong:2020rao}%
  \BibitemOpen
  \bibfield  {author} {\bibinfo {author} {\bibfnamefont {Dhong~Yeon}\
  \bibnamefont {Cheong}}, \bibinfo {author} {\bibfnamefont {Hyun~Min}\
  \bibnamefont {Lee}}, \ and\ \bibinfo {author} {\bibfnamefont {Seong~Chan}\
  \bibnamefont {Park}},\ }\bibfield  {title} {\enquote {\bibinfo {title}
  {{Beyond the Starobinsky model for inflation}},}\ }\href {\doibase
  10.1016/j.physletb.2020.135453} {\bibfield  {journal} {\bibinfo  {journal}
  {Phys. Lett. B}\ }\textbf {\bibinfo {volume} {805}},\ \bibinfo {pages}
  {135453} (\bibinfo {year} {2020})},\ \Eprint
  {http://arxiv.org/abs/2002.07981} {arXiv:2002.07981 [hep-ph]} \BibitemShut
  {NoStop}%
\bibitem [{\citenamefont {Dalianis}\ \emph {et~al.}(2020)\citenamefont
  {Dalianis}, \citenamefont {Karydas},\ and\ \citenamefont
  {Papantonopoulos}}]{Dalianis:2019vit}%
  \BibitemOpen
  \bibfield  {author} {\bibinfo {author} {\bibfnamefont {Ioannis}\ \bibnamefont
  {Dalianis}}, \bibinfo {author} {\bibfnamefont {Stelios}\ \bibnamefont
  {Karydas}}, \ and\ \bibinfo {author} {\bibfnamefont {Eleftherios}\
  \bibnamefont {Papantonopoulos}},\ }\bibfield  {title} {\enquote {\bibinfo
  {title} {{Generalized Non-Minimal Derivative Coupling: Application to
  Inflation and Primordial Black Hole Production}},}\ }\href {\doibase
  10.1088/1475-7516/2020/06/040} {\bibfield  {journal} {\bibinfo  {journal}
  {JCAP}\ }\textbf {\bibinfo {volume} {06}},\ \bibinfo {pages} {040} (\bibinfo
  {year} {2020})},\ \Eprint {http://arxiv.org/abs/1910.00622} {arXiv:1910.00622
  [astro-ph.CO]} \BibitemShut {NoStop}%
\bibitem [{\citenamefont {Martin}\ \emph
  {et~al.}(2020{\natexlab{a}})\citenamefont {Martin}, \citenamefont
  {Papanikolaou},\ and\ \citenamefont {Vennin}}]{Martin:2019nuw}%
  \BibitemOpen
  \bibfield  {author} {\bibinfo {author} {\bibfnamefont {J\'er\^ome}\
  \bibnamefont {Martin}}, \bibinfo {author} {\bibfnamefont {Theodoros}\
  \bibnamefont {Papanikolaou}}, \ and\ \bibinfo {author} {\bibfnamefont
  {Vincent}\ \bibnamefont {Vennin}},\ }\bibfield  {title} {\enquote {\bibinfo
  {title} {{Primordial black holes from the preheating instability in
  single-field inflation}},}\ }\href {\doibase 10.1088/1475-7516/2020/01/024}
  {\bibfield  {journal} {\bibinfo  {journal} {JCAP}\ }\textbf {\bibinfo
  {volume} {01}},\ \bibinfo {pages} {024} (\bibinfo {year}
  {2020}{\natexlab{a}})},\ \Eprint {http://arxiv.org/abs/1907.04236}
  {arXiv:1907.04236 [astro-ph.CO]} \BibitemShut {NoStop}%
\bibitem [{\citenamefont {Martin}\ \emph
  {et~al.}(2020{\natexlab{b}})\citenamefont {Martin}, \citenamefont
  {Papanikolaou}, \citenamefont {Pinol},\ and\ \citenamefont
  {Vennin}}]{Martin:2020fgl}%
  \BibitemOpen
  \bibfield  {author} {\bibinfo {author} {\bibfnamefont {J\'er\^ome}\
  \bibnamefont {Martin}}, \bibinfo {author} {\bibfnamefont {Theodoros}\
  \bibnamefont {Papanikolaou}}, \bibinfo {author} {\bibfnamefont {Lucas}\
  \bibnamefont {Pinol}}, \ and\ \bibinfo {author} {\bibfnamefont {Vincent}\
  \bibnamefont {Vennin}},\ }\bibfield  {title} {\enquote {\bibinfo {title}
  {{Metric preheating and radiative decay in single-field inflation}},}\ }\href
  {\doibase 10.1088/1475-7516/2020/05/003} {\bibfield  {journal} {\bibinfo
  {journal} {JCAP}\ }\textbf {\bibinfo {volume} {05}},\ \bibinfo {pages} {003}
  (\bibinfo {year} {2020}{\natexlab{b}})},\ \Eprint
  {http://arxiv.org/abs/2002.01820} {arXiv:2002.01820 [astro-ph.CO]}
  \BibitemShut {NoStop}%
\bibitem [{\citenamefont {Lin}\ \emph {et~al.}(2020)\citenamefont {Lin},
  \citenamefont {Gao}, \citenamefont {Gong}, \citenamefont {Lu}, \citenamefont
  {Zhang},\ and\ \citenamefont {Zhang}}]{Lin:2020goi}%
  \BibitemOpen
  \bibfield  {author} {\bibinfo {author} {\bibfnamefont {Jiong}\ \bibnamefont
  {Lin}}, \bibinfo {author} {\bibfnamefont {Qing}\ \bibnamefont {Gao}},
  \bibinfo {author} {\bibfnamefont {Yungui}\ \bibnamefont {Gong}}, \bibinfo
  {author} {\bibfnamefont {Yizhou}\ \bibnamefont {Lu}}, \bibinfo {author}
  {\bibfnamefont {Chao}\ \bibnamefont {Zhang}}, \ and\ \bibinfo {author}
  {\bibfnamefont {Fengge}\ \bibnamefont {Zhang}},\ }\bibfield  {title}
  {\enquote {\bibinfo {title} {{Primordial black holes and secondary
  gravitational waves from $k$ and $G$ inflation}},}\ }\href {\doibase
  10.1103/PhysRevD.101.103515} {\bibfield  {journal} {\bibinfo  {journal}
  {Phys. Rev. D}\ }\textbf {\bibinfo {volume} {101}},\ \bibinfo {pages}
  {103515} (\bibinfo {year} {2020})},\ \Eprint
  {http://arxiv.org/abs/2001.05909} {arXiv:2001.05909 [gr-qc]} \BibitemShut
  {NoStop}%
\bibitem [{\citenamefont {Yi}\ \emph {et~al.}(2020)\citenamefont {Yi},
  \citenamefont {Gao}, \citenamefont {Gong},\ and\ \citenamefont
  {Zhu}}]{Yi:2020cut}%
  \BibitemOpen
  \bibfield  {author} {\bibinfo {author} {\bibfnamefont {Zhu}\ \bibnamefont
  {Yi}}, \bibinfo {author} {\bibfnamefont {Qing}\ \bibnamefont {Gao}}, \bibinfo
  {author} {\bibfnamefont {Yungui}\ \bibnamefont {Gong}}, \ and\ \bibinfo
  {author} {\bibfnamefont {Zong-hong}\ \bibnamefont {Zhu}},\ }\bibfield
  {title} {\enquote {\bibinfo {title} {{Primordial black holes and secondary
  gravitational waves from inflationary model with a non-canonical kinetic
  term}},}\ }\href@noop {} {\  (\bibinfo {year} {2020})},\ \Eprint
  {http://arxiv.org/abs/2011.10606} {arXiv:2011.10606 [astro-ph.CO]}
  \BibitemShut {NoStop}%
\bibitem [{\citenamefont {Gao}\ \emph {et~al.}(2020)\citenamefont {Gao},
  \citenamefont {Gong},\ and\ \citenamefont {Yi}}]{Gao:2020tsa}%
  \BibitemOpen
  \bibfield  {author} {\bibinfo {author} {\bibfnamefont {Qing}\ \bibnamefont
  {Gao}}, \bibinfo {author} {\bibfnamefont {Yungui}\ \bibnamefont {Gong}}, \
  and\ \bibinfo {author} {\bibfnamefont {Zhu}\ \bibnamefont {Yi}},\ }\bibfield
  {title} {\enquote {\bibinfo {title} {{Primordial black holes and secondary
  gravitational waves from natural inflation}},}\ }\href@noop {} {\  (\bibinfo
  {year} {2020})},\ \Eprint {http://arxiv.org/abs/2012.03856} {arXiv:2012.03856
  [gr-qc]} \BibitemShut {NoStop}%
\bibitem [{\citenamefont {Gao}(2021)}]{Gao:2021vxb}%
  \BibitemOpen
  \bibfield  {author} {\bibinfo {author} {\bibfnamefont {Qing}\ \bibnamefont
  {Gao}},\ }\bibfield  {title} {\enquote {\bibinfo {title} {{Primordial black
  holes and secondary gravitational waves from chaotic inflation}},}\
  }\href@noop {} {\  (\bibinfo {year} {2021})},\ \Eprint
  {http://arxiv.org/abs/2102.07369} {arXiv:2102.07369 [gr-qc]} \BibitemShut
  {NoStop}%
\bibitem [{\citenamefont {Wu}\ \emph {et~al.}(2021)\citenamefont {Wu},
  \citenamefont {Gong},\ and\ \citenamefont {Li}}]{Wu:2021zta}%
  \BibitemOpen
  \bibfield  {author} {\bibinfo {author} {\bibfnamefont {Lina}\ \bibnamefont
  {Wu}}, \bibinfo {author} {\bibfnamefont {Yungui}\ \bibnamefont {Gong}}, \
  and\ \bibinfo {author} {\bibfnamefont {Tianjun}\ \bibnamefont {Li}},\
  }\bibfield  {title} {\enquote {\bibinfo {title} {{Primordial black holes and
  secondary gravitational waves from string inspired general no-scale
  supergravity}},}\ }\href {\doibase 10.1103/PhysRevD.104.123544} {\bibfield
  {journal} {\bibinfo  {journal} {Phys. Rev. D}\ }\textbf {\bibinfo {volume}
  {104}},\ \bibinfo {pages} {123544} (\bibinfo {year} {2021})},\ \Eprint
  {http://arxiv.org/abs/2105.07694} {arXiv:2105.07694 [gr-qc]} \BibitemShut
  {NoStop}%
\bibitem [{\citenamefont {Teimoori}\ \emph {et~al.}(2021)\citenamefont
  {Teimoori}, \citenamefont {Rezazadeh},\ and\ \citenamefont
  {Karami}}]{Teimoori:2021thk}%
  \BibitemOpen
  \bibfield  {author} {\bibinfo {author} {\bibfnamefont {Zeinab}\ \bibnamefont
  {Teimoori}}, \bibinfo {author} {\bibfnamefont {Kazem}\ \bibnamefont
  {Rezazadeh}}, \ and\ \bibinfo {author} {\bibfnamefont {Kayoomars}\
  \bibnamefont {Karami}},\ }\bibfield  {title} {\enquote {\bibinfo {title}
  {{Primordial Black Holes Formation and Secondary Gravitational Waves in
  Nonminimal Derivative Coupling Inflation}},}\ }\href {\doibase
  10.3847/1538-4357/ac01cf} {\bibfield  {journal} {\bibinfo  {journal}
  {Astrophys. J.}\ }\textbf {\bibinfo {volume} {915}},\ \bibinfo {pages} {118}
  (\bibinfo {year} {2021})},\ \Eprint {http://arxiv.org/abs/2107.08048}
  {arXiv:2107.08048 [gr-qc]} \BibitemShut {NoStop}%
\bibitem [{\citenamefont {Kawai}\ and\ \citenamefont
  {Kim}(2021)}]{Kawai:2021edk}%
  \BibitemOpen
  \bibfield  {author} {\bibinfo {author} {\bibfnamefont {Shinsuke}\
  \bibnamefont {Kawai}}\ and\ \bibinfo {author} {\bibfnamefont {Jinsu}\
  \bibnamefont {Kim}},\ }\bibfield  {title} {\enquote {\bibinfo {title}
  {{Primordial black holes from Gauss-Bonnet-corrected single field
  inflation}},}\ }\href {\doibase 10.1103/PhysRevD.104.083545} {\bibfield
  {journal} {\bibinfo  {journal} {Phys. Rev. D}\ }\textbf {\bibinfo {volume}
  {104}},\ \bibinfo {pages} {083545} (\bibinfo {year} {2021})},\ \Eprint
  {http://arxiv.org/abs/2108.01340} {arXiv:2108.01340 [astro-ph.CO]}
  \BibitemShut {NoStop}%
\bibitem [{\citenamefont {Zhang}(2022)}]{Zhang:2021rqs}%
  \BibitemOpen
  \bibfield  {author} {\bibinfo {author} {\bibfnamefont {Fengge}\ \bibnamefont
  {Zhang}},\ }\bibfield  {title} {\enquote {\bibinfo {title} {{Primordial black
  holes and scalar induced gravitational waves from the E model with a
  Gauss-Bonnet term}},}\ }\href {\doibase 10.1103/PhysRevD.105.063539}
  {\bibfield  {journal} {\bibinfo  {journal} {Phys. Rev. D}\ }\textbf {\bibinfo
  {volume} {105}},\ \bibinfo {pages} {063539} (\bibinfo {year} {2022})},\
  \Eprint {http://arxiv.org/abs/2112.10516} {arXiv:2112.10516 [gr-qc]}
  \BibitemShut {NoStop}%
\bibitem [{\citenamefont {Yi}(2022)}]{Yi:2022anu}%
  \BibitemOpen
  \bibfield  {author} {\bibinfo {author} {\bibfnamefont {Zhu}\ \bibnamefont
  {Yi}},\ }\bibfield  {title} {\enquote {\bibinfo {title} {{Primordial black
  holes and scalar-induced gravitational waves from scalar-tensor
  inflation}},}\ }\href@noop {} {\  (\bibinfo {year} {2022})},\ \Eprint
  {http://arxiv.org/abs/2206.01039} {arXiv:2206.01039 [gr-qc]} \BibitemShut
  {NoStop}%
\bibitem [{\citenamefont {Gu}\ \emph {et~al.}(2022)\citenamefont {Gu},
  \citenamefont {Shu}, \citenamefont {Yang},\ and\ \citenamefont
  {Zhang}}]{Gu:2022pbo}%
  \BibitemOpen
  \bibfield  {author} {\bibinfo {author} {\bibfnamefont {Bao-Min}\ \bibnamefont
  {Gu}}, \bibinfo {author} {\bibfnamefont {Fu-Wen}\ \bibnamefont {Shu}},
  \bibinfo {author} {\bibfnamefont {Ke}~\bibnamefont {Yang}}, \ and\ \bibinfo
  {author} {\bibfnamefont {Yu-Peng}\ \bibnamefont {Zhang}},\ }\bibfield
  {title} {\enquote {\bibinfo {title} {{Primordial black holes from valley}},}\
  }\href@noop {} {\  (\bibinfo {year} {2022})},\ \Eprint
  {http://arxiv.org/abs/2207.09968} {arXiv:2207.09968 [astro-ph.CO]}
  \BibitemShut {NoStop}%
\bibitem [{\citenamefont {Cook}(2022)}]{Cook:2022zol}%
  \BibitemOpen
  \bibfield  {author} {\bibinfo {author} {\bibfnamefont {Jessica~L.}\
  \bibnamefont {Cook}},\ }\bibfield  {title} {\enquote {\bibinfo {title}
  {{Primordial Black Hole Production in Natural and Hilltop Inflation}},}\
  }\href@noop {} {\  (\bibinfo {year} {2022})},\ \Eprint
  {http://arxiv.org/abs/2209.05674} {arXiv:2209.05674 [astro-ph.CO]}
  \BibitemShut {NoStop}%
\bibitem [{\citenamefont {Hidalgo}\ \emph {et~al.}(2022)\citenamefont
  {Hidalgo}, \citenamefont {Padilla},\ and\ \citenamefont
  {German}}]{Hidalgo:2022yed}%
  \BibitemOpen
  \bibfield  {author} {\bibinfo {author} {\bibfnamefont {Juan~Carlos}\
  \bibnamefont {Hidalgo}}, \bibinfo {author} {\bibfnamefont {Luis~E.}\
  \bibnamefont {Padilla}}, \ and\ \bibinfo {author} {\bibfnamefont {Gabriel}\
  \bibnamefont {German}},\ }\bibfield  {title} {\enquote {\bibinfo {title}
  {{Production of PBHs from inflaton structure}},}\ }\href@noop {} {\
  (\bibinfo {year} {2022})},\ \Eprint {http://arxiv.org/abs/2208.09462}
  {arXiv:2208.09462 [astro-ph.CO]} \BibitemShut {NoStop}%
\bibitem [{\citenamefont {Animali}\ and\ \citenamefont
  {Vennin}(2022)}]{Animali:2022otk}%
  \BibitemOpen
  \bibfield  {author} {\bibinfo {author} {\bibfnamefont {Chiara}\ \bibnamefont
  {Animali}}\ and\ \bibinfo {author} {\bibfnamefont {Vincent}\ \bibnamefont
  {Vennin}},\ }\bibfield  {title} {\enquote {\bibinfo {title} {{Primordial
  black holes from stochastic tunnelling}},}\ }\href@noop {} {\  (\bibinfo
  {year} {2022})},\ \Eprint {http://arxiv.org/abs/2210.03812} {arXiv:2210.03812
  [astro-ph.CO]} \BibitemShut {NoStop}%
\bibitem [{\citenamefont {Papanikolaou}\ \emph {et~al.}(2022)\citenamefont
  {Papanikolaou}, \citenamefont {Lymperis}, \citenamefont {Lola},\ and\
  \citenamefont {Saridakis}}]{Papanikolaou:2022did}%
  \BibitemOpen
  \bibfield  {author} {\bibinfo {author} {\bibfnamefont {Theodoros}\
  \bibnamefont {Papanikolaou}}, \bibinfo {author} {\bibfnamefont {Andreas}\
  \bibnamefont {Lymperis}}, \bibinfo {author} {\bibfnamefont {Smaragda}\
  \bibnamefont {Lola}}, \ and\ \bibinfo {author} {\bibfnamefont {Emmanuel~N.}\
  \bibnamefont {Saridakis}},\ }\bibfield  {title} {\enquote {\bibinfo {title}
  {{Primordial black holes and gravitational waves from non-canonical
  inflation}},}\ }\href@noop {} {\  (\bibinfo {year} {2022})},\ \Eprint
  {http://arxiv.org/abs/2211.14900} {arXiv:2211.14900 [astro-ph.CO]}
  \BibitemShut {NoStop}%
\bibitem [{\citenamefont {Braglia}\ \emph {et~al.}(2022)\citenamefont
  {Braglia}, \citenamefont {Linde}, \citenamefont {Kallosh},\ and\
  \citenamefont {Finelli}}]{Braglia:2022phb}%
  \BibitemOpen
  \bibfield  {author} {\bibinfo {author} {\bibfnamefont {Matteo}\ \bibnamefont
  {Braglia}}, \bibinfo {author} {\bibfnamefont {Andrei}\ \bibnamefont {Linde}},
  \bibinfo {author} {\bibfnamefont {Renata}\ \bibnamefont {Kallosh}}, \ and\
  \bibinfo {author} {\bibfnamefont {Fabio}\ \bibnamefont {Finelli}},\
  }\bibfield  {title} {\enquote {\bibinfo {title} {{Hybrid $\alpha$-attractors,
  primordial black holes and gravitational wave backgrounds}},}\ }\href@noop {}
  {\  (\bibinfo {year} {2022})},\ \Eprint {http://arxiv.org/abs/2211.14262}
  {arXiv:2211.14262 [astro-ph.CO]} \BibitemShut {NoStop}%
\bibitem [{\citenamefont {Kawaguchi}\ and\ \citenamefont
  {Tsujikawa}(2022)}]{Kawaguchi:2022nku}%
  \BibitemOpen
  \bibfield  {author} {\bibinfo {author} {\bibfnamefont {Ryodai}\ \bibnamefont
  {Kawaguchi}}\ and\ \bibinfo {author} {\bibfnamefont {Shinji}\ \bibnamefont
  {Tsujikawa}},\ }\bibfield  {title} {\enquote {\bibinfo {title} {{Primordial
  black holes from Higgs inflation with a Gauss-Bonnet coupling}},}\
  }\href@noop {} {\  (\bibinfo {year} {2022})},\ \Eprint
  {http://arxiv.org/abs/2211.13364} {arXiv:2211.13364 [astro-ph.CO]}
  \BibitemShut {NoStop}%
\bibitem [{\citenamefont {Ashoorioon}\ \emph
  {et~al.}(2021{\natexlab{a}})\citenamefont {Ashoorioon}, \citenamefont
  {Rostami},\ and\ \citenamefont {Firouzjaee}}]{Ashoorioon:2019xqc}%
  \BibitemOpen
  \bibfield  {author} {\bibinfo {author} {\bibfnamefont {Amjad}\ \bibnamefont
  {Ashoorioon}}, \bibinfo {author} {\bibfnamefont {Abasalt}\ \bibnamefont
  {Rostami}}, \ and\ \bibinfo {author} {\bibfnamefont {Javad~T.}\ \bibnamefont
  {Firouzjaee}},\ }\bibfield  {title} {\enquote {\bibinfo {title} {{EFT
  compatible PBHs: effective spawning of the seeds for primordial black holes
  during inflation}},}\ }\href {\doibase 10.1007/JHEP07(2021)087} {\bibfield
  {journal} {\bibinfo  {journal} {JHEP}\ }\textbf {\bibinfo {volume} {07}},\
  \bibinfo {pages} {087} (\bibinfo {year} {2021}{\natexlab{a}})},\ \Eprint
  {http://arxiv.org/abs/1912.13326} {arXiv:1912.13326 [astro-ph.CO]}
  \BibitemShut {NoStop}%
\bibitem [{\citenamefont {Fu}\ and\ \citenamefont {Chen}(2022)}]{Fu:2022ssq}%
  \BibitemOpen
  \bibfield  {author} {\bibinfo {author} {\bibfnamefont {Chengjie}\
  \bibnamefont {Fu}}\ and\ \bibinfo {author} {\bibfnamefont {Chao}\
  \bibnamefont {Chen}},\ }\bibfield  {title} {\enquote {\bibinfo {title}
  {{Sudden braking and turning in the single/multi-stream inflation: primordial
  black hole formation}},}\ }\href@noop {} {\  (\bibinfo {year} {2022})},\
  \Eprint {http://arxiv.org/abs/2211.11387} {arXiv:2211.11387 [astro-ph.CO]}
  \BibitemShut {NoStop}%
\bibitem [{\citenamefont {Ahmed}\ \emph {et~al.}(2022)\citenamefont {Ahmed},
  \citenamefont {Junaid},\ and\ \citenamefont {Zubair}}]{Ahmed:2021ucx}%
  \BibitemOpen
  \bibfield  {author} {\bibinfo {author} {\bibfnamefont {Waqas}\ \bibnamefont
  {Ahmed}}, \bibinfo {author} {\bibfnamefont {M.}~\bibnamefont {Junaid}}, \
  and\ \bibinfo {author} {\bibfnamefont {Umer}\ \bibnamefont {Zubair}},\
  }\bibfield  {title} {\enquote {\bibinfo {title} {{Primordial black holes and
  gravitational waves in hybrid inflation with chaotic potentials}},}\ }\href
  {\doibase 10.1016/j.nuclphysb.2022.115968} {\bibfield  {journal} {\bibinfo
  {journal} {Nucl. Phys. B}\ }\textbf {\bibinfo {volume} {984}},\ \bibinfo
  {pages} {115968} (\bibinfo {year} {2022})},\ \Eprint
  {http://arxiv.org/abs/2109.14838} {arXiv:2109.14838 [astro-ph.CO]}
  \BibitemShut {NoStop}%
\bibitem [{\citenamefont {Garcia-Bellido}\ \emph {et~al.}(1996)\citenamefont
  {Garcia-Bellido}, \citenamefont {Linde},\ and\ \citenamefont
  {Wands}}]{Garcia-Bellido:1996mdl}%
  \BibitemOpen
  \bibfield  {author} {\bibinfo {author} {\bibfnamefont {Juan}\ \bibnamefont
  {Garcia-Bellido}}, \bibinfo {author} {\bibfnamefont {Andrei~D.}\ \bibnamefont
  {Linde}}, \ and\ \bibinfo {author} {\bibfnamefont {David}\ \bibnamefont
  {Wands}},\ }\bibfield  {title} {\enquote {\bibinfo {title} {{Density
  perturbations and black hole formation in hybrid inflation}},}\ }\href
  {\doibase 10.1103/PhysRevD.54.6040} {\bibfield  {journal} {\bibinfo
  {journal} {Phys. Rev. D}\ }\textbf {\bibinfo {volume} {54}},\ \bibinfo
  {pages} {6040--6058} (\bibinfo {year} {1996})},\ \Eprint
  {http://arxiv.org/abs/astro-ph/9605094} {arXiv:astro-ph/9605094} \BibitemShut
  {NoStop}%
\bibitem [{\citenamefont {Kawasaki}\ \emph {et~al.}(1998)\citenamefont
  {Kawasaki}, \citenamefont {Sugiyama},\ and\ \citenamefont
  {Yanagida}}]{Kawasaki:1997ju}%
  \BibitemOpen
  \bibfield  {author} {\bibinfo {author} {\bibfnamefont {M.}~\bibnamefont
  {Kawasaki}}, \bibinfo {author} {\bibfnamefont {N.}~\bibnamefont {Sugiyama}},
  \ and\ \bibinfo {author} {\bibfnamefont {T.}~\bibnamefont {Yanagida}},\
  }\bibfield  {title} {\enquote {\bibinfo {title} {{Primordial black hole
  formation in a double inflation model in supergravity}},}\ }\href {\doibase
  10.1103/PhysRevD.57.6050} {\bibfield  {journal} {\bibinfo  {journal} {Phys.
  Rev. D}\ }\textbf {\bibinfo {volume} {57}},\ \bibinfo {pages} {6050--6056}
  (\bibinfo {year} {1998})},\ \Eprint {http://arxiv.org/abs/hep-ph/9710259}
  {arXiv:hep-ph/9710259} \BibitemShut {NoStop}%
\bibitem [{\citenamefont {Yokoyama}(1997)}]{Yokoyama:1995ex}%
  \BibitemOpen
  \bibfield  {author} {\bibinfo {author} {\bibfnamefont {Junichi}\ \bibnamefont
  {Yokoyama}},\ }\bibfield  {title} {\enquote {\bibinfo {title} {{Formation of
  MACHO primordial black holes in inflationary cosmology}},}\ }\href@noop {}
  {\bibfield  {journal} {\bibinfo  {journal} {Astron. Astrophys.}\ }\textbf
  {\bibinfo {volume} {318}},\ \bibinfo {pages} {673} (\bibinfo {year}
  {1997})},\ \Eprint {http://arxiv.org/abs/astro-ph/9509027}
  {arXiv:astro-ph/9509027} \BibitemShut {NoStop}%
\bibitem [{\citenamefont {Frampton}\ \emph {et~al.}(2010)\citenamefont
  {Frampton}, \citenamefont {Kawasaki}, \citenamefont {Takahashi},\ and\
  \citenamefont {Yanagida}}]{Frampton:2010sw}%
  \BibitemOpen
  \bibfield  {author} {\bibinfo {author} {\bibfnamefont {Paul~H.}\ \bibnamefont
  {Frampton}}, \bibinfo {author} {\bibfnamefont {Masahiro}\ \bibnamefont
  {Kawasaki}}, \bibinfo {author} {\bibfnamefont {Fuminobu}\ \bibnamefont
  {Takahashi}}, \ and\ \bibinfo {author} {\bibfnamefont {Tsutomu~T.}\
  \bibnamefont {Yanagida}},\ }\bibfield  {title} {\enquote {\bibinfo {title}
  {{Primordial Black Holes as All Dark Matter}},}\ }\href {\doibase
  10.1088/1475-7516/2010/04/023} {\bibfield  {journal} {\bibinfo  {journal}
  {JCAP}\ }\textbf {\bibinfo {volume} {04}},\ \bibinfo {pages} {023} (\bibinfo
  {year} {2010})},\ \Eprint {http://arxiv.org/abs/1001.2308} {arXiv:1001.2308
  [hep-ph]} \BibitemShut {NoStop}%
\bibitem [{\citenamefont {Giovannini}(2010)}]{Giovannini:2010tk}%
  \BibitemOpen
  \bibfield  {author} {\bibinfo {author} {\bibfnamefont {Massimo}\ \bibnamefont
  {Giovannini}},\ }\bibfield  {title} {\enquote {\bibinfo {title} {{Secondary
  graviton spectra and waterfall-like fields}},}\ }\href {\doibase
  10.1103/PhysRevD.82.083523} {\bibfield  {journal} {\bibinfo  {journal} {Phys.
  Rev. D}\ }\textbf {\bibinfo {volume} {82}},\ \bibinfo {pages} {083523}
  (\bibinfo {year} {2010})},\ \Eprint {http://arxiv.org/abs/1008.1164}
  {arXiv:1008.1164 [astro-ph.CO]} \BibitemShut {NoStop}%
\bibitem [{\citenamefont {Clesse}\ and\ \citenamefont
  {Garc\'\i{}a-Bellido}(2015)}]{Clesse:2015wea}%
  \BibitemOpen
  \bibfield  {author} {\bibinfo {author} {\bibfnamefont {S\'ebastien}\
  \bibnamefont {Clesse}}\ and\ \bibinfo {author} {\bibfnamefont {Juan}\
  \bibnamefont {Garc\'\i{}a-Bellido}},\ }\bibfield  {title} {\enquote {\bibinfo
  {title} {{Massive Primordial Black Holes from Hybrid Inflation as Dark Matter
  and the seeds of Galaxies}},}\ }\href {\doibase 10.1103/PhysRevD.92.023524}
  {\bibfield  {journal} {\bibinfo  {journal} {Phys. Rev. D}\ }\textbf {\bibinfo
  {volume} {92}},\ \bibinfo {pages} {023524} (\bibinfo {year} {2015})},\
  \Eprint {http://arxiv.org/abs/1501.07565} {arXiv:1501.07565 [astro-ph.CO]}
  \BibitemShut {NoStop}%
\bibitem [{\citenamefont {Inomata}\ \emph
  {et~al.}(2017{\natexlab{a}})\citenamefont {Inomata}, \citenamefont
  {Kawasaki}, \citenamefont {Mukaida}, \citenamefont {Tada},\ and\
  \citenamefont {Yanagida}}]{Inomata:2017okj}%
  \BibitemOpen
  \bibfield  {author} {\bibinfo {author} {\bibfnamefont {Keisuke}\ \bibnamefont
  {Inomata}}, \bibinfo {author} {\bibfnamefont {Masahiro}\ \bibnamefont
  {Kawasaki}}, \bibinfo {author} {\bibfnamefont {Kyohei}\ \bibnamefont
  {Mukaida}}, \bibinfo {author} {\bibfnamefont {Yuichiro}\ \bibnamefont
  {Tada}}, \ and\ \bibinfo {author} {\bibfnamefont {Tsutomu~T.}\ \bibnamefont
  {Yanagida}},\ }\bibfield  {title} {\enquote {\bibinfo {title} {{Inflationary
  Primordial Black Holes as All Dark Matter}},}\ }\href {\doibase
  10.1103/PhysRevD.96.043504} {\bibfield  {journal} {\bibinfo  {journal} {Phys.
  Rev. D}\ }\textbf {\bibinfo {volume} {96}},\ \bibinfo {pages} {043504}
  (\bibinfo {year} {2017}{\natexlab{a}})},\ \Eprint
  {http://arxiv.org/abs/1701.02544} {arXiv:1701.02544 [astro-ph.CO]}
  \BibitemShut {NoStop}%
\bibitem [{\citenamefont {Inomata}\ \emph {et~al.}(2018)\citenamefont
  {Inomata}, \citenamefont {Kawasaki}, \citenamefont {Mukaida},\ and\
  \citenamefont {Yanagida}}]{Inomata:2018cht}%
  \BibitemOpen
  \bibfield  {author} {\bibinfo {author} {\bibfnamefont {Keisuke}\ \bibnamefont
  {Inomata}}, \bibinfo {author} {\bibfnamefont {Masahiro}\ \bibnamefont
  {Kawasaki}}, \bibinfo {author} {\bibfnamefont {Kyohei}\ \bibnamefont
  {Mukaida}}, \ and\ \bibinfo {author} {\bibfnamefont {Tsutomu~T.}\
  \bibnamefont {Yanagida}},\ }\bibfield  {title} {\enquote {\bibinfo {title}
  {{Double inflation as a single origin of primordial black holes for all dark
  matter and LIGO observations}},}\ }\href {\doibase
  10.1103/PhysRevD.97.043514} {\bibfield  {journal} {\bibinfo  {journal} {Phys.
  Rev. D}\ }\textbf {\bibinfo {volume} {97}},\ \bibinfo {pages} {043514}
  (\bibinfo {year} {2018})},\ \Eprint {http://arxiv.org/abs/1711.06129}
  {arXiv:1711.06129 [astro-ph.CO]} \BibitemShut {NoStop}%
\bibitem [{\citenamefont {Espinosa}\ \emph
  {et~al.}(2018{\natexlab{a}})\citenamefont {Espinosa}, \citenamefont {Racco},\
  and\ \citenamefont {Riotto}}]{Espinosa:2017sgp}%
  \BibitemOpen
  \bibfield  {author} {\bibinfo {author} {\bibfnamefont {J.~R.}\ \bibnamefont
  {Espinosa}}, \bibinfo {author} {\bibfnamefont {D.}~\bibnamefont {Racco}}, \
  and\ \bibinfo {author} {\bibfnamefont {A.}~\bibnamefont {Riotto}},\
  }\bibfield  {title} {\enquote {\bibinfo {title} {{Cosmological Signature of
  the Standard Model Higgs Vacuum Instability: Primordial Black Holes as Dark
  Matter}},}\ }\href {\doibase 10.1103/PhysRevLett.120.121301} {\bibfield
  {journal} {\bibinfo  {journal} {Phys. Rev. Lett.}\ }\textbf {\bibinfo
  {volume} {120}},\ \bibinfo {pages} {121301} (\bibinfo {year}
  {2018}{\natexlab{a}})},\ \Eprint {http://arxiv.org/abs/1710.11196}
  {arXiv:1710.11196 [hep-ph]} \BibitemShut {NoStop}%
\bibitem [{\citenamefont {Kawasaki}\ \emph {et~al.}(2020)\citenamefont
  {Kawasaki}, \citenamefont {Nakatsuka},\ and\ \citenamefont
  {Obata}}]{Kawasaki:2019hvt}%
  \BibitemOpen
  \bibfield  {author} {\bibinfo {author} {\bibfnamefont {Masahiro}\
  \bibnamefont {Kawasaki}}, \bibinfo {author} {\bibfnamefont {Hiromasa}\
  \bibnamefont {Nakatsuka}}, \ and\ \bibinfo {author} {\bibfnamefont {Ippei}\
  \bibnamefont {Obata}},\ }\bibfield  {title} {\enquote {\bibinfo {title}
  {{Generation of Primordial Black Holes and Gravitational Waves from
  Dilaton-Gauge Field Dynamics}},}\ }\href {\doibase
  10.1088/1475-7516/2020/05/007} {\bibfield  {journal} {\bibinfo  {journal}
  {JCAP}\ }\textbf {\bibinfo {volume} {2005}},\ \bibinfo {pages} {007}
  (\bibinfo {year} {2020})},\ \Eprint {http://arxiv.org/abs/1912.09111}
  {arXiv:1912.09111 [astro-ph.CO]} \BibitemShut {NoStop}%
\bibitem [{\citenamefont {Palma}\ \emph {et~al.}(2020)\citenamefont {Palma},
  \citenamefont {Sypsas},\ and\ \citenamefont {Zenteno}}]{Palma:2020ejf}%
  \BibitemOpen
  \bibfield  {author} {\bibinfo {author} {\bibfnamefont {Gonzalo~A.}\
  \bibnamefont {Palma}}, \bibinfo {author} {\bibfnamefont {Spyros}\
  \bibnamefont {Sypsas}}, \ and\ \bibinfo {author} {\bibfnamefont {Cristobal}\
  \bibnamefont {Zenteno}},\ }\bibfield  {title} {\enquote {\bibinfo {title}
  {{Seeding primordial black holes in multifield inflation}},}\ }\href
  {\doibase 10.1103/PhysRevLett.125.121301} {\bibfield  {journal} {\bibinfo
  {journal} {Phys. Rev. Lett.}\ }\textbf {\bibinfo {volume} {125}},\ \bibinfo
  {pages} {121301} (\bibinfo {year} {2020})},\ \Eprint
  {http://arxiv.org/abs/2004.06106} {arXiv:2004.06106 [astro-ph.CO]}
  \BibitemShut {NoStop}%
\bibitem [{\citenamefont {Fumagalli}\ \emph {et~al.}(2020)\citenamefont
  {Fumagalli}, \citenamefont {Renaux-Petel}, \citenamefont {Ronayne},\ and\
  \citenamefont {Witkowski}}]{Fumagalli:2020adf}%
  \BibitemOpen
  \bibfield  {author} {\bibinfo {author} {\bibfnamefont {Jacopo}\ \bibnamefont
  {Fumagalli}}, \bibinfo {author} {\bibfnamefont {S\'ebastien}\ \bibnamefont
  {Renaux-Petel}}, \bibinfo {author} {\bibfnamefont {John~W.}\ \bibnamefont
  {Ronayne}}, \ and\ \bibinfo {author} {\bibfnamefont {Lukas~T.}\ \bibnamefont
  {Witkowski}},\ }\bibfield  {title} {\enquote {\bibinfo {title} {{Turning in
  the landscape: a new mechanism for generating Primordial Black Holes}},}\
  }\href@noop {} {\  (\bibinfo {year} {2020})},\ \Eprint
  {http://arxiv.org/abs/2004.08369} {arXiv:2004.08369 [hep-th]} \BibitemShut
  {NoStop}%
\bibitem [{\citenamefont {Braglia}\ \emph
  {et~al.}(2020{\natexlab{a}})\citenamefont {Braglia}, \citenamefont {Hazra},
  \citenamefont {Finelli}, \citenamefont {Smoot}, \citenamefont {Sriramkumar},\
  and\ \citenamefont {Starobinsky}}]{Braglia:2020eai}%
  \BibitemOpen
  \bibfield  {author} {\bibinfo {author} {\bibfnamefont {Matteo}\ \bibnamefont
  {Braglia}}, \bibinfo {author} {\bibfnamefont {Dhiraj~Kumar}\ \bibnamefont
  {Hazra}}, \bibinfo {author} {\bibfnamefont {Fabio}\ \bibnamefont {Finelli}},
  \bibinfo {author} {\bibfnamefont {George~F.}\ \bibnamefont {Smoot}}, \bibinfo
  {author} {\bibfnamefont {L.}~\bibnamefont {Sriramkumar}}, \ and\ \bibinfo
  {author} {\bibfnamefont {Alexei~A.}\ \bibnamefont {Starobinsky}},\ }\bibfield
   {title} {\enquote {\bibinfo {title} {{Generating PBHs and small-scale GWs in
  two-field models of inflation}},}\ }\href {\doibase
  10.1088/1475-7516/2020/08/001} {\bibfield  {journal} {\bibinfo  {journal}
  {JCAP}\ }\textbf {\bibinfo {volume} {08}},\ \bibinfo {pages} {001} (\bibinfo
  {year} {2020}{\natexlab{a}})},\ \Eprint {http://arxiv.org/abs/2005.02895}
  {arXiv:2005.02895 [astro-ph.CO]} \BibitemShut {NoStop}%
\bibitem [{\citenamefont {Anguelova}(2021)}]{Anguelova:2020nzl}%
  \BibitemOpen
  \bibfield  {author} {\bibinfo {author} {\bibfnamefont {Lilia}\ \bibnamefont
  {Anguelova}},\ }\bibfield  {title} {\enquote {\bibinfo {title} {{On
  Primordial Black Holes from Rapid Turns in Two-field Models}},}\ }\href
  {\doibase 10.1088/1475-7516/2021/06/004} {\bibfield  {journal} {\bibinfo
  {journal} {JCAP}\ }\textbf {\bibinfo {volume} {06}},\ \bibinfo {pages} {004}
  (\bibinfo {year} {2021})},\ \Eprint {http://arxiv.org/abs/2012.03705}
  {arXiv:2012.03705 [hep-th]} \BibitemShut {NoStop}%
\bibitem [{\citenamefont {Romano}(2020)}]{Romano:2020gtn}%
  \BibitemOpen
  \bibfield  {author} {\bibinfo {author} {\bibfnamefont {Antonio~Enea}\
  \bibnamefont {Romano}},\ }\bibfield  {title} {\enquote {\bibinfo {title}
  {{Sound speed induced production of primordial black holes}},}\ }\href@noop
  {} {\  (\bibinfo {year} {2020})},\ \Eprint {http://arxiv.org/abs/2006.07321}
  {arXiv:2006.07321 [astro-ph.CO]} \BibitemShut {NoStop}%
\bibitem [{\citenamefont {Gundhi}\ and\ \citenamefont
  {Steinwachs}(2021)}]{Gundhi:2020zvb}%
  \BibitemOpen
  \bibfield  {author} {\bibinfo {author} {\bibfnamefont {Anirudh}\ \bibnamefont
  {Gundhi}}\ and\ \bibinfo {author} {\bibfnamefont {Christian~F.}\ \bibnamefont
  {Steinwachs}},\ }\bibfield  {title} {\enquote {\bibinfo {title}
  {{Scalaron\textendash{}Higgs inflation reloaded: Higgs-dependent scalaron
  mass and primordial black hole dark matter}},}\ }\href {\doibase
  10.1140/epjc/s10052-021-09225-2} {\bibfield  {journal} {\bibinfo  {journal}
  {Eur. Phys. J. C}\ }\textbf {\bibinfo {volume} {81}},\ \bibinfo {pages} {460}
  (\bibinfo {year} {2021})},\ \Eprint {http://arxiv.org/abs/2011.09485}
  {arXiv:2011.09485 [hep-th]} \BibitemShut {NoStop}%
\bibitem [{\citenamefont {Gundhi}\ \emph {et~al.}(2021)\citenamefont {Gundhi},
  \citenamefont {Ketov},\ and\ \citenamefont {Steinwachs}}]{Gundhi:2020kzm}%
  \BibitemOpen
  \bibfield  {author} {\bibinfo {author} {\bibfnamefont {Anirudh}\ \bibnamefont
  {Gundhi}}, \bibinfo {author} {\bibfnamefont {Sergei~V.}\ \bibnamefont
  {Ketov}}, \ and\ \bibinfo {author} {\bibfnamefont {Christian~F.}\
  \bibnamefont {Steinwachs}},\ }\bibfield  {title} {\enquote {\bibinfo {title}
  {{Primordial black hole dark matter in dilaton-extended two-field Starobinsky
  inflation}},}\ }\href {\doibase 10.1103/PhysRevD.103.083518} {\bibfield
  {journal} {\bibinfo  {journal} {Phys. Rev. D}\ }\textbf {\bibinfo {volume}
  {103}},\ \bibinfo {pages} {083518} (\bibinfo {year} {2021})},\ \Eprint
  {http://arxiv.org/abs/2011.05999} {arXiv:2011.05999 [hep-th]} \BibitemShut
  {NoStop}%
\bibitem [{\citenamefont {Cai}\ \emph {et~al.}(2021)\citenamefont {Cai},
  \citenamefont {Chen},\ and\ \citenamefont {Fu}}]{Cai:2021wzd}%
  \BibitemOpen
  \bibfield  {author} {\bibinfo {author} {\bibfnamefont {Rong-Gen}\
  \bibnamefont {Cai}}, \bibinfo {author} {\bibfnamefont {Chao}\ \bibnamefont
  {Chen}}, \ and\ \bibinfo {author} {\bibfnamefont {Chengjie}\ \bibnamefont
  {Fu}},\ }\bibfield  {title} {\enquote {\bibinfo {title} {{Primordial black
  holes and stochastic gravitational wave background from inflation with a
  noncanonical spectator field}},}\ }\href {\doibase
  10.1103/PhysRevD.104.083537} {\bibfield  {journal} {\bibinfo  {journal}
  {Phys. Rev. D}\ }\textbf {\bibinfo {volume} {104}},\ \bibinfo {pages}
  {083537} (\bibinfo {year} {2021})},\ \Eprint
  {http://arxiv.org/abs/2108.03422} {arXiv:2108.03422 [astro-ph.CO]}
  \BibitemShut {NoStop}%
\bibitem [{\citenamefont {Ishikawa}\ and\ \citenamefont
  {Ketov}(2022)}]{Ishikawa:2021xya}%
  \BibitemOpen
  \bibfield  {author} {\bibinfo {author} {\bibfnamefont {Ryotaro}\ \bibnamefont
  {Ishikawa}}\ and\ \bibinfo {author} {\bibfnamefont {Sergei~V.}\ \bibnamefont
  {Ketov}},\ }\bibfield  {title} {\enquote {\bibinfo {title} {{Exploring the
  parameter space of modified supergravity for double inflation and primordial
  black hole formation}},}\ }\href {\doibase 10.1088/1361-6382/ac3bd9}
  {\bibfield  {journal} {\bibinfo  {journal} {Class. Quant. Grav.}\ }\textbf
  {\bibinfo {volume} {39}},\ \bibinfo {pages} {015016} (\bibinfo {year}
  {2022})},\ \Eprint {http://arxiv.org/abs/2108.04408} {arXiv:2108.04408
  [astro-ph.CO]} \BibitemShut {NoStop}%
\bibitem [{\citenamefont {Spanos}\ and\ \citenamefont
  {Stamou}(2021)}]{Spanos:2021hpk}%
  \BibitemOpen
  \bibfield  {author} {\bibinfo {author} {\bibfnamefont {Vassilis~C.}\
  \bibnamefont {Spanos}}\ and\ \bibinfo {author} {\bibfnamefont {Ioanna~D.}\
  \bibnamefont {Stamou}},\ }\bibfield  {title} {\enquote {\bibinfo {title}
  {{Gravitational waves and primordial black holes from supersymmetric hybrid
  inflation}},}\ }\href {\doibase 10.1103/PhysRevD.104.123537} {\bibfield
  {journal} {\bibinfo  {journal} {Phys. Rev. D}\ }\textbf {\bibinfo {volume}
  {104}},\ \bibinfo {pages} {123537} (\bibinfo {year} {2021})},\ \Eprint
  {http://arxiv.org/abs/2108.05671} {arXiv:2108.05671 [astro-ph.CO]}
  \BibitemShut {NoStop}%
\bibitem [{\citenamefont {Hooshangi}\ \emph {et~al.}(2022)\citenamefont
  {Hooshangi}, \citenamefont {Talebian}, \citenamefont {Namjoo},\ and\
  \citenamefont {Firouzjahi}}]{Hooshangi:2022lao}%
  \BibitemOpen
  \bibfield  {author} {\bibinfo {author} {\bibfnamefont {Sina}\ \bibnamefont
  {Hooshangi}}, \bibinfo {author} {\bibfnamefont {Alireza}\ \bibnamefont
  {Talebian}}, \bibinfo {author} {\bibfnamefont {Mohammad~Hossein}\
  \bibnamefont {Namjoo}}, \ and\ \bibinfo {author} {\bibfnamefont {Hassan}\
  \bibnamefont {Firouzjahi}},\ }\bibfield  {title} {\enquote {\bibinfo {title}
  {{Multiple field ultraslow-roll inflation: Primordial black holes from
  straight bulk and distorted boundary}},}\ }\href {\doibase
  10.1103/PhysRevD.105.083525} {\bibfield  {journal} {\bibinfo  {journal}
  {Phys. Rev. D}\ }\textbf {\bibinfo {volume} {105}},\ \bibinfo {pages}
  {083525} (\bibinfo {year} {2022})},\ \Eprint
  {http://arxiv.org/abs/2201.07258} {arXiv:2201.07258 [astro-ph.CO]}
  \BibitemShut {NoStop}%
\bibitem [{\citenamefont {Chen}\ and\ \citenamefont
  {Cai}(2019)}]{Chen:2019zza}%
  \BibitemOpen
  \bibfield  {author} {\bibinfo {author} {\bibfnamefont {Chao}\ \bibnamefont
  {Chen}}\ and\ \bibinfo {author} {\bibfnamefont {Yi-Fu}\ \bibnamefont {Cai}},\
  }\bibfield  {title} {\enquote {\bibinfo {title} {{Primordial black holes from
  sound speed resonance in the inflaton-curvaton mixed scenario}},}\
  }\href@noop {} {\  (\bibinfo {year} {2019})},\ \Eprint
  {http://arxiv.org/abs/1908.03942} {arXiv:1908.03942 [astro-ph.CO]}
  \BibitemShut {NoStop}%
\bibitem [{\citenamefont {Kohri}\ \emph {et~al.}(2013)\citenamefont {Kohri},
  \citenamefont {Lin},\ and\ \citenamefont {Matsuda}}]{Kohri:2012yw}%
  \BibitemOpen
  \bibfield  {author} {\bibinfo {author} {\bibfnamefont {Kazunori}\
  \bibnamefont {Kohri}}, \bibinfo {author} {\bibfnamefont {Chia-Min}\
  \bibnamefont {Lin}}, \ and\ \bibinfo {author} {\bibfnamefont {Tomohiro}\
  \bibnamefont {Matsuda}},\ }\bibfield  {title} {\enquote {\bibinfo {title}
  {{Primordial black holes from the inflating curvaton}},}\ }\href {\doibase
  10.1103/PhysRevD.87.103527} {\bibfield  {journal} {\bibinfo  {journal} {Phys.
  Rev. D}\ }\textbf {\bibinfo {volume} {87}},\ \bibinfo {pages} {103527}
  (\bibinfo {year} {2013})},\ \Eprint {http://arxiv.org/abs/1211.2371}
  {arXiv:1211.2371 [hep-ph]} \BibitemShut {NoStop}%
\bibitem [{\citenamefont {Kawasaki}\ \emph {et~al.}(2013)\citenamefont
  {Kawasaki}, \citenamefont {Kitajima},\ and\ \citenamefont
  {Yanagida}}]{Kawasaki:2012wr}%
  \BibitemOpen
  \bibfield  {author} {\bibinfo {author} {\bibfnamefont {Masahiro}\
  \bibnamefont {Kawasaki}}, \bibinfo {author} {\bibfnamefont {Naoya}\
  \bibnamefont {Kitajima}}, \ and\ \bibinfo {author} {\bibfnamefont
  {Tsutomu~T.}\ \bibnamefont {Yanagida}},\ }\bibfield  {title} {\enquote
  {\bibinfo {title} {{Primordial black hole formation from an axionlike
  curvaton model}},}\ }\href {\doibase 10.1103/PhysRevD.87.063519} {\bibfield
  {journal} {\bibinfo  {journal} {Phys. Rev. D}\ }\textbf {\bibinfo {volume}
  {87}},\ \bibinfo {pages} {063519} (\bibinfo {year} {2013})},\ \Eprint
  {http://arxiv.org/abs/1207.2550} {arXiv:1207.2550 [hep-ph]} \BibitemShut
  {NoStop}%
\bibitem [{\citenamefont {Pi}\ \emph {et~al.}(2018)\citenamefont {Pi},
  \citenamefont {Zhang}, \citenamefont {Huang},\ and\ \citenamefont
  {Sasaki}}]{Pi:2017gih}%
  \BibitemOpen
  \bibfield  {author} {\bibinfo {author} {\bibfnamefont {Shi}\ \bibnamefont
  {Pi}}, \bibinfo {author} {\bibfnamefont {Ying-li}\ \bibnamefont {Zhang}},
  \bibinfo {author} {\bibfnamefont {Qing-Guo}\ \bibnamefont {Huang}}, \ and\
  \bibinfo {author} {\bibfnamefont {Misao}\ \bibnamefont {Sasaki}},\ }\bibfield
   {title} {\enquote {\bibinfo {title} {{Scalaron from $R^2$-gravity as a heavy
  field}},}\ }\href {\doibase 10.1088/1475-7516/2018/05/042} {\bibfield
  {journal} {\bibinfo  {journal} {JCAP}\ }\textbf {\bibinfo {volume} {05}},\
  \bibinfo {pages} {042} (\bibinfo {year} {2018})},\ \Eprint
  {http://arxiv.org/abs/1712.09896} {arXiv:1712.09896 [astro-ph.CO]}
  \BibitemShut {NoStop}%
\bibitem [{\citenamefont {Liu}(2021)}]{Liu:2021rgq}%
  \BibitemOpen
  \bibfield  {author} {\bibinfo {author} {\bibfnamefont {Lei-Hua}\ \bibnamefont
  {Liu}},\ }\bibfield  {title} {\enquote {\bibinfo {title} {{The primordial
  black hole from running curvaton}},}\ }\href@noop {} {\  (\bibinfo {year}
  {2021})},\ \Eprint {http://arxiv.org/abs/2107.07310} {arXiv:2107.07310
  [astro-ph.CO]} \BibitemShut {NoStop}%
\bibitem [{\citenamefont {Pi}\ and\ \citenamefont {Sasaki}(2021)}]{Pi:2021dft}%
  \BibitemOpen
  \bibfield  {author} {\bibinfo {author} {\bibfnamefont {Shi}\ \bibnamefont
  {Pi}}\ and\ \bibinfo {author} {\bibfnamefont {Misao}\ \bibnamefont
  {Sasaki}},\ }\bibfield  {title} {\enquote {\bibinfo {title} {{Primordial
  Black Hole Formation in Non-Minimal Curvaton Scenario}},}\ }\href@noop {} {\
  (\bibinfo {year} {2021})},\ \Eprint {http://arxiv.org/abs/2112.12680}
  {arXiv:2112.12680 [astro-ph.CO]} \BibitemShut {NoStop}%
\bibitem [{\citenamefont {Kawai}\ and\ \citenamefont
  {Kim}(2022)}]{Kawai:2022emp}%
  \BibitemOpen
  \bibfield  {author} {\bibinfo {author} {\bibfnamefont {Shinsuke}\
  \bibnamefont {Kawai}}\ and\ \bibinfo {author} {\bibfnamefont {Jinsu}\
  \bibnamefont {Kim}},\ }\bibfield  {title} {\enquote {\bibinfo {title}
  {{Primordial black holes and gravitational waves from nonminimally coupled
  supergravity inflation}},}\ }\href@noop {} {\  (\bibinfo {year} {2022})},\
  \Eprint {http://arxiv.org/abs/2209.15343} {arXiv:2209.15343 [astro-ph.CO]}
  \BibitemShut {NoStop}%
\bibitem [{\citenamefont {Ashoorioon}\ \emph
  {et~al.}(2021{\natexlab{b}})\citenamefont {Ashoorioon}, \citenamefont
  {Rostami},\ and\ \citenamefont {Firouzjaee}}]{Ashoorioon:2020hln}%
  \BibitemOpen
  \bibfield  {author} {\bibinfo {author} {\bibfnamefont {Amjad}\ \bibnamefont
  {Ashoorioon}}, \bibinfo {author} {\bibfnamefont {Abasalt}\ \bibnamefont
  {Rostami}}, \ and\ \bibinfo {author} {\bibfnamefont {Javad~T.}\ \bibnamefont
  {Firouzjaee}},\ }\bibfield  {title} {\enquote {\bibinfo {title} {{Examining
  the end of inflation with primordial black holes mass distribution and
  gravitational waves}},}\ }\href {\doibase 10.1103/PhysRevD.103.123512}
  {\bibfield  {journal} {\bibinfo  {journal} {Phys. Rev. D}\ }\textbf {\bibinfo
  {volume} {103}},\ \bibinfo {pages} {123512} (\bibinfo {year}
  {2021}{\natexlab{b}})},\ \Eprint {http://arxiv.org/abs/2012.02817}
  {arXiv:2012.02817 [astro-ph.CO]} \BibitemShut {NoStop}%
\bibitem [{\citenamefont {Ashoorioon}\ \emph {et~al.}(2022)\citenamefont
  {Ashoorioon}, \citenamefont {Rezazadeh},\ and\ \citenamefont
  {Rostami}}]{Ashoorioon:2022raz}%
  \BibitemOpen
  \bibfield  {author} {\bibinfo {author} {\bibfnamefont {Amjad}\ \bibnamefont
  {Ashoorioon}}, \bibinfo {author} {\bibfnamefont {Kazem}\ \bibnamefont
  {Rezazadeh}}, \ and\ \bibinfo {author} {\bibfnamefont {Abasalt}\ \bibnamefont
  {Rostami}},\ }\bibfield  {title} {\enquote {\bibinfo {title} {{NANOGrav
  signal from the end of inflation and the LIGO mass and heavier primordial
  black holes}},}\ }\href {\doibase 10.1016/j.physletb.2022.137542} {\bibfield
  {journal} {\bibinfo  {journal} {Phys. Lett. B}\ }\textbf {\bibinfo {volume}
  {835}},\ \bibinfo {pages} {137542} (\bibinfo {year} {2022})},\ \Eprint
  {http://arxiv.org/abs/2202.01131} {arXiv:2202.01131 [astro-ph.CO]}
  \BibitemShut {NoStop}%
\bibitem [{\citenamefont {Meng}\ \emph {et~al.}(2022)\citenamefont {Meng},
  \citenamefont {Yuan},\ and\ \citenamefont {Huang}}]{Meng:2022ixx}%
  \BibitemOpen
  \bibfield  {author} {\bibinfo {author} {\bibfnamefont {De-Shuang}\
  \bibnamefont {Meng}}, \bibinfo {author} {\bibfnamefont {Chen}\ \bibnamefont
  {Yuan}}, \ and\ \bibinfo {author} {\bibfnamefont {Qing-guo}\ \bibnamefont
  {Huang}},\ }\bibfield  {title} {\enquote {\bibinfo {title} {{One-loop
  correction to the enhanced curvature perturbation with local-type
  non-Gaussianity for the formation of primordial black holes}},}\ }\href@noop
  {} {\  (\bibinfo {year} {2022})},\ \Eprint {http://arxiv.org/abs/2207.07668}
  {arXiv:2207.07668 [astro-ph.CO]} \BibitemShut {NoStop}%
\bibitem [{\citenamefont {Kristiano}\ and\ \citenamefont
  {Yokoyama}(2022{\natexlab{a}})}]{Kristiano:2022maq}%
  \BibitemOpen
  \bibfield  {author} {\bibinfo {author} {\bibfnamefont {Jason}\ \bibnamefont
  {Kristiano}}\ and\ \bibinfo {author} {\bibfnamefont {Jun'ichi}\ \bibnamefont
  {Yokoyama}},\ }\bibfield  {title} {\enquote {\bibinfo {title} {{Ruling Out
  Primordial Black Hole Formation From Single-Field Inflation}},}\ }\href@noop
  {} {\  (\bibinfo {year} {2022}{\natexlab{a}})},\ \Eprint
  {http://arxiv.org/abs/2211.03395} {arXiv:2211.03395 [hep-th]} \BibitemShut
  {NoStop}%
\bibitem [{\citenamefont {Inomata}\ \emph
  {et~al.}(2022{\natexlab{b}})\citenamefont {Inomata}, \citenamefont
  {Braglia},\ and\ \citenamefont {Chen}}]{Inomata:2022yte}%
  \BibitemOpen
  \bibfield  {author} {\bibinfo {author} {\bibfnamefont {Keisuke}\ \bibnamefont
  {Inomata}}, \bibinfo {author} {\bibfnamefont {Matteo}\ \bibnamefont
  {Braglia}}, \ and\ \bibinfo {author} {\bibfnamefont {Xingang}\ \bibnamefont
  {Chen}},\ }\bibfield  {title} {\enquote {\bibinfo {title} {{Questions on
  calculation of primordial power spectrum with large spikes: the resonance
  model case}},}\ }\href@noop {} {\  (\bibinfo {year} {2022}{\natexlab{b}})},\
  \Eprint {http://arxiv.org/abs/2211.02586} {arXiv:2211.02586 [astro-ph.CO]}
  \BibitemShut {NoStop}%
\bibitem [{\citenamefont {Riotto}(2023)}]{Riotto:2023hoz}%
  \BibitemOpen
  \bibfield  {author} {\bibinfo {author} {\bibfnamefont {Antonio}\ \bibnamefont
  {Riotto}},\ }\bibfield  {title} {\enquote {\bibinfo {title} {{The Primordial
  Black Hole Formation from Single-Field Inflation is Not Ruled Out}},}\
  }\href@noop {} {\  (\bibinfo {year} {2023})},\ \Eprint
  {http://arxiv.org/abs/2301.00599} {arXiv:2301.00599 [astro-ph.CO]}
  \BibitemShut {NoStop}%
\bibitem [{\citenamefont {Choudhury}\ \emph {et~al.}(2023)\citenamefont
  {Choudhury}, \citenamefont {Gangopadhyay},\ and\ \citenamefont
  {Sami}}]{Choudhury:2023vuj}%
  \BibitemOpen
  \bibfield  {author} {\bibinfo {author} {\bibfnamefont {Sayantan}\
  \bibnamefont {Choudhury}}, \bibinfo {author} {\bibfnamefont {Mayukh~R.}\
  \bibnamefont {Gangopadhyay}}, \ and\ \bibinfo {author} {\bibfnamefont
  {M.}~\bibnamefont {Sami}},\ }\bibfield  {title} {\enquote {\bibinfo {title}
  {{No-go for the formation of heavy mass Primordial Black Holes in Single
  Field Inflation}},}\ }\href@noop {} {\  (\bibinfo {year} {2023})},\ \Eprint
  {http://arxiv.org/abs/2301.10000} {arXiv:2301.10000 [astro-ph.CO]}
  \BibitemShut {NoStop}%
\bibitem [{\citenamefont {Lalak}\ \emph {et~al.}(2007)\citenamefont {Lalak},
  \citenamefont {Langlois}, \citenamefont {Pokorski},\ and\ \citenamefont
  {Turzynski}}]{Lalak:2007vi}%
  \BibitemOpen
  \bibfield  {author} {\bibinfo {author} {\bibfnamefont {Z.}~\bibnamefont
  {Lalak}}, \bibinfo {author} {\bibfnamefont {D.}~\bibnamefont {Langlois}},
  \bibinfo {author} {\bibfnamefont {S.}~\bibnamefont {Pokorski}}, \ and\
  \bibinfo {author} {\bibfnamefont {K.}~\bibnamefont {Turzynski}},\ }\bibfield
  {title} {\enquote {\bibinfo {title} {{Curvature and isocurvature
  perturbations in two-field inflation}},}\ }\href {\doibase
  10.1088/1475-7516/2007/07/014} {\bibfield  {journal} {\bibinfo  {journal}
  {JCAP}\ }\textbf {\bibinfo {volume} {07}},\ \bibinfo {pages} {014} (\bibinfo
  {year} {2007})},\ \Eprint {http://arxiv.org/abs/0704.0212} {arXiv:0704.0212
  [hep-th]} \BibitemShut {NoStop}%
\bibitem [{\citenamefont {van~de Bruck}\ and\ \citenamefont
  {Robinson}(2014)}]{vandeBruck:2014ata}%
  \BibitemOpen
  \bibfield  {author} {\bibinfo {author} {\bibfnamefont {Carsten}\ \bibnamefont
  {van~de Bruck}}\ and\ \bibinfo {author} {\bibfnamefont {Mathew}\ \bibnamefont
  {Robinson}},\ }\bibfield  {title} {\enquote {\bibinfo {title} {{Power Spectra
  beyond the Slow Roll Approximation in Theories with Non-Canonical Kinetic
  Terms}},}\ }\href {\doibase 10.1088/1475-7516/2014/08/024} {\bibfield
  {journal} {\bibinfo  {journal} {JCAP}\ }\textbf {\bibinfo {volume} {08}},\
  \bibinfo {pages} {024} (\bibinfo {year} {2014})},\ \Eprint
  {http://arxiv.org/abs/1404.7806} {arXiv:1404.7806 [astro-ph.CO]} \BibitemShut
  {NoStop}%
\bibitem [{\citenamefont {Braglia}\ \emph
  {et~al.}(2020{\natexlab{b}})\citenamefont {Braglia}, \citenamefont {Hazra},
  \citenamefont {Sriramkumar},\ and\ \citenamefont
  {Finelli}}]{Braglia:2020fms}%
  \BibitemOpen
  \bibfield  {author} {\bibinfo {author} {\bibfnamefont {Matteo}\ \bibnamefont
  {Braglia}}, \bibinfo {author} {\bibfnamefont {Dhiraj~Kumar}\ \bibnamefont
  {Hazra}}, \bibinfo {author} {\bibfnamefont {L.}~\bibnamefont {Sriramkumar}},
  \ and\ \bibinfo {author} {\bibfnamefont {Fabio}\ \bibnamefont {Finelli}},\
  }\bibfield  {title} {\enquote {\bibinfo {title} {{Generating primordial
  features at large scales in two field models of inflation}},}\ }\href
  {\doibase 10.1088/1475-7516/2020/08/025} {\bibfield  {journal} {\bibinfo
  {journal} {JCAP}\ }\textbf {\bibinfo {volume} {08}},\ \bibinfo {pages} {025}
  (\bibinfo {year} {2020}{\natexlab{b}})},\ \Eprint
  {http://arxiv.org/abs/2004.00672} {arXiv:2004.00672 [astro-ph.CO]}
  \BibitemShut {NoStop}%
\bibitem [{\citenamefont {Sasaki}(2008)}]{Sasaki:2008uc}%
  \BibitemOpen
  \bibfield  {author} {\bibinfo {author} {\bibfnamefont {Misao}\ \bibnamefont
  {Sasaki}},\ }\bibfield  {title} {\enquote {\bibinfo {title} {{Multi-brid
  inflation and non-Gaussianity}},}\ }\href {\doibase 10.1143/PTP.120.159}
  {\bibfield  {journal} {\bibinfo  {journal} {Prog. Theor. Phys.}\ }\textbf
  {\bibinfo {volume} {120}},\ \bibinfo {pages} {159--174} (\bibinfo {year}
  {2008})},\ \Eprint {http://arxiv.org/abs/0805.0974} {arXiv:0805.0974
  [astro-ph]} \BibitemShut {NoStop}%
\bibitem [{\citenamefont {Huang}(2009)}]{Huang:2009vk}%
  \BibitemOpen
  \bibfield  {author} {\bibinfo {author} {\bibfnamefont {Qing-Guo}\
  \bibnamefont {Huang}},\ }\bibfield  {title} {\enquote {\bibinfo {title} {{A
  Geometric description of the non-Gaussianity generated at the end of
  multi-field inflation}},}\ }\href {\doibase 10.1088/1475-7516/2009/06/035}
  {\bibfield  {journal} {\bibinfo  {journal} {JCAP}\ }\textbf {\bibinfo
  {volume} {06}},\ \bibinfo {pages} {035} (\bibinfo {year} {2009})},\ \Eprint
  {http://arxiv.org/abs/0904.2649} {arXiv:0904.2649 [hep-th]} \BibitemShut
  {NoStop}%
\bibitem [{\citenamefont {Suyama}\ and\ \citenamefont
  {Yamaguchi}(2008)}]{Suyama:2007bg}%
  \BibitemOpen
  \bibfield  {author} {\bibinfo {author} {\bibfnamefont {Teruaki}\ \bibnamefont
  {Suyama}}\ and\ \bibinfo {author} {\bibfnamefont {Masahide}\ \bibnamefont
  {Yamaguchi}},\ }\bibfield  {title} {\enquote {\bibinfo {title}
  {{Non-Gaussianity in the modulated reheating scenario}},}\ }\href {\doibase
  10.1103/PhysRevD.77.023505} {\bibfield  {journal} {\bibinfo  {journal} {Phys.
  Rev. D}\ }\textbf {\bibinfo {volume} {77}},\ \bibinfo {pages} {023505}
  (\bibinfo {year} {2008})},\ \Eprint {http://arxiv.org/abs/0709.2545}
  {arXiv:0709.2545 [astro-ph]} \BibitemShut {NoStop}%
\bibitem [{\citenamefont {Dvali}\ \emph {et~al.}(2004)\citenamefont {Dvali},
  \citenamefont {Gruzinov},\ and\ \citenamefont {Zaldarriaga}}]{Dvali:2003em}%
  \BibitemOpen
  \bibfield  {author} {\bibinfo {author} {\bibfnamefont {Gia}\ \bibnamefont
  {Dvali}}, \bibinfo {author} {\bibfnamefont {Andrei}\ \bibnamefont
  {Gruzinov}}, \ and\ \bibinfo {author} {\bibfnamefont {Matias}\ \bibnamefont
  {Zaldarriaga}},\ }\bibfield  {title} {\enquote {\bibinfo {title} {{A new
  mechanism for generating density perturbations from inflation}},}\ }\href
  {\doibase 10.1103/PhysRevD.69.023505} {\bibfield  {journal} {\bibinfo
  {journal} {Phys. Rev. D}\ }\textbf {\bibinfo {volume} {69}},\ \bibinfo
  {pages} {023505} (\bibinfo {year} {2004})},\ \Eprint
  {http://arxiv.org/abs/astro-ph/0303591} {arXiv:astro-ph/0303591} \BibitemShut
  {NoStop}%
\bibitem [{\citenamefont {Kofman}(2003)}]{Kofman:2003nx}%
  \BibitemOpen
  \bibfield  {author} {\bibinfo {author} {\bibfnamefont {Lev}\ \bibnamefont
  {Kofman}},\ }\bibfield  {title} {\enquote {\bibinfo {title} {{Probing string
  theory with modulated cosmological fluctuations}},}\ }\href@noop {} {\
  (\bibinfo {year} {2003})},\ \Eprint {http://arxiv.org/abs/astro-ph/0303614}
  {arXiv:astro-ph/0303614} \BibitemShut {NoStop}%
\bibitem [{\citenamefont {Mollerach}(1990)}]{Mollerach:1989hu}%
  \BibitemOpen
  \bibfield  {author} {\bibinfo {author} {\bibfnamefont {Silvia}\ \bibnamefont
  {Mollerach}},\ }\bibfield  {title} {\enquote {\bibinfo {title} {{Isocurvature
  Baryon Perturbations and Inflation}},}\ }\href {\doibase
  10.1103/PhysRevD.42.313} {\bibfield  {journal} {\bibinfo  {journal} {Phys.
  Rev. D}\ }\textbf {\bibinfo {volume} {42}},\ \bibinfo {pages} {313--325}
  (\bibinfo {year} {1990})}\BibitemShut {NoStop}%
\bibitem [{\citenamefont {Linde}\ and\ \citenamefont
  {Mukhanov}(1997)}]{Linde:1996gt}%
  \BibitemOpen
  \bibfield  {author} {\bibinfo {author} {\bibfnamefont {Andrei~D.}\
  \bibnamefont {Linde}}\ and\ \bibinfo {author} {\bibfnamefont
  {Viatcheslav~F.}\ \bibnamefont {Mukhanov}},\ }\bibfield  {title} {\enquote
  {\bibinfo {title} {{Nongaussian isocurvature perturbations from
  inflation}},}\ }\href {\doibase 10.1103/PhysRevD.56.R535} {\bibfield
  {journal} {\bibinfo  {journal} {Phys. Rev. D}\ }\textbf {\bibinfo {volume}
  {56}},\ \bibinfo {pages} {R535--R539} (\bibinfo {year} {1997})},\ \Eprint
  {http://arxiv.org/abs/astro-ph/9610219} {arXiv:astro-ph/9610219} \BibitemShut
  {NoStop}%
\bibitem [{\citenamefont {Enqvist}\ and\ \citenamefont
  {Sloth}(2002)}]{Enqvist:2001zp}%
  \BibitemOpen
  \bibfield  {author} {\bibinfo {author} {\bibfnamefont {Kari}\ \bibnamefont
  {Enqvist}}\ and\ \bibinfo {author} {\bibfnamefont {Martin~S.}\ \bibnamefont
  {Sloth}},\ }\bibfield  {title} {\enquote {\bibinfo {title} {{Adiabatic CMB
  perturbations in pre - big bang string cosmology}},}\ }\href {\doibase
  10.1016/S0550-3213(02)00043-3} {\bibfield  {journal} {\bibinfo  {journal}
  {Nucl. Phys. B}\ }\textbf {\bibinfo {volume} {626}},\ \bibinfo {pages}
  {395--409} (\bibinfo {year} {2002})},\ \Eprint
  {http://arxiv.org/abs/hep-ph/0109214} {arXiv:hep-ph/0109214} \BibitemShut
  {NoStop}%
\bibitem [{\citenamefont {Lyth}\ and\ \citenamefont
  {Wands}(2002)}]{Lyth:2001nq}%
  \BibitemOpen
  \bibfield  {author} {\bibinfo {author} {\bibfnamefont {David~H.}\
  \bibnamefont {Lyth}}\ and\ \bibinfo {author} {\bibfnamefont {David}\
  \bibnamefont {Wands}},\ }\bibfield  {title} {\enquote {\bibinfo {title}
  {{Generating the curvature perturbation without an inflaton}},}\ }\href
  {\doibase 10.1016/S0370-2693(01)01366-1} {\bibfield  {journal} {\bibinfo
  {journal} {Phys. Lett. B}\ }\textbf {\bibinfo {volume} {524}},\ \bibinfo
  {pages} {5--14} (\bibinfo {year} {2002})},\ \Eprint
  {http://arxiv.org/abs/hep-ph/0110002} {arXiv:hep-ph/0110002} \BibitemShut
  {NoStop}%
\bibitem [{\citenamefont {Moroi}\ and\ \citenamefont
  {Takahashi}(2001)}]{Moroi:2001ct}%
  \BibitemOpen
  \bibfield  {author} {\bibinfo {author} {\bibfnamefont {Takeo}\ \bibnamefont
  {Moroi}}\ and\ \bibinfo {author} {\bibfnamefont {Tomo}\ \bibnamefont
  {Takahashi}},\ }\bibfield  {title} {\enquote {\bibinfo {title} {{Effects of
  cosmological moduli fields on cosmic microwave background}},}\ }\href
  {\doibase 10.1016/S0370-2693(01)01295-3} {\bibfield  {journal} {\bibinfo
  {journal} {Phys. Lett. B}\ }\textbf {\bibinfo {volume} {522}},\ \bibinfo
  {pages} {215--221} (\bibinfo {year} {2001})},\ \bibinfo {note} {[Erratum:
  Phys.Lett.B 539, 303--303 (2002)]},\ \Eprint
  {http://arxiv.org/abs/hep-ph/0110096} {arXiv:hep-ph/0110096} \BibitemShut
  {NoStop}%
\bibitem [{\citenamefont {Sasaki}\ \emph {et~al.}(2006)\citenamefont {Sasaki},
  \citenamefont {Valiviita},\ and\ \citenamefont {Wands}}]{Sasaki:2006kq}%
  \BibitemOpen
  \bibfield  {author} {\bibinfo {author} {\bibfnamefont {Misao}\ \bibnamefont
  {Sasaki}}, \bibinfo {author} {\bibfnamefont {Jussi}\ \bibnamefont
  {Valiviita}}, \ and\ \bibinfo {author} {\bibfnamefont {David}\ \bibnamefont
  {Wands}},\ }\bibfield  {title} {\enquote {\bibinfo {title} {{Non-Gaussianity
  of the primordial perturbation in the curvaton model}},}\ }\href {\doibase
  10.1103/PhysRevD.74.103003} {\bibfield  {journal} {\bibinfo  {journal} {Phys.
  Rev. D}\ }\textbf {\bibinfo {volume} {74}},\ \bibinfo {pages} {103003}
  (\bibinfo {year} {2006})},\ \Eprint {http://arxiv.org/abs/astro-ph/0607627}
  {arXiv:astro-ph/0607627} \BibitemShut {NoStop}%
\bibitem [{\citenamefont {Enqvist}\ and\ \citenamefont
  {Nurmi}(2005)}]{Enqvist:2005pg}%
  \BibitemOpen
  \bibfield  {author} {\bibinfo {author} {\bibfnamefont {Kari}\ \bibnamefont
  {Enqvist}}\ and\ \bibinfo {author} {\bibfnamefont {Sami}\ \bibnamefont
  {Nurmi}},\ }\bibfield  {title} {\enquote {\bibinfo {title} {{Non-gaussianity
  in curvaton models with nearly quadratic potential}},}\ }\href {\doibase
  10.1088/1475-7516/2005/10/013} {\bibfield  {journal} {\bibinfo  {journal}
  {JCAP}\ }\textbf {\bibinfo {volume} {10}},\ \bibinfo {pages} {013} (\bibinfo
  {year} {2005})},\ \Eprint {http://arxiv.org/abs/astro-ph/0508573}
  {arXiv:astro-ph/0508573} \BibitemShut {NoStop}%
\bibitem [{\citenamefont {Huang}\ and\ \citenamefont
  {Wang}(2008)}]{Huang:2008bg}%
  \BibitemOpen
  \bibfield  {author} {\bibinfo {author} {\bibfnamefont {Qing-Guo}\
  \bibnamefont {Huang}}\ and\ \bibinfo {author} {\bibfnamefont
  {Yi}~\bibnamefont {Wang}},\ }\bibfield  {title} {\enquote {\bibinfo {title}
  {{Curvaton Dynamics and the Non-Linearity Parameters in Curvaton Model}},}\
  }\href {\doibase 10.1088/1475-7516/2008/09/025} {\bibfield  {journal}
  {\bibinfo  {journal} {JCAP}\ }\textbf {\bibinfo {volume} {09}},\ \bibinfo
  {pages} {025} (\bibinfo {year} {2008})},\ \Eprint
  {http://arxiv.org/abs/0808.1168} {arXiv:0808.1168 [hep-th]} \BibitemShut
  {NoStop}%
\bibitem [{\citenamefont {Huang}(2008)}]{Huang:2008zj}%
  \BibitemOpen
  \bibfield  {author} {\bibinfo {author} {\bibfnamefont {Qing-Guo}\
  \bibnamefont {Huang}},\ }\bibfield  {title} {\enquote {\bibinfo {title} {{A
  Curvaton with a Polynomial Potential}},}\ }\href {\doibase
  10.1088/1475-7516/2008/11/005} {\bibfield  {journal} {\bibinfo  {journal}
  {JCAP}\ }\textbf {\bibinfo {volume} {11}},\ \bibinfo {pages} {005} (\bibinfo
  {year} {2008})},\ \Eprint {http://arxiv.org/abs/0808.1793} {arXiv:0808.1793
  [hep-th]} \BibitemShut {NoStop}%
\bibitem [{\citenamefont {Chingangbam}\ and\ \citenamefont
  {Huang}(2009)}]{Chingangbam:2009xi}%
  \BibitemOpen
  \bibfield  {author} {\bibinfo {author} {\bibfnamefont {Pravabati}\
  \bibnamefont {Chingangbam}}\ and\ \bibinfo {author} {\bibfnamefont
  {Qing-Guo}\ \bibnamefont {Huang}},\ }\bibfield  {title} {\enquote {\bibinfo
  {title} {{The Curvature Perturbation in the Axion-type Curvaton Model}},}\
  }\href {\doibase 10.1088/1475-7516/2009/04/031} {\bibfield  {journal}
  {\bibinfo  {journal} {JCAP}\ }\textbf {\bibinfo {volume} {04}},\ \bibinfo
  {pages} {031} (\bibinfo {year} {2009})},\ \Eprint
  {http://arxiv.org/abs/0902.2619} {arXiv:0902.2619 [astro-ph.CO]} \BibitemShut
  {NoStop}%
\bibitem [{\citenamefont {Chingangbam}\ and\ \citenamefont
  {Huang}(2011)}]{Chingangbam:2010xn}%
  \BibitemOpen
  \bibfield  {author} {\bibinfo {author} {\bibfnamefont {Pravabati}\
  \bibnamefont {Chingangbam}}\ and\ \bibinfo {author} {\bibfnamefont
  {Qing-Guo}\ \bibnamefont {Huang}},\ }\bibfield  {title} {\enquote {\bibinfo
  {title} {{New features in curvaton model}},}\ }\href {\doibase
  10.1103/PhysRevD.83.023527} {\bibfield  {journal} {\bibinfo  {journal} {Phys.
  Rev. D}\ }\textbf {\bibinfo {volume} {83}},\ \bibinfo {pages} {023527}
  (\bibinfo {year} {2011})},\ \Eprint {http://arxiv.org/abs/1006.4006}
  {arXiv:1006.4006 [astro-ph.CO]} \BibitemShut {NoStop}%
\bibitem [{\citenamefont {Kawasaki}\ \emph {et~al.}(2011)\citenamefont
  {Kawasaki}, \citenamefont {Kobayashi},\ and\ \citenamefont
  {Takahashi}}]{Kawasaki:2011pd}%
  \BibitemOpen
  \bibfield  {author} {\bibinfo {author} {\bibfnamefont {Masahiro}\
  \bibnamefont {Kawasaki}}, \bibinfo {author} {\bibfnamefont {Takeshi}\
  \bibnamefont {Kobayashi}}, \ and\ \bibinfo {author} {\bibfnamefont
  {Fuminobu}\ \bibnamefont {Takahashi}},\ }\bibfield  {title} {\enquote
  {\bibinfo {title} {{Non-Gaussianity from Curvatons Revisited}},}\ }\href
  {\doibase 10.1103/PhysRevD.84.123506} {\bibfield  {journal} {\bibinfo
  {journal} {Phys. Rev. D}\ }\textbf {\bibinfo {volume} {84}},\ \bibinfo
  {pages} {123506} (\bibinfo {year} {2011})},\ \Eprint
  {http://arxiv.org/abs/1107.6011} {arXiv:1107.6011 [astro-ph.CO]} \BibitemShut
  {NoStop}%
\bibitem [{\citenamefont {Kristiano}\ and\ \citenamefont
  {Yokoyama}(2022{\natexlab{b}})}]{Kristiano:2021urj}%
  \BibitemOpen
  \bibfield  {author} {\bibinfo {author} {\bibfnamefont {Jason}\ \bibnamefont
  {Kristiano}}\ and\ \bibinfo {author} {\bibfnamefont {Jun'ichi}\ \bibnamefont
  {Yokoyama}},\ }\bibfield  {title} {\enquote {\bibinfo {title} {{Why Must
  Primordial Non-Gaussianity Be Very Small?}}}\ }\href {\doibase
  10.1103/PhysRevLett.128.061301} {\bibfield  {journal} {\bibinfo  {journal}
  {Phys. Rev. Lett.}\ }\textbf {\bibinfo {volume} {128}},\ \bibinfo {pages}
  {061301} (\bibinfo {year} {2022}{\natexlab{b}})},\ \Eprint
  {http://arxiv.org/abs/2104.01953} {arXiv:2104.01953 [hep-th]} \BibitemShut
  {NoStop}%
\bibitem [{\citenamefont {Press}\ and\ \citenamefont
  {Schechter}(1974)}]{Press:1973iz}%
  \BibitemOpen
  \bibfield  {author} {\bibinfo {author} {\bibfnamefont {William~H.}\
  \bibnamefont {Press}}\ and\ \bibinfo {author} {\bibfnamefont {Paul}\
  \bibnamefont {Schechter}},\ }\bibfield  {title} {\enquote {\bibinfo {title}
  {{Formation of galaxies and clusters of galaxies by selfsimilar gravitational
  condensation}},}\ }\href {\doibase 10.1086/152650} {\bibfield  {journal}
  {\bibinfo  {journal} {Astrophys. J.}\ }\textbf {\bibinfo {volume} {187}},\
  \bibinfo {pages} {425--438} (\bibinfo {year} {1974})}\BibitemShut {NoStop}%
\bibitem [{\citenamefont {Harada}\ \emph {et~al.}(2013)\citenamefont {Harada},
  \citenamefont {Yoo},\ and\ \citenamefont {Kohri}}]{Harada:2013epa}%
  \BibitemOpen
  \bibfield  {author} {\bibinfo {author} {\bibfnamefont {Tomohiro}\
  \bibnamefont {Harada}}, \bibinfo {author} {\bibfnamefont {Chul-Moon}\
  \bibnamefont {Yoo}}, \ and\ \bibinfo {author} {\bibfnamefont {Kazunori}\
  \bibnamefont {Kohri}},\ }\bibfield  {title} {\enquote {\bibinfo {title}
  {{Threshold of primordial black hole formation}},}\ }\href {\doibase
  10.1103/PhysRevD.88.084051, 10.1103/PhysRevD.89.029903} {\bibfield  {journal}
  {\bibinfo  {journal} {Phys. Rev.}\ }\textbf {\bibinfo {volume} {D88}},\
  \bibinfo {pages} {084051} (\bibinfo {year} {2013})},\ \bibinfo {note}
  {[Erratum: Phys. Rev.D89,no.2,029903(2014)]},\ \Eprint
  {http://arxiv.org/abs/1309.4201} {arXiv:1309.4201 [astro-ph.CO]} \BibitemShut
  {NoStop}%
\bibitem [{\citenamefont {Choptuik}(1993)}]{Choptuik:1992jv}%
  \BibitemOpen
  \bibfield  {author} {\bibinfo {author} {\bibfnamefont {Matthew~W.}\
  \bibnamefont {Choptuik}},\ }\bibfield  {title} {\enquote {\bibinfo {title}
  {{Universality and scaling in gravitational collapse of a massless scalar
  field}},}\ }\href {\doibase 10.1103/PhysRevLett.70.9} {\bibfield  {journal}
  {\bibinfo  {journal} {Phys. Rev. Lett.}\ }\textbf {\bibinfo {volume} {70}},\
  \bibinfo {pages} {9--12} (\bibinfo {year} {1993})}\BibitemShut {NoStop}%
\bibitem [{\citenamefont {Evans}\ and\ \citenamefont
  {Coleman}(1994)}]{Evans:1994pj}%
  \BibitemOpen
  \bibfield  {author} {\bibinfo {author} {\bibfnamefont {Charles~R.}\
  \bibnamefont {Evans}}\ and\ \bibinfo {author} {\bibfnamefont {Jason~S.}\
  \bibnamefont {Coleman}},\ }\bibfield  {title} {\enquote {\bibinfo {title}
  {{Observation of critical phenomena and selfsimilarity in the gravitational
  collapse of radiation fluid}},}\ }\href {\doibase
  10.1103/PhysRevLett.72.1782} {\bibfield  {journal} {\bibinfo  {journal}
  {Phys. Rev. Lett.}\ }\textbf {\bibinfo {volume} {72}},\ \bibinfo {pages}
  {1782--1785} (\bibinfo {year} {1994})},\ \Eprint
  {http://arxiv.org/abs/gr-qc/9402041} {arXiv:gr-qc/9402041} \BibitemShut
  {NoStop}%
\bibitem [{\citenamefont {Niemeyer}\ and\ \citenamefont
  {Jedamzik}(1998)}]{Niemeyer:1997mt}%
  \BibitemOpen
  \bibfield  {author} {\bibinfo {author} {\bibfnamefont {Jens~C.}\ \bibnamefont
  {Niemeyer}}\ and\ \bibinfo {author} {\bibfnamefont {K.}~\bibnamefont
  {Jedamzik}},\ }\bibfield  {title} {\enquote {\bibinfo {title} {{Near-critical
  gravitational collapse and the initial mass function of primordial black
  holes}},}\ }\href {\doibase 10.1103/PhysRevLett.80.5481} {\bibfield
  {journal} {\bibinfo  {journal} {Phys. Rev. Lett.}\ }\textbf {\bibinfo
  {volume} {80}},\ \bibinfo {pages} {5481--5484} (\bibinfo {year} {1998})},\
  \Eprint {http://arxiv.org/abs/astro-ph/9709072} {arXiv:astro-ph/9709072}
  \BibitemShut {NoStop}%
\bibitem [{\citenamefont {Koike}\ \emph {et~al.}(1995)\citenamefont {Koike},
  \citenamefont {Hara},\ and\ \citenamefont {Adachi}}]{Koike:1995jm}%
  \BibitemOpen
  \bibfield  {author} {\bibinfo {author} {\bibfnamefont {Tatsuhiko}\
  \bibnamefont {Koike}}, \bibinfo {author} {\bibfnamefont {Takashi}\
  \bibnamefont {Hara}}, \ and\ \bibinfo {author} {\bibfnamefont {Satoshi}\
  \bibnamefont {Adachi}},\ }\bibfield  {title} {\enquote {\bibinfo {title}
  {{Critical behavior in gravitational collapse of radiation fluid: A
  Renormalization group (linear perturbation) analysis}},}\ }\href {\doibase
  10.1103/PhysRevLett.74.5170} {\bibfield  {journal} {\bibinfo  {journal}
  {Phys. Rev. Lett.}\ }\textbf {\bibinfo {volume} {74}},\ \bibinfo {pages}
  {5170--5173} (\bibinfo {year} {1995})},\ \Eprint
  {http://arxiv.org/abs/gr-qc/9503007} {arXiv:gr-qc/9503007} \BibitemShut
  {NoStop}%
\bibitem [{\citenamefont {Young}\ \emph {et~al.}(2019)\citenamefont {Young},
  \citenamefont {Musco},\ and\ \citenamefont {Byrnes}}]{Young:2019yug}%
  \BibitemOpen
  \bibfield  {author} {\bibinfo {author} {\bibfnamefont {Sam}\ \bibnamefont
  {Young}}, \bibinfo {author} {\bibfnamefont {Ilia}\ \bibnamefont {Musco}}, \
  and\ \bibinfo {author} {\bibfnamefont {Christian~T.}\ \bibnamefont
  {Byrnes}},\ }\bibfield  {title} {\enquote {\bibinfo {title} {{Primordial
  black hole formation and abundance: contribution from the non-linear relation
  between the density and curvature perturbation}},}\ }\href {\doibase
  10.1088/1475-7516/2019/11/012} {\bibfield  {journal} {\bibinfo  {journal}
  {JCAP}\ }\textbf {\bibinfo {volume} {11}},\ \bibinfo {pages} {012} (\bibinfo
  {year} {2019})},\ \Eprint {http://arxiv.org/abs/1904.00984} {arXiv:1904.00984
  [astro-ph.CO]} \BibitemShut {NoStop}%
\bibitem [{\citenamefont {De~Luca}\ \emph
  {et~al.}(2019{\natexlab{a}})\citenamefont {De~Luca}, \citenamefont
  {Franciolini}, \citenamefont {Kehagias}, \citenamefont {Peloso},
  \citenamefont {Riotto},\ and\ \citenamefont {\"Unal}}]{DeLuca:2019qsy}%
  \BibitemOpen
  \bibfield  {author} {\bibinfo {author} {\bibfnamefont {V.}~\bibnamefont
  {De~Luca}}, \bibinfo {author} {\bibfnamefont {G.}~\bibnamefont
  {Franciolini}}, \bibinfo {author} {\bibfnamefont {A.}~\bibnamefont
  {Kehagias}}, \bibinfo {author} {\bibfnamefont {M.}~\bibnamefont {Peloso}},
  \bibinfo {author} {\bibfnamefont {A.}~\bibnamefont {Riotto}}, \ and\ \bibinfo
  {author} {\bibfnamefont {C.}~\bibnamefont {\"Unal}},\ }\bibfield  {title}
  {\enquote {\bibinfo {title} {{The Ineludible non-Gaussianity of the
  Primordial Black Hole Abundance}},}\ }\href {\doibase
  10.1088/1475-7516/2019/07/048} {\bibfield  {journal} {\bibinfo  {journal}
  {JCAP}\ }\textbf {\bibinfo {volume} {07}},\ \bibinfo {pages} {048} (\bibinfo
  {year} {2019}{\natexlab{a}})},\ \Eprint {http://arxiv.org/abs/1904.00970}
  {arXiv:1904.00970 [astro-ph.CO]} \BibitemShut {NoStop}%
\bibitem [{\citenamefont {Kawasaki}\ and\ \citenamefont
  {Nakatsuka}(2019)}]{Kawasaki:2019mbl}%
  \BibitemOpen
  \bibfield  {author} {\bibinfo {author} {\bibfnamefont {Masahiro}\
  \bibnamefont {Kawasaki}}\ and\ \bibinfo {author} {\bibfnamefont {Hiromasa}\
  \bibnamefont {Nakatsuka}},\ }\bibfield  {title} {\enquote {\bibinfo {title}
  {{Effect of nonlinearity between density and curvature perturbations on the
  primordial black hole formation}},}\ }\href {\doibase
  10.1103/PhysRevD.99.123501} {\bibfield  {journal} {\bibinfo  {journal} {Phys.
  Rev. D}\ }\textbf {\bibinfo {volume} {99}},\ \bibinfo {pages} {123501}
  (\bibinfo {year} {2019})},\ \Eprint {http://arxiv.org/abs/1903.02994}
  {arXiv:1903.02994 [astro-ph.CO]} \BibitemShut {NoStop}%
\bibitem [{\citenamefont {Tomita}(1967)}]{tomita1967non}%
  \BibitemOpen
  \bibfield  {author} {\bibinfo {author} {\bibfnamefont {Kenji}\ \bibnamefont
  {Tomita}},\ }\bibfield  {title} {\enquote {\bibinfo {title} {Non-linear
  theory of gravitational instability in the expanding universe},}\ }\href@noop
  {} {\bibfield  {journal} {\bibinfo  {journal} {Progress of Theoretical
  Physics}\ }\textbf {\bibinfo {volume} {37}},\ \bibinfo {pages} {831--846}
  (\bibinfo {year} {1967})}\BibitemShut {NoStop}%
\bibitem [{\citenamefont {Matarrese}\ \emph {et~al.}(1993)\citenamefont
  {Matarrese}, \citenamefont {Pantano},\ and\ \citenamefont
  {Saez}}]{Matarrese:1992rp}%
  \BibitemOpen
  \bibfield  {author} {\bibinfo {author} {\bibfnamefont {Sabino}\ \bibnamefont
  {Matarrese}}, \bibinfo {author} {\bibfnamefont {Ornella}\ \bibnamefont
  {Pantano}}, \ and\ \bibinfo {author} {\bibfnamefont {Diego}\ \bibnamefont
  {Saez}},\ }\bibfield  {title} {\enquote {\bibinfo {title} {{A General
  relativistic approach to the nonlinear evolution of collisionless matter}},}\
  }\href {\doibase 10.1103/PhysRevD.47.1311} {\bibfield  {journal} {\bibinfo
  {journal} {Phys. Rev.}\ }\textbf {\bibinfo {volume} {D47}},\ \bibinfo {pages}
  {1311--1323} (\bibinfo {year} {1993})}\BibitemShut {NoStop}%
\bibitem [{\citenamefont {Matarrese}\ \emph {et~al.}(1994)\citenamefont
  {Matarrese}, \citenamefont {Pantano},\ and\ \citenamefont
  {Saez}}]{Matarrese:1993zf}%
  \BibitemOpen
  \bibfield  {author} {\bibinfo {author} {\bibfnamefont {Sabino}\ \bibnamefont
  {Matarrese}}, \bibinfo {author} {\bibfnamefont {Ornella}\ \bibnamefont
  {Pantano}}, \ and\ \bibinfo {author} {\bibfnamefont {Diego}\ \bibnamefont
  {Saez}},\ }\bibfield  {title} {\enquote {\bibinfo {title} {{General
  relativistic dynamics of irrotational dust: Cosmological implications}},}\
  }\href {\doibase 10.1103/PhysRevLett.72.320} {\bibfield  {journal} {\bibinfo
  {journal} {Phys. Rev. Lett.}\ }\textbf {\bibinfo {volume} {72}},\ \bibinfo
  {pages} {320--323} (\bibinfo {year} {1994})},\ \Eprint
  {http://arxiv.org/abs/astro-ph/9310036} {arXiv:astro-ph/9310036 [astro-ph]}
  \BibitemShut {NoStop}%
\bibitem [{\citenamefont {Matarrese}\ \emph {et~al.}(1998)\citenamefont
  {Matarrese}, \citenamefont {Mollerach},\ and\ \citenamefont
  {Bruni}}]{Matarrese:1997ay}%
  \BibitemOpen
  \bibfield  {author} {\bibinfo {author} {\bibfnamefont {Sabino}\ \bibnamefont
  {Matarrese}}, \bibinfo {author} {\bibfnamefont {Silvia}\ \bibnamefont
  {Mollerach}}, \ and\ \bibinfo {author} {\bibfnamefont {Marco}\ \bibnamefont
  {Bruni}},\ }\bibfield  {title} {\enquote {\bibinfo {title} {{Second order
  perturbations of the Einstein-de Sitter universe}},}\ }\href {\doibase
  10.1103/PhysRevD.58.043504} {\bibfield  {journal} {\bibinfo  {journal} {Phys.
  Rev.}\ }\textbf {\bibinfo {volume} {D58}},\ \bibinfo {pages} {043504}
  (\bibinfo {year} {1998})},\ \Eprint {http://arxiv.org/abs/astro-ph/9707278}
  {arXiv:astro-ph/9707278 [astro-ph]} \BibitemShut {NoStop}%
\bibitem [{\citenamefont {Noh}\ and\ \citenamefont {Hwang}(2004)}]{Noh:2004bc}%
  \BibitemOpen
  \bibfield  {author} {\bibinfo {author} {\bibfnamefont {Hyerim}\ \bibnamefont
  {Noh}}\ and\ \bibinfo {author} {\bibfnamefont {Jai-chan}\ \bibnamefont
  {Hwang}},\ }\bibfield  {title} {\enquote {\bibinfo {title} {{Second-order
  perturbations of the Friedmann world model}},}\ }\href {\doibase
  10.1103/PhysRevD.69.104011} {\bibfield  {journal} {\bibinfo  {journal} {Phys.
  Rev.}\ }\textbf {\bibinfo {volume} {D69}},\ \bibinfo {pages} {104011}
  (\bibinfo {year} {2004})}\BibitemShut {NoStop}%
\bibitem [{\citenamefont {Carbone}\ and\ \citenamefont
  {Matarrese}(2005)}]{Carbone:2004iv}%
  \BibitemOpen
  \bibfield  {author} {\bibinfo {author} {\bibfnamefont {Carmelita}\
  \bibnamefont {Carbone}}\ and\ \bibinfo {author} {\bibfnamefont {Sabino}\
  \bibnamefont {Matarrese}},\ }\bibfield  {title} {\enquote {\bibinfo {title}
  {{A Unified treatment of cosmological perturbations from super-horizon to
  small scales}},}\ }\href {\doibase 10.1103/PhysRevD.71.043508} {\bibfield
  {journal} {\bibinfo  {journal} {Phys. Rev.}\ }\textbf {\bibinfo {volume}
  {D71}},\ \bibinfo {pages} {043508} (\bibinfo {year} {2005})},\ \Eprint
  {http://arxiv.org/abs/astro-ph/0407611} {arXiv:astro-ph/0407611 [astro-ph]}
  \BibitemShut {NoStop}%
\bibitem [{\citenamefont {Nakamura}(2007)}]{Nakamura:2004rm}%
  \BibitemOpen
  \bibfield  {author} {\bibinfo {author} {\bibfnamefont {Kouji}\ \bibnamefont
  {Nakamura}},\ }\bibfield  {title} {\enquote {\bibinfo {title} {{Second-order
  gauge invariant cosmological perturbation theory: Einstein equations in terms
  of gauge invariant variables}},}\ }\href {\doibase 10.1143/PTP.117.17}
  {\bibfield  {journal} {\bibinfo  {journal} {Prog. Theor. Phys.}\ }\textbf
  {\bibinfo {volume} {117}},\ \bibinfo {pages} {17--74} (\bibinfo {year}
  {2007})},\ \Eprint {http://arxiv.org/abs/gr-qc/0605108} {arXiv:gr-qc/0605108
  [gr-qc]} \BibitemShut {NoStop}%
\bibitem [{\citenamefont {Ananda}\ \emph {et~al.}(2007)\citenamefont {Ananda},
  \citenamefont {Clarkson},\ and\ \citenamefont {Wands}}]{Ananda:2006af}%
  \BibitemOpen
  \bibfield  {author} {\bibinfo {author} {\bibfnamefont {Kishore~N.}\
  \bibnamefont {Ananda}}, \bibinfo {author} {\bibfnamefont {Chris}\
  \bibnamefont {Clarkson}}, \ and\ \bibinfo {author} {\bibfnamefont {David}\
  \bibnamefont {Wands}},\ }\bibfield  {title} {\enquote {\bibinfo {title} {{The
  Cosmological gravitational wave background from primordial density
  perturbations}},}\ }\href {\doibase 10.1103/PhysRevD.75.123518} {\bibfield
  {journal} {\bibinfo  {journal} {Phys. Rev.}\ }\textbf {\bibinfo {volume}
  {D75}},\ \bibinfo {pages} {123518} (\bibinfo {year} {2007})},\ \Eprint
  {http://arxiv.org/abs/gr-qc/0612013} {arXiv:gr-qc/0612013 [gr-qc]}
  \BibitemShut {NoStop}%
\bibitem [{\citenamefont {Baumann}\ \emph {et~al.}(2007)\citenamefont
  {Baumann}, \citenamefont {Steinhardt}, \citenamefont {Takahashi},\ and\
  \citenamefont {Ichiki}}]{Baumann:2007zm}%
  \BibitemOpen
  \bibfield  {author} {\bibinfo {author} {\bibfnamefont {Daniel}\ \bibnamefont
  {Baumann}}, \bibinfo {author} {\bibfnamefont {Paul~J.}\ \bibnamefont
  {Steinhardt}}, \bibinfo {author} {\bibfnamefont {Keitaro}\ \bibnamefont
  {Takahashi}}, \ and\ \bibinfo {author} {\bibfnamefont {Kiyotomo}\
  \bibnamefont {Ichiki}},\ }\bibfield  {title} {\enquote {\bibinfo {title}
  {{Gravitational Wave Spectrum Induced by Primordial Scalar Perturbations}},}\
  }\href {\doibase 10.1103/PhysRevD.76.084019} {\bibfield  {journal} {\bibinfo
  {journal} {Phys. Rev.}\ }\textbf {\bibinfo {volume} {D76}},\ \bibinfo {pages}
  {084019} (\bibinfo {year} {2007})},\ \Eprint
  {http://arxiv.org/abs/hep-th/0703290} {arXiv:hep-th/0703290 [hep-th]}
  \BibitemShut {NoStop}%
\bibitem [{\citenamefont {Saito}\ and\ \citenamefont
  {Yokoyama}(2009)}]{Saito:2008jc}%
  \BibitemOpen
  \bibfield  {author} {\bibinfo {author} {\bibfnamefont {Ryo}\ \bibnamefont
  {Saito}}\ and\ \bibinfo {author} {\bibfnamefont {Jun'ichi}\ \bibnamefont
  {Yokoyama}},\ }\bibfield  {title} {\enquote {\bibinfo {title} {{Gravitational
  wave background as a probe of the primordial black hole abundance}},}\ }\href
  {\doibase 10.1103/PhysRevLett.102.161101, 10.1103/PhysRevLett.107.069901}
  {\bibfield  {journal} {\bibinfo  {journal} {Phys. Rev. Lett.}\ }\textbf
  {\bibinfo {volume} {102}},\ \bibinfo {pages} {161101} (\bibinfo {year}
  {2009})},\ \bibinfo {note} {[Erratum: Phys. Rev. Lett.107,069901(2011)]},\
  \Eprint {http://arxiv.org/abs/0812.4339} {arXiv:0812.4339 [astro-ph]}
  \BibitemShut {NoStop}%
\bibitem [{\citenamefont {Arroja}\ \emph {et~al.}(2009)\citenamefont {Arroja},
  \citenamefont {Assadullahi}, \citenamefont {Koyama},\ and\ \citenamefont
  {Wands}}]{Arroja:2009sh}%
  \BibitemOpen
  \bibfield  {author} {\bibinfo {author} {\bibfnamefont {Frederico}\
  \bibnamefont {Arroja}}, \bibinfo {author} {\bibfnamefont {Hooshyar}\
  \bibnamefont {Assadullahi}}, \bibinfo {author} {\bibfnamefont {Kazuya}\
  \bibnamefont {Koyama}}, \ and\ \bibinfo {author} {\bibfnamefont {David}\
  \bibnamefont {Wands}},\ }\bibfield  {title} {\enquote {\bibinfo {title}
  {{Cosmological matching conditions for gravitational waves at second
  order}},}\ }\href {\doibase 10.1103/PhysRevD.80.123526} {\bibfield  {journal}
  {\bibinfo  {journal} {Phys. Rev. D}\ }\textbf {\bibinfo {volume} {80}},\
  \bibinfo {pages} {123526} (\bibinfo {year} {2009})},\ \Eprint
  {http://arxiv.org/abs/0907.3618} {arXiv:0907.3618 [astro-ph.CO]} \BibitemShut
  {NoStop}%
\bibitem [{\citenamefont {Assadullahi}\ and\ \citenamefont
  {Wands}(2010)}]{Assadullahi:2009jc}%
  \BibitemOpen
  \bibfield  {author} {\bibinfo {author} {\bibfnamefont {Hooshyar}\
  \bibnamefont {Assadullahi}}\ and\ \bibinfo {author} {\bibfnamefont {David}\
  \bibnamefont {Wands}},\ }\bibfield  {title} {\enquote {\bibinfo {title}
  {{Constraints on primordial density perturbations from induced gravitational
  waves}},}\ }\href {\doibase 10.1103/PhysRevD.81.023527} {\bibfield  {journal}
  {\bibinfo  {journal} {Phys. Rev.}\ }\textbf {\bibinfo {volume} {D81}},\
  \bibinfo {pages} {023527} (\bibinfo {year} {2010})},\ \Eprint
  {http://arxiv.org/abs/0907.4073} {arXiv:0907.4073 [astro-ph.CO]} \BibitemShut
  {NoStop}%
\bibitem [{\citenamefont {Bugaev}\ and\ \citenamefont
  {Klimai}(2010{\natexlab{a}})}]{Bugaev:2009kq}%
  \BibitemOpen
  \bibfield  {author} {\bibinfo {author} {\bibfnamefont {E.~V.}\ \bibnamefont
  {Bugaev}}\ and\ \bibinfo {author} {\bibfnamefont {P.~A.}\ \bibnamefont
  {Klimai}},\ }\bibfield  {title} {\enquote {\bibinfo {title} {{Bound on
  induced gravitational wave background from primordial black holes}},}\ }\href
  {\doibase 10.1134/S0021364010010017} {\bibfield  {journal} {\bibinfo
  {journal} {JETP Lett.}\ }\textbf {\bibinfo {volume} {91}},\ \bibinfo {pages}
  {1--5} (\bibinfo {year} {2010}{\natexlab{a}})},\ \Eprint
  {http://arxiv.org/abs/0911.0611} {arXiv:0911.0611 [astro-ph.CO]} \BibitemShut
  {NoStop}%
\bibitem [{\citenamefont {Bugaev}\ and\ \citenamefont
  {Klimai}(2010{\natexlab{b}})}]{Bugaev:2009zh}%
  \BibitemOpen
  \bibfield  {author} {\bibinfo {author} {\bibfnamefont {Edgar}\ \bibnamefont
  {Bugaev}}\ and\ \bibinfo {author} {\bibfnamefont {Peter}\ \bibnamefont
  {Klimai}},\ }\bibfield  {title} {\enquote {\bibinfo {title} {{Induced
  gravitational wave background and primordial black holes}},}\ }\href
  {\doibase 10.1103/PhysRevD.81.023517} {\bibfield  {journal} {\bibinfo
  {journal} {Phys. Rev.}\ }\textbf {\bibinfo {volume} {D81}},\ \bibinfo {pages}
  {023517} (\bibinfo {year} {2010}{\natexlab{b}})},\ \Eprint
  {http://arxiv.org/abs/0908.0664} {arXiv:0908.0664 [astro-ph.CO]} \BibitemShut
  {NoStop}%
\bibitem [{\citenamefont {Saito}\ and\ \citenamefont
  {Yokoyama}(2010)}]{Saito:2009jt}%
  \BibitemOpen
  \bibfield  {author} {\bibinfo {author} {\bibfnamefont {Ryo}\ \bibnamefont
  {Saito}}\ and\ \bibinfo {author} {\bibfnamefont {Jun'ichi}\ \bibnamefont
  {Yokoyama}},\ }\bibfield  {title} {\enquote {\bibinfo {title}
  {{Gravitational-Wave Constraints on the Abundance of Primordial Black
  Holes}},}\ }\href {\doibase 10.1143/PTP.126.351, 10.1143/PTP.123.867}
  {\bibfield  {journal} {\bibinfo  {journal} {Prog. Theor. Phys.}\ }\textbf
  {\bibinfo {volume} {123}},\ \bibinfo {pages} {867--886} (\bibinfo {year}
  {2010})},\ \bibinfo {note} {[Erratum: Prog. Theor. Phys.126,351(2011)]},\
  \Eprint {http://arxiv.org/abs/0912.5317} {arXiv:0912.5317 [astro-ph.CO]}
  \BibitemShut {NoStop}%
\bibitem [{\citenamefont {Bugaev}\ and\ \citenamefont
  {Klimai}(2011)}]{Bugaev:2010bb}%
  \BibitemOpen
  \bibfield  {author} {\bibinfo {author} {\bibfnamefont {Edgar}\ \bibnamefont
  {Bugaev}}\ and\ \bibinfo {author} {\bibfnamefont {Peter}\ \bibnamefont
  {Klimai}},\ }\bibfield  {title} {\enquote {\bibinfo {title} {{Constraints on
  the induced gravitational wave background from primordial black holes}},}\
  }\href {\doibase 10.1103/PhysRevD.83.083521} {\bibfield  {journal} {\bibinfo
  {journal} {Phys. Rev.}\ }\textbf {\bibinfo {volume} {D83}},\ \bibinfo {pages}
  {083521} (\bibinfo {year} {2011})},\ \Eprint {http://arxiv.org/abs/1012.4697}
  {arXiv:1012.4697 [astro-ph.CO]} \BibitemShut {NoStop}%
\bibitem [{\citenamefont {Alabidi}\ \emph {et~al.}(2013)\citenamefont
  {Alabidi}, \citenamefont {Kohri}, \citenamefont {Sasaki},\ and\ \citenamefont
  {Sendouda}}]{Alabidi:2013lya}%
  \BibitemOpen
  \bibfield  {author} {\bibinfo {author} {\bibfnamefont {Laila}\ \bibnamefont
  {Alabidi}}, \bibinfo {author} {\bibfnamefont {Kazunori}\ \bibnamefont
  {Kohri}}, \bibinfo {author} {\bibfnamefont {Misao}\ \bibnamefont {Sasaki}}, \
  and\ \bibinfo {author} {\bibfnamefont {Yuuiti}\ \bibnamefont {Sendouda}},\
  }\bibfield  {title} {\enquote {\bibinfo {title} {{Observable induced
  gravitational waves from an early matter phase}},}\ }\href {\doibase
  10.1088/1475-7516/2013/05/033} {\bibfield  {journal} {\bibinfo  {journal}
  {JCAP}\ }\textbf {\bibinfo {volume} {05}},\ \bibinfo {pages} {033} (\bibinfo
  {year} {2013})},\ \Eprint {http://arxiv.org/abs/1303.4519} {arXiv:1303.4519
  [astro-ph.CO]} \BibitemShut {NoStop}%
\bibitem [{\citenamefont {Nakama}\ and\ \citenamefont
  {Suyama}(2016)}]{Nakama:2016enz}%
  \BibitemOpen
  \bibfield  {author} {\bibinfo {author} {\bibfnamefont {Tomohiro}\
  \bibnamefont {Nakama}}\ and\ \bibinfo {author} {\bibfnamefont {Teruaki}\
  \bibnamefont {Suyama}},\ }\bibfield  {title} {\enquote {\bibinfo {title}
  {{Primordial black holes as a novel probe of primordial gravitational waves.
  II: Detailed analysis}},}\ }\href {\doibase 10.1103/PhysRevD.94.043507}
  {\bibfield  {journal} {\bibinfo  {journal} {Phys. Rev.}\ }\textbf {\bibinfo
  {volume} {D94}},\ \bibinfo {pages} {043507} (\bibinfo {year} {2016})},\
  \Eprint {http://arxiv.org/abs/1605.04482} {arXiv:1605.04482 [gr-qc]}
  \BibitemShut {NoStop}%
\bibitem [{\citenamefont {Nakama}\ \emph {et~al.}(2017)\citenamefont {Nakama},
  \citenamefont {Silk},\ and\ \citenamefont {Kamionkowski}}]{Nakama:2016gzw}%
  \BibitemOpen
  \bibfield  {author} {\bibinfo {author} {\bibfnamefont {Tomohiro}\
  \bibnamefont {Nakama}}, \bibinfo {author} {\bibfnamefont {Joseph}\
  \bibnamefont {Silk}}, \ and\ \bibinfo {author} {\bibfnamefont {Marc}\
  \bibnamefont {Kamionkowski}},\ }\bibfield  {title} {\enquote {\bibinfo
  {title} {{Stochastic gravitational waves associated with the formation of
  primordial black holes}},}\ }\href {\doibase 10.1103/PhysRevD.95.043511}
  {\bibfield  {journal} {\bibinfo  {journal} {Phys. Rev.}\ }\textbf {\bibinfo
  {volume} {D95}},\ \bibinfo {pages} {043511} (\bibinfo {year} {2017})},\
  \Eprint {http://arxiv.org/abs/1612.06264} {arXiv:1612.06264 [astro-ph.CO]}
  \BibitemShut {NoStop}%
\bibitem [{\citenamefont {Inomata}\ \emph
  {et~al.}(2017{\natexlab{b}})\citenamefont {Inomata}, \citenamefont
  {Kawasaki}, \citenamefont {Mukaida}, \citenamefont {Tada},\ and\
  \citenamefont {Yanagida}}]{Inomata:2016rbd}%
  \BibitemOpen
  \bibfield  {author} {\bibinfo {author} {\bibfnamefont {Keisuke}\ \bibnamefont
  {Inomata}}, \bibinfo {author} {\bibfnamefont {Masahiro}\ \bibnamefont
  {Kawasaki}}, \bibinfo {author} {\bibfnamefont {Kyohei}\ \bibnamefont
  {Mukaida}}, \bibinfo {author} {\bibfnamefont {Yuichiro}\ \bibnamefont
  {Tada}}, \ and\ \bibinfo {author} {\bibfnamefont {Tsutomu~T.}\ \bibnamefont
  {Yanagida}},\ }\bibfield  {title} {\enquote {\bibinfo {title} {{Inflationary
  primordial black holes for the LIGO gravitational wave events and pulsar
  timing array experiments}},}\ }\href {\doibase 10.1103/PhysRevD.95.123510}
  {\bibfield  {journal} {\bibinfo  {journal} {Phys. Rev. D}\ }\textbf {\bibinfo
  {volume} {95}},\ \bibinfo {pages} {123510} (\bibinfo {year}
  {2017}{\natexlab{b}})},\ \Eprint {http://arxiv.org/abs/1611.06130}
  {arXiv:1611.06130 [astro-ph.CO]} \BibitemShut {NoStop}%
\bibitem [{\citenamefont {Orlofsky}\ \emph {et~al.}(2017)\citenamefont
  {Orlofsky}, \citenamefont {Pierce},\ and\ \citenamefont
  {Wells}}]{Orlofsky:2016vbd}%
  \BibitemOpen
  \bibfield  {author} {\bibinfo {author} {\bibfnamefont {Nicholas}\
  \bibnamefont {Orlofsky}}, \bibinfo {author} {\bibfnamefont {Aaron}\
  \bibnamefont {Pierce}}, \ and\ \bibinfo {author} {\bibfnamefont {James~D.}\
  \bibnamefont {Wells}},\ }\bibfield  {title} {\enquote {\bibinfo {title}
  {{Inflationary theory and pulsar timing investigations of primordial black
  holes and gravitational waves}},}\ }\href {\doibase
  10.1103/PhysRevD.95.063518} {\bibfield  {journal} {\bibinfo  {journal} {Phys.
  Rev. D}\ }\textbf {\bibinfo {volume} {95}},\ \bibinfo {pages} {063518}
  (\bibinfo {year} {2017})},\ \Eprint {http://arxiv.org/abs/1612.05279}
  {arXiv:1612.05279 [astro-ph.CO]} \BibitemShut {NoStop}%
\bibitem [{\citenamefont {Garcia-Bellido}\ \emph {et~al.}(2017)\citenamefont
  {Garcia-Bellido}, \citenamefont {Peloso},\ and\ \citenamefont
  {Unal}}]{Garcia-Bellido:2017aan}%
  \BibitemOpen
  \bibfield  {author} {\bibinfo {author} {\bibfnamefont {Juan}\ \bibnamefont
  {Garcia-Bellido}}, \bibinfo {author} {\bibfnamefont {Marco}\ \bibnamefont
  {Peloso}}, \ and\ \bibinfo {author} {\bibfnamefont {Caner}\ \bibnamefont
  {Unal}},\ }\bibfield  {title} {\enquote {\bibinfo {title} {{Gravitational
  Wave signatures of inflationary models from Primordial Black Hole Dark
  Matter}},}\ }\href {\doibase 10.1088/1475-7516/2017/09/013} {\bibfield
  {journal} {\bibinfo  {journal} {JCAP}\ }\textbf {\bibinfo {volume} {09}},\
  \bibinfo {pages} {013} (\bibinfo {year} {2017})},\ \Eprint
  {http://arxiv.org/abs/1707.02441} {arXiv:1707.02441 [astro-ph.CO]}
  \BibitemShut {NoStop}%
\bibitem [{\citenamefont {Sasaki}\ \emph {et~al.}(2018)\citenamefont {Sasaki},
  \citenamefont {Suyama}, \citenamefont {Tanaka},\ and\ \citenamefont
  {Yokoyama}}]{Sasaki:2018dmp}%
  \BibitemOpen
  \bibfield  {author} {\bibinfo {author} {\bibfnamefont {Misao}\ \bibnamefont
  {Sasaki}}, \bibinfo {author} {\bibfnamefont {Teruaki}\ \bibnamefont
  {Suyama}}, \bibinfo {author} {\bibfnamefont {Takahiro}\ \bibnamefont
  {Tanaka}}, \ and\ \bibinfo {author} {\bibfnamefont {Shuichiro}\ \bibnamefont
  {Yokoyama}},\ }\bibfield  {title} {\enquote {\bibinfo {title} {{Primordial
  black holes—perspectives in gravitational wave astronomy}},}\ }\href
  {\doibase 10.1088/1361-6382/aaa7b4} {\bibfield  {journal} {\bibinfo
  {journal} {Class. Quant. Grav.}\ }\textbf {\bibinfo {volume} {35}},\ \bibinfo
  {pages} {063001} (\bibinfo {year} {2018})},\ \Eprint
  {http://arxiv.org/abs/1801.05235} {arXiv:1801.05235 [astro-ph.CO]}
  \BibitemShut {NoStop}%
\bibitem [{\citenamefont {Espinosa}\ \emph
  {et~al.}(2018{\natexlab{b}})\citenamefont {Espinosa}, \citenamefont {Racco},\
  and\ \citenamefont {Riotto}}]{Espinosa:2018eve}%
  \BibitemOpen
  \bibfield  {author} {\bibinfo {author} {\bibfnamefont {José~Ramón}\
  \bibnamefont {Espinosa}}, \bibinfo {author} {\bibfnamefont {Davide}\
  \bibnamefont {Racco}}, \ and\ \bibinfo {author} {\bibfnamefont {Antonio}\
  \bibnamefont {Riotto}},\ }\bibfield  {title} {\enquote {\bibinfo {title} {{A
  Cosmological Signature of the SM Higgs Instability: Gravitational Waves}},}\
  }\href {\doibase 10.1088/1475-7516/2018/09/012} {\bibfield  {journal}
  {\bibinfo  {journal} {JCAP}\ }\textbf {\bibinfo {volume} {1809}},\ \bibinfo
  {pages} {012} (\bibinfo {year} {2018}{\natexlab{b}})},\ \Eprint
  {http://arxiv.org/abs/1804.07732} {arXiv:1804.07732 [hep-ph]} \BibitemShut
  {NoStop}%
\bibitem [{\citenamefont {Kohri}\ and\ \citenamefont
  {Terada}(2018)}]{Kohri:2018awv}%
  \BibitemOpen
  \bibfield  {author} {\bibinfo {author} {\bibfnamefont {Kazunori}\
  \bibnamefont {Kohri}}\ and\ \bibinfo {author} {\bibfnamefont {Takahiro}\
  \bibnamefont {Terada}},\ }\bibfield  {title} {\enquote {\bibinfo {title}
  {{Semianalytic calculation of gravitational wave spectrum nonlinearly induced
  from primordial curvature perturbations}},}\ }\href {\doibase
  10.1103/PhysRevD.97.123532} {\bibfield  {journal} {\bibinfo  {journal} {Phys.
  Rev. D}\ }\textbf {\bibinfo {volume} {97}},\ \bibinfo {pages} {123532}
  (\bibinfo {year} {2018})},\ \Eprint {http://arxiv.org/abs/1804.08577}
  {arXiv:1804.08577 [gr-qc]} \BibitemShut {NoStop}%
\bibitem [{\citenamefont {Cai}\ \emph {et~al.}(2019{\natexlab{b}})\citenamefont
  {Cai}, \citenamefont {Pi},\ and\ \citenamefont {Sasaki}}]{Cai:2018dig}%
  \BibitemOpen
  \bibfield  {author} {\bibinfo {author} {\bibfnamefont {Rong-gen}\
  \bibnamefont {Cai}}, \bibinfo {author} {\bibfnamefont {Shi}\ \bibnamefont
  {Pi}}, \ and\ \bibinfo {author} {\bibfnamefont {Misao}\ \bibnamefont
  {Sasaki}},\ }\bibfield  {title} {\enquote {\bibinfo {title} {{Gravitational
  Waves Induced by non-Gaussian Scalar Perturbations}},}\ }\href {\doibase
  10.1103/PhysRevLett.122.201101} {\bibfield  {journal} {\bibinfo  {journal}
  {Phys. Rev. Lett.}\ }\textbf {\bibinfo {volume} {122}},\ \bibinfo {pages}
  {201101} (\bibinfo {year} {2019}{\natexlab{b}})},\ \Eprint
  {http://arxiv.org/abs/1810.11000} {arXiv:1810.11000 [astro-ph.CO]}
  \BibitemShut {NoStop}%
\bibitem [{\citenamefont {Bartolo}\ \emph
  {et~al.}(2019{\natexlab{a}})\citenamefont {Bartolo}, \citenamefont {De~Luca},
  \citenamefont {Franciolini}, \citenamefont {Lewis}, \citenamefont {Peloso},\
  and\ \citenamefont {Riotto}}]{Bartolo:2018evs}%
  \BibitemOpen
  \bibfield  {author} {\bibinfo {author} {\bibfnamefont {N.}~\bibnamefont
  {Bartolo}}, \bibinfo {author} {\bibfnamefont {V.}~\bibnamefont {De~Luca}},
  \bibinfo {author} {\bibfnamefont {G.}~\bibnamefont {Franciolini}}, \bibinfo
  {author} {\bibfnamefont {A.}~\bibnamefont {Lewis}}, \bibinfo {author}
  {\bibfnamefont {M.}~\bibnamefont {Peloso}}, \ and\ \bibinfo {author}
  {\bibfnamefont {A.}~\bibnamefont {Riotto}},\ }\bibfield  {title} {\enquote
  {\bibinfo {title} {{Primordial Black Hole Dark Matter: LISA Serendipity}},}\
  }\href {\doibase 10.1103/PhysRevLett.122.211301} {\bibfield  {journal}
  {\bibinfo  {journal} {Phys. Rev. Lett.}\ }\textbf {\bibinfo {volume} {122}},\
  \bibinfo {pages} {211301} (\bibinfo {year} {2019}{\natexlab{a}})},\ \Eprint
  {http://arxiv.org/abs/1810.12218} {arXiv:1810.12218 [astro-ph.CO]}
  \BibitemShut {NoStop}%
\bibitem [{\citenamefont {Bartolo}\ \emph
  {et~al.}(2019{\natexlab{b}})\citenamefont {Bartolo}, \citenamefont {De~Luca},
  \citenamefont {Franciolini}, \citenamefont {Peloso}, \citenamefont {Racco},\
  and\ \citenamefont {Riotto}}]{Bartolo:2018rku}%
  \BibitemOpen
  \bibfield  {author} {\bibinfo {author} {\bibfnamefont {N.}~\bibnamefont
  {Bartolo}}, \bibinfo {author} {\bibfnamefont {V.}~\bibnamefont {De~Luca}},
  \bibinfo {author} {\bibfnamefont {G.}~\bibnamefont {Franciolini}}, \bibinfo
  {author} {\bibfnamefont {M.}~\bibnamefont {Peloso}}, \bibinfo {author}
  {\bibfnamefont {D.}~\bibnamefont {Racco}}, \ and\ \bibinfo {author}
  {\bibfnamefont {A.}~\bibnamefont {Riotto}},\ }\bibfield  {title} {\enquote
  {\bibinfo {title} {{Testing primordial black holes as dark matter with
  LISA}},}\ }\href {\doibase 10.1103/PhysRevD.99.103521} {\bibfield  {journal}
  {\bibinfo  {journal} {Phys. Rev.}\ }\textbf {\bibinfo {volume} {D99}},\
  \bibinfo {pages} {103521} (\bibinfo {year} {2019}{\natexlab{b}})},\ \Eprint
  {http://arxiv.org/abs/1810.12224} {arXiv:1810.12224 [astro-ph.CO]}
  \BibitemShut {NoStop}%
\bibitem [{\citenamefont {Unal}(2019)}]{Unal:2018yaa}%
  \BibitemOpen
  \bibfield  {author} {\bibinfo {author} {\bibfnamefont {Caner}\ \bibnamefont
  {Unal}},\ }\bibfield  {title} {\enquote {\bibinfo {title} {{Imprints of
  Primordial Non-Gaussianity on Gravitational Wave Spectrum}},}\ }\href
  {\doibase 10.1103/PhysRevD.99.041301} {\bibfield  {journal} {\bibinfo
  {journal} {Phys. Rev.}\ }\textbf {\bibinfo {volume} {D99}},\ \bibinfo {pages}
  {041301} (\bibinfo {year} {2019})},\ \Eprint
  {http://arxiv.org/abs/1811.09151} {arXiv:1811.09151 [astro-ph.CO]}
  \BibitemShut {NoStop}%
\bibitem [{\citenamefont {Inomata}\ and\ \citenamefont
  {Nakama}(2019)}]{Inomata:2018epa}%
  \BibitemOpen
  \bibfield  {author} {\bibinfo {author} {\bibfnamefont {Keisuke}\ \bibnamefont
  {Inomata}}\ and\ \bibinfo {author} {\bibfnamefont {Tomohiro}\ \bibnamefont
  {Nakama}},\ }\bibfield  {title} {\enquote {\bibinfo {title} {{Gravitational
  waves induced by scalar perturbations as probes of the small-scale primordial
  spectrum}},}\ }\href {\doibase 10.1103/PhysRevD.99.043511} {\bibfield
  {journal} {\bibinfo  {journal} {Phys. Rev.}\ }\textbf {\bibinfo {volume}
  {D99}},\ \bibinfo {pages} {043511} (\bibinfo {year} {2019})},\ \Eprint
  {http://arxiv.org/abs/1812.00674} {arXiv:1812.00674 [astro-ph.CO]}
  \BibitemShut {NoStop}%
\bibitem [{\citenamefont {Clesse}\ \emph {et~al.}(2018)\citenamefont {Clesse},
  \citenamefont {García-Bellido},\ and\ \citenamefont
  {Orani}}]{Clesse:2018ogk}%
  \BibitemOpen
  \bibfield  {author} {\bibinfo {author} {\bibfnamefont {Sebastien}\
  \bibnamefont {Clesse}}, \bibinfo {author} {\bibfnamefont {Juan}\ \bibnamefont
  {García-Bellido}}, \ and\ \bibinfo {author} {\bibfnamefont {Stefano}\
  \bibnamefont {Orani}},\ }\bibfield  {title} {\enquote {\bibinfo {title}
  {{Detecting the Stochastic Gravitational Wave Background from Primordial
  Black Hole Formation}},}\ }\href@noop {} {\  (\bibinfo {year} {2018})},\
  \Eprint {http://arxiv.org/abs/1812.11011} {arXiv:1812.11011 [astro-ph.CO]}
  \BibitemShut {NoStop}%
\bibitem [{\citenamefont {Cai}\ \emph {et~al.}(2019{\natexlab{c}})\citenamefont
  {Cai}, \citenamefont {Pi}, \citenamefont {Wang},\ and\ \citenamefont
  {Yang}}]{Cai:2019amo}%
  \BibitemOpen
  \bibfield  {author} {\bibinfo {author} {\bibfnamefont {Rong-Gen}\
  \bibnamefont {Cai}}, \bibinfo {author} {\bibfnamefont {Shi}\ \bibnamefont
  {Pi}}, \bibinfo {author} {\bibfnamefont {Shao-Jiang}\ \bibnamefont {Wang}}, \
  and\ \bibinfo {author} {\bibfnamefont {Xing-Yu}\ \bibnamefont {Yang}},\
  }\bibfield  {title} {\enquote {\bibinfo {title} {{Resonant multiple peaks in
  the induced gravitational waves}},}\ }\href {\doibase
  10.1088/1475-7516/2019/05/013} {\bibfield  {journal} {\bibinfo  {journal}
  {JCAP}\ }\textbf {\bibinfo {volume} {1905}},\ \bibinfo {pages} {013}
  (\bibinfo {year} {2019}{\natexlab{c}})},\ \Eprint
  {http://arxiv.org/abs/1901.10152} {arXiv:1901.10152 [astro-ph.CO]}
  \BibitemShut {NoStop}%
\bibitem [{\citenamefont {Inomata}\ \emph
  {et~al.}(2019{\natexlab{a}})\citenamefont {Inomata}, \citenamefont {Kohri},
  \citenamefont {Nakama},\ and\ \citenamefont {Terada}}]{Inomata:2019zqy}%
  \BibitemOpen
  \bibfield  {author} {\bibinfo {author} {\bibfnamefont {Keisuke}\ \bibnamefont
  {Inomata}}, \bibinfo {author} {\bibfnamefont {Kazunori}\ \bibnamefont
  {Kohri}}, \bibinfo {author} {\bibfnamefont {Tomohiro}\ \bibnamefont
  {Nakama}}, \ and\ \bibinfo {author} {\bibfnamefont {Takahiro}\ \bibnamefont
  {Terada}},\ }\bibfield  {title} {\enquote {\bibinfo {title} {{Gravitational
  Waves Induced by Scalar Perturbations during a Gradual Transition from an
  Early Matter Era to the Radiation Era}},}\ }\href@noop {} {\  (\bibinfo
  {year} {2019}{\natexlab{a}})},\ \Eprint {http://arxiv.org/abs/1904.12878}
  {arXiv:1904.12878 [astro-ph.CO]} \BibitemShut {NoStop}%
\bibitem [{\citenamefont {Inomata}\ \emph
  {et~al.}(2019{\natexlab{b}})\citenamefont {Inomata}, \citenamefont {Kohri},
  \citenamefont {Nakama},\ and\ \citenamefont {Terada}}]{Inomata:2019ivs}%
  \BibitemOpen
  \bibfield  {author} {\bibinfo {author} {\bibfnamefont {Keisuke}\ \bibnamefont
  {Inomata}}, \bibinfo {author} {\bibfnamefont {Kazunori}\ \bibnamefont
  {Kohri}}, \bibinfo {author} {\bibfnamefont {Tomohiro}\ \bibnamefont
  {Nakama}}, \ and\ \bibinfo {author} {\bibfnamefont {Takahiro}\ \bibnamefont
  {Terada}},\ }\bibfield  {title} {\enquote {\bibinfo {title} {{Enhancement of
  Gravitational Waves Induced by Scalar Perturbations due to a Sudden
  Transition from an Early Matter Era to the Radiation Era}},}\ }\href
  {\doibase 10.1103/PhysRevD.100.043532} {\bibfield  {journal} {\bibinfo
  {journal} {Phys. Rev.}\ }\textbf {\bibinfo {volume} {D100}},\ \bibinfo
  {pages} {043532} (\bibinfo {year} {2019}{\natexlab{b}})},\ \Eprint
  {http://arxiv.org/abs/1904.12879} {arXiv:1904.12879 [astro-ph.CO]}
  \BibitemShut {NoStop}%
\bibitem [{\citenamefont {Cai}\ \emph {et~al.}(2019{\natexlab{d}})\citenamefont
  {Cai}, \citenamefont {Pi}, \citenamefont {Wang},\ and\ \citenamefont
  {Yang}}]{Cai:2019elf}%
  \BibitemOpen
  \bibfield  {author} {\bibinfo {author} {\bibfnamefont {Rong-Gen}\
  \bibnamefont {Cai}}, \bibinfo {author} {\bibfnamefont {Shi}\ \bibnamefont
  {Pi}}, \bibinfo {author} {\bibfnamefont {Shao-Jiang}\ \bibnamefont {Wang}}, \
  and\ \bibinfo {author} {\bibfnamefont {Xing-Yu}\ \bibnamefont {Yang}},\
  }\bibfield  {title} {\enquote {\bibinfo {title} {{Pulsar Timing Array
  Constraints on the Induced Gravitational Waves}},}\ }\href@noop {} {\
  (\bibinfo {year} {2019}{\natexlab{d}})},\ \Eprint
  {http://arxiv.org/abs/1907.06372} {arXiv:1907.06372 [astro-ph.CO]}
  \BibitemShut {NoStop}%
\bibitem [{\citenamefont {Cai}\ \emph {et~al.}(2019{\natexlab{e}})\citenamefont
  {Cai}, \citenamefont {Pi},\ and\ \citenamefont {Sasaki}}]{Cai:2019cdl}%
  \BibitemOpen
  \bibfield  {author} {\bibinfo {author} {\bibfnamefont {Rong-Gen}\
  \bibnamefont {Cai}}, \bibinfo {author} {\bibfnamefont {Shi}\ \bibnamefont
  {Pi}}, \ and\ \bibinfo {author} {\bibfnamefont {Misao}\ \bibnamefont
  {Sasaki}},\ }\bibfield  {title} {\enquote {\bibinfo {title} {{Universal
  infrared scaling of gravitational wave background spectra}},}\ }\href@noop {}
  {\  (\bibinfo {year} {2019}{\natexlab{e}})},\ \Eprint
  {http://arxiv.org/abs/1909.13728} {arXiv:1909.13728 [astro-ph.CO]}
  \BibitemShut {NoStop}%
\bibitem [{\citenamefont {Lu}\ \emph {et~al.}(2019)\citenamefont {Lu},
  \citenamefont {Gong}, \citenamefont {Yi},\ and\ \citenamefont
  {Zhang}}]{Lu:2019sti}%
  \BibitemOpen
  \bibfield  {author} {\bibinfo {author} {\bibfnamefont {Yizhou}\ \bibnamefont
  {Lu}}, \bibinfo {author} {\bibfnamefont {Yungui}\ \bibnamefont {Gong}},
  \bibinfo {author} {\bibfnamefont {Zhu}\ \bibnamefont {Yi}}, \ and\ \bibinfo
  {author} {\bibfnamefont {Fengge}\ \bibnamefont {Zhang}},\ }\bibfield  {title}
  {\enquote {\bibinfo {title} {{Constraints on primordial curvature
  perturbations from primordial black hole dark matter and secondary
  gravitational waves}},}\ }\href {\doibase 10.1088/1475-7516/2019/12/031}
  {\bibfield  {journal} {\bibinfo  {journal} {JCAP}\ }\textbf {\bibinfo
  {volume} {12}},\ \bibinfo {pages} {031} (\bibinfo {year} {2019})},\ \Eprint
  {http://arxiv.org/abs/1907.11896} {arXiv:1907.11896 [gr-qc]} \BibitemShut
  {NoStop}%
\bibitem [{\citenamefont {Yuan}\ \emph
  {et~al.}(2019{\natexlab{b}})\citenamefont {Yuan}, \citenamefont {Chen},\ and\
  \citenamefont {Huang}}]{Yuan:2019wwo}%
  \BibitemOpen
  \bibfield  {author} {\bibinfo {author} {\bibfnamefont {Chen}\ \bibnamefont
  {Yuan}}, \bibinfo {author} {\bibfnamefont {Zu-Cheng}\ \bibnamefont {Chen}}, \
  and\ \bibinfo {author} {\bibfnamefont {Qing-Guo}\ \bibnamefont {Huang}},\
  }\bibfield  {title} {\enquote {\bibinfo {title} {{Log-dependent slope of
  scalar induced gravitational waves in the infrared regions}},}\ }\href@noop
  {} {\  (\bibinfo {year} {2019}{\natexlab{b}})},\ \Eprint
  {http://arxiv.org/abs/1910.09099} {arXiv:1910.09099 [astro-ph.CO]}
  \BibitemShut {NoStop}%
\bibitem [{\citenamefont {Tomikawa}\ and\ \citenamefont
  {Kobayashi}(2019)}]{Tomikawa:2019tvi}%
  \BibitemOpen
  \bibfield  {author} {\bibinfo {author} {\bibfnamefont {Keitaro}\ \bibnamefont
  {Tomikawa}}\ and\ \bibinfo {author} {\bibfnamefont {Tsutomu}\ \bibnamefont
  {Kobayashi}},\ }\bibfield  {title} {\enquote {\bibinfo {title} {{On the gauge
  dependence of gravitational waves generated at second order from scalar
  perturbations}},}\ }\href@noop {} {\  (\bibinfo {year} {2019})},\ \Eprint
  {http://arxiv.org/abs/1910.01880} {arXiv:1910.01880 [gr-qc]} \BibitemShut
  {NoStop}%
\bibitem [{\citenamefont {De~Luca}\ \emph
  {et~al.}(2019{\natexlab{b}})\citenamefont {De~Luca}, \citenamefont
  {Franciolini}, \citenamefont {Kehagias},\ and\ \citenamefont
  {Riotto}}]{DeLuca:2019ufz}%
  \BibitemOpen
  \bibfield  {author} {\bibinfo {author} {\bibfnamefont {V.}~\bibnamefont
  {De~Luca}}, \bibinfo {author} {\bibfnamefont {G.}~\bibnamefont
  {Franciolini}}, \bibinfo {author} {\bibfnamefont {A.}~\bibnamefont
  {Kehagias}}, \ and\ \bibinfo {author} {\bibfnamefont {A.}~\bibnamefont
  {Riotto}},\ }\bibfield  {title} {\enquote {\bibinfo {title} {{On the Gauge
  Invariance of Cosmological Gravitational Waves}},}\ }\href@noop {} {\
  (\bibinfo {year} {2019}{\natexlab{b}})},\ \Eprint
  {http://arxiv.org/abs/1911.09689} {arXiv:1911.09689 [gr-qc]} \BibitemShut
  {NoStop}%
\bibitem [{\citenamefont {Yuan}\ \emph {et~al.}(2020)\citenamefont {Yuan},
  \citenamefont {Chen},\ and\ \citenamefont {Huang}}]{Yuan:2019fwv}%
  \BibitemOpen
  \bibfield  {author} {\bibinfo {author} {\bibfnamefont {Chen}\ \bibnamefont
  {Yuan}}, \bibinfo {author} {\bibfnamefont {Zu-Cheng}\ \bibnamefont {Chen}}, \
  and\ \bibinfo {author} {\bibfnamefont {Qing-Guo}\ \bibnamefont {Huang}},\
  }\bibfield  {title} {\enquote {\bibinfo {title} {{Scalar induced
  gravitational waves in different gauges}},}\ }\href {\doibase
  10.1103/PhysRevD.101.063018} {\bibfield  {journal} {\bibinfo  {journal}
  {Phys. Rev. D}\ }\textbf {\bibinfo {volume} {101}},\ \bibinfo {pages}
  {063018} (\bibinfo {year} {2020})},\ \Eprint
  {http://arxiv.org/abs/1912.00885} {arXiv:1912.00885 [astro-ph.CO]}
  \BibitemShut {NoStop}%
\bibitem [{\citenamefont {Inomata}\ \emph
  {et~al.}(2020{\natexlab{a}})\citenamefont {Inomata}, \citenamefont {Kohri},
  \citenamefont {Nakama},\ and\ \citenamefont {Terada}}]{Inomata:2020tkl}%
  \BibitemOpen
  \bibfield  {author} {\bibinfo {author} {\bibfnamefont {Keisuke}\ \bibnamefont
  {Inomata}}, \bibinfo {author} {\bibfnamefont {Kazunori}\ \bibnamefont
  {Kohri}}, \bibinfo {author} {\bibfnamefont {Tomohiro}\ \bibnamefont
  {Nakama}}, \ and\ \bibinfo {author} {\bibfnamefont {Takahiro}\ \bibnamefont
  {Terada}},\ }\bibfield  {title} {\enquote {\bibinfo {title} {{Enhancement of
  gravitational waves induced by scalar perturbations due to a sudden
  transition from an early matter era to the radiation era}},}\ }\href
  {\doibase 10.1088/1742-6596/1468/1/012002} {\bibfield  {journal} {\bibinfo
  {journal} {J. Phys. Conf. Ser.}\ }\textbf {\bibinfo {volume} {1468}},\
  \bibinfo {pages} {012002} (\bibinfo {year} {2020}{\natexlab{a}})}\BibitemShut
  {NoStop}%
\bibitem [{\citenamefont {Inomata}\ \emph
  {et~al.}(2020{\natexlab{b}})\citenamefont {Inomata}, \citenamefont {Kohri},
  \citenamefont {Nakama},\ and\ \citenamefont {Terada}}]{Inomata:2020yqv}%
  \BibitemOpen
  \bibfield  {author} {\bibinfo {author} {\bibfnamefont {Keisuke}\ \bibnamefont
  {Inomata}}, \bibinfo {author} {\bibfnamefont {Kazunori}\ \bibnamefont
  {Kohri}}, \bibinfo {author} {\bibfnamefont {Tomohiro}\ \bibnamefont
  {Nakama}}, \ and\ \bibinfo {author} {\bibfnamefont {Takahiro}\ \bibnamefont
  {Terada}},\ }\bibfield  {title} {\enquote {\bibinfo {title} {{Gravitational
  waves induced by scalar perturbations during a gradual transition from an
  early matter era to the radiation era}},}\ }\href {\doibase
  10.1088/1742-6596/1468/1/012001} {\bibfield  {journal} {\bibinfo  {journal}
  {J. Phys. Conf. Ser.}\ }\textbf {\bibinfo {volume} {1468}},\ \bibinfo {pages}
  {012001} (\bibinfo {year} {2020}{\natexlab{b}})}\BibitemShut {NoStop}%
\bibitem [{\citenamefont {Inomata}\ \emph
  {et~al.}(2020{\natexlab{c}})\citenamefont {Inomata}, \citenamefont
  {Kawasaki}, \citenamefont {Mukaida}, \citenamefont {Terada},\ and\
  \citenamefont {Yanagida}}]{Inomata:2020lmk}%
  \BibitemOpen
  \bibfield  {author} {\bibinfo {author} {\bibfnamefont {Keisuke}\ \bibnamefont
  {Inomata}}, \bibinfo {author} {\bibfnamefont {Masahiro}\ \bibnamefont
  {Kawasaki}}, \bibinfo {author} {\bibfnamefont {Kyohei}\ \bibnamefont
  {Mukaida}}, \bibinfo {author} {\bibfnamefont {Takahiro}\ \bibnamefont
  {Terada}}, \ and\ \bibinfo {author} {\bibfnamefont {Tsutomu~T.}\ \bibnamefont
  {Yanagida}},\ }\bibfield  {title} {\enquote {\bibinfo {title} {{Gravitational
  Wave Production right after a Primordial Black Hole Evaporation}},}\ }\href
  {\doibase 10.1103/PhysRevD.101.123533} {\bibfield  {journal} {\bibinfo
  {journal} {Phys. Rev. D}\ }\textbf {\bibinfo {volume} {101}},\ \bibinfo
  {pages} {123533} (\bibinfo {year} {2020}{\natexlab{c}})},\ \Eprint
  {http://arxiv.org/abs/2003.10455} {arXiv:2003.10455 [astro-ph.CO]}
  \BibitemShut {NoStop}%
\bibitem [{\citenamefont {Yuan}\ and\ \citenamefont
  {Huang}(2020)}]{Yuan:2020iwf}%
  \BibitemOpen
  \bibfield  {author} {\bibinfo {author} {\bibfnamefont {Chen}\ \bibnamefont
  {Yuan}}\ and\ \bibinfo {author} {\bibfnamefont {Qing-Guo}\ \bibnamefont
  {Huang}},\ }\bibfield  {title} {\enquote {\bibinfo {title} {{Gravitational
  waves induced by the local-type non-Gaussian curvature perturbations}},}\
  }\href@noop {} {\  (\bibinfo {year} {2020})},\ \Eprint
  {http://arxiv.org/abs/2007.10686} {arXiv:2007.10686 [astro-ph.CO]}
  \BibitemShut {NoStop}%
\bibitem [{\citenamefont {Papanikolaou}\ \emph {et~al.}(2020)\citenamefont
  {Papanikolaou}, \citenamefont {Vennin},\ and\ \citenamefont
  {Langlois}}]{Papanikolaou:2020qtd}%
  \BibitemOpen
  \bibfield  {author} {\bibinfo {author} {\bibfnamefont {Theodoros}\
  \bibnamefont {Papanikolaou}}, \bibinfo {author} {\bibfnamefont {Vincent}\
  \bibnamefont {Vennin}}, \ and\ \bibinfo {author} {\bibfnamefont {David}\
  \bibnamefont {Langlois}},\ }\bibfield  {title} {\enquote {\bibinfo {title}
  {{Gravitational waves from a universe filled with primordial black holes}},}\
  }\href@noop {} {\  (\bibinfo {year} {2020})},\ \Eprint
  {http://arxiv.org/abs/2010.11573} {arXiv:2010.11573 [astro-ph.CO]}
  \BibitemShut {NoStop}%
\bibitem [{\citenamefont {Zhang}\ \emph
  {et~al.}(2020{\natexlab{a}})\citenamefont {Zhang}, \citenamefont {Ali},
  \citenamefont {Gong}, \citenamefont {Lin},\ and\ \citenamefont
  {Lu}}]{Zhang:2020ptw}%
  \BibitemOpen
  \bibfield  {author} {\bibinfo {author} {\bibfnamefont {Fengge}\ \bibnamefont
  {Zhang}}, \bibinfo {author} {\bibfnamefont {Arshad}\ \bibnamefont {Ali}},
  \bibinfo {author} {\bibfnamefont {Yungui}\ \bibnamefont {Gong}}, \bibinfo
  {author} {\bibfnamefont {Jiong}\ \bibnamefont {Lin}}, \ and\ \bibinfo
  {author} {\bibfnamefont {Yizhou}\ \bibnamefont {Lu}},\ }\bibfield  {title}
  {\enquote {\bibinfo {title} {{On the waveform of the scalar induced
  gravitational waves}},}\ }\href@noop {} {\  (\bibinfo {year}
  {2020}{\natexlab{a}})},\ \Eprint {http://arxiv.org/abs/2008.12961}
  {arXiv:2008.12961 [gr-qc]} \BibitemShut {NoStop}%
\bibitem [{\citenamefont {Kapadia}\ \emph {et~al.}(2020)\citenamefont
  {Kapadia}, \citenamefont {Pandey}, \citenamefont {Suyama}, \citenamefont
  {Kandhasamy},\ and\ \citenamefont {Ajith}}]{Kapadia:2020pnr}%
  \BibitemOpen
  \bibfield  {author} {\bibinfo {author} {\bibfnamefont {Shasvath~J.}\
  \bibnamefont {Kapadia}}, \bibinfo {author} {\bibfnamefont {Kanhaiya~Lal}\
  \bibnamefont {Pandey}}, \bibinfo {author} {\bibfnamefont {Teruaki}\
  \bibnamefont {Suyama}}, \bibinfo {author} {\bibfnamefont {Shivaraj}\
  \bibnamefont {Kandhasamy}}, \ and\ \bibinfo {author} {\bibfnamefont
  {Parameswaran}\ \bibnamefont {Ajith}},\ }\bibfield  {title} {\enquote
  {\bibinfo {title} {{Search for the stochastic gravitational-wave background
  induced by primordial curvature perturbations in LIGO's second observing
  run}},}\ }\href@noop {} {\  (\bibinfo {year} {2020})},\ \Eprint
  {http://arxiv.org/abs/2009.05514} {arXiv:2009.05514 [gr-qc]} \BibitemShut
  {NoStop}%
\bibitem [{\citenamefont {Zhang}\ \emph
  {et~al.}(2020{\natexlab{b}})\citenamefont {Zhang}, \citenamefont {Gong},
  \citenamefont {Lin}, \citenamefont {Lu},\ and\ \citenamefont
  {Yi}}]{Zhang:2020uek}%
  \BibitemOpen
  \bibfield  {author} {\bibinfo {author} {\bibfnamefont {Fengge}\ \bibnamefont
  {Zhang}}, \bibinfo {author} {\bibfnamefont {Yungui}\ \bibnamefont {Gong}},
  \bibinfo {author} {\bibfnamefont {Jiong}\ \bibnamefont {Lin}}, \bibinfo
  {author} {\bibfnamefont {Yizhou}\ \bibnamefont {Lu}}, \ and\ \bibinfo
  {author} {\bibfnamefont {Zhu}\ \bibnamefont {Yi}},\ }\bibfield  {title}
  {\enquote {\bibinfo {title} {{Primordial Non-Gaussianity from k/G
  inflation}},}\ }\href@noop {} {\  (\bibinfo {year} {2020}{\natexlab{b}})},\
  \Eprint {http://arxiv.org/abs/2012.06960} {arXiv:2012.06960 [astro-ph.CO]}
  \BibitemShut {NoStop}%
\bibitem [{\citenamefont {Dom\`enech}\ \emph {et~al.}(2020)\citenamefont
  {Dom\`enech}, \citenamefont {Lin},\ and\ \citenamefont
  {Sasaki}}]{Domenech:2020ssp}%
  \BibitemOpen
  \bibfield  {author} {\bibinfo {author} {\bibfnamefont {Guillem}\ \bibnamefont
  {Dom\`enech}}, \bibinfo {author} {\bibfnamefont {Chunshan}\ \bibnamefont
  {Lin}}, \ and\ \bibinfo {author} {\bibfnamefont {Misao}\ \bibnamefont
  {Sasaki}},\ }\bibfield  {title} {\enquote {\bibinfo {title} {{Gravitational
  wave constraints on the primordial black hole dominated early universe}},}\
  }\href@noop {} {\  (\bibinfo {year} {2020})},\ \Eprint
  {http://arxiv.org/abs/2012.08151} {arXiv:2012.08151 [gr-qc]} \BibitemShut
  {NoStop}%
\bibitem [{\citenamefont {Dalianis}\ and\ \citenamefont
  {Kouvaris}(2020)}]{Dalianis:2020gup}%
  \BibitemOpen
  \bibfield  {author} {\bibinfo {author} {\bibfnamefont {Ioannis}\ \bibnamefont
  {Dalianis}}\ and\ \bibinfo {author} {\bibfnamefont {Chris}\ \bibnamefont
  {Kouvaris}},\ }\bibfield  {title} {\enquote {\bibinfo {title} {{Gravitational
  Waves from Density Perturbations in an Early Matter Domination Era}},}\
  }\href@noop {} {\  (\bibinfo {year} {2020})},\ \Eprint
  {http://arxiv.org/abs/2012.09255} {arXiv:2012.09255 [astro-ph.CO]}
  \BibitemShut {NoStop}%
\bibitem [{\citenamefont {Atal}\ and\ \citenamefont
  {Dom\`enech}(2021)}]{Atal:2021jyo}%
  \BibitemOpen
  \bibfield  {author} {\bibinfo {author} {\bibfnamefont {Vicente}\ \bibnamefont
  {Atal}}\ and\ \bibinfo {author} {\bibfnamefont {Guillem}\ \bibnamefont
  {Dom\`enech}},\ }\bibfield  {title} {\enquote {\bibinfo {title} {{Probing
  non-Gaussianities with the high frequency tail of induced gravitational
  waves}},}\ }\href@noop {} {\  (\bibinfo {year} {2021})},\ \Eprint
  {http://arxiv.org/abs/2103.01056} {arXiv:2103.01056 [astro-ph.CO]}
  \BibitemShut {NoStop}%
\bibitem [{\citenamefont {Franciolini}(2021)}]{Franciolini:2021nvv}%
  \BibitemOpen
  \bibfield  {author} {\bibinfo {author} {\bibfnamefont {Gabriele}\
  \bibnamefont {Franciolini}},\ }\emph {\bibinfo {title} {{Primordial Black
  Holes: from Theory to Gravitational Wave Observations}}},\ \href {\doibase
  10.13097/archive-ouverte/unige:156136} {Ph.D. thesis},\ \bibinfo  {school}
  {Geneva U., Dept. Theor. Phys.} (\bibinfo {year} {2021}),\ \Eprint
  {http://arxiv.org/abs/2110.06815} {arXiv:2110.06815 [astro-ph.CO]}
  \BibitemShut {NoStop}%
\bibitem [{\citenamefont {Witkowski}\ \emph {et~al.}(2022)\citenamefont
  {Witkowski}, \citenamefont {Dom\`enech}, \citenamefont {Fumagalli},\ and\
  \citenamefont {Renaux-Petel}}]{Witkowski:2021raz}%
  \BibitemOpen
  \bibfield  {author} {\bibinfo {author} {\bibfnamefont {Lukas~T.}\
  \bibnamefont {Witkowski}}, \bibinfo {author} {\bibfnamefont {Guillem}\
  \bibnamefont {Dom\`enech}}, \bibinfo {author} {\bibfnamefont {Jacopo}\
  \bibnamefont {Fumagalli}}, \ and\ \bibinfo {author} {\bibfnamefont
  {S\'ebastien}\ \bibnamefont {Renaux-Petel}},\ }\bibfield  {title} {\enquote
  {\bibinfo {title} {{Expansion history-dependent oscillations in the
  scalar-induced gravitational wave background}},}\ }\href {\doibase
  10.1088/1475-7516/2022/05/028} {\bibfield  {journal} {\bibinfo  {journal}
  {JCAP}\ }\textbf {\bibinfo {volume} {05}},\ \bibinfo {pages} {028} (\bibinfo
  {year} {2022})},\ \Eprint {http://arxiv.org/abs/2110.09480} {arXiv:2110.09480
  [astro-ph.CO]} \BibitemShut {NoStop}%
\bibitem [{\citenamefont {Balaji}\ \emph {et~al.}(2022)\citenamefont {Balaji},
  \citenamefont {Domenech},\ and\ \citenamefont {Silk}}]{Balaji:2022dbi}%
  \BibitemOpen
  \bibfield  {author} {\bibinfo {author} {\bibfnamefont {Shyam}\ \bibnamefont
  {Balaji}}, \bibinfo {author} {\bibfnamefont {Guillem}\ \bibnamefont
  {Domenech}}, \ and\ \bibinfo {author} {\bibfnamefont {Joseph}\ \bibnamefont
  {Silk}},\ }\bibfield  {title} {\enquote {\bibinfo {title} {{Induced
  gravitational waves from slow-roll inflation after an enhancing phase}},}\
  }\href@noop {} {\  (\bibinfo {year} {2022})},\ \Eprint
  {http://arxiv.org/abs/2205.01696} {arXiv:2205.01696 [astro-ph.CO]}
  \BibitemShut {NoStop}%
\bibitem [{\citenamefont {Cang}\ \emph {et~al.}(2022)\citenamefont {Cang},
  \citenamefont {Ma},\ and\ \citenamefont {Gao}}]{Cang:2022oia}%
  \BibitemOpen
  \bibfield  {author} {\bibinfo {author} {\bibfnamefont {Junsong}\ \bibnamefont
  {Cang}}, \bibinfo {author} {\bibfnamefont {Yin-Zhe}\ \bibnamefont {Ma}}, \
  and\ \bibinfo {author} {\bibfnamefont {Yu}~\bibnamefont {Gao}},\ }\bibfield
  {title} {\enquote {\bibinfo {title} {{Constraining primordial black holes
  with relativistic degrees of freedom}},}\ }\href@noop {} {\  (\bibinfo {year}
  {2022})},\ \Eprint {http://arxiv.org/abs/2210.03476} {arXiv:2210.03476
  [astro-ph.CO]} \BibitemShut {NoStop}%
\bibitem [{\citenamefont {Gehrman}\ \emph {et~al.}(2022)\citenamefont
  {Gehrman}, \citenamefont {Shams Es~Haghi}, \citenamefont {Sinha},\ and\
  \citenamefont {Xu}}]{Gehrman:2022imk}%
  \BibitemOpen
  \bibfield  {author} {\bibinfo {author} {\bibfnamefont {Thomas~C.}\
  \bibnamefont {Gehrman}}, \bibinfo {author} {\bibfnamefont {Barmak}\
  \bibnamefont {Shams Es~Haghi}}, \bibinfo {author} {\bibfnamefont {Kuver}\
  \bibnamefont {Sinha}}, \ and\ \bibinfo {author} {\bibfnamefont {Tao}\
  \bibnamefont {Xu}},\ }\bibfield  {title} {\enquote {\bibinfo {title}
  {{Baryogenesis, Primordial Black Holes and MHz-GHz Gravitational Waves}},}\
  }\href@noop {} {\  (\bibinfo {year} {2022})},\ \Eprint
  {http://arxiv.org/abs/2211.08431} {arXiv:2211.08431 [hep-ph]} \BibitemShut
  {NoStop}%
\bibitem [{\citenamefont {Braglia}\ \emph {et~al.}(2021)\citenamefont
  {Braglia}, \citenamefont {Chen},\ and\ \citenamefont
  {Hazra}}]{Braglia:2020taf}%
  \BibitemOpen
  \bibfield  {author} {\bibinfo {author} {\bibfnamefont {Matteo}\ \bibnamefont
  {Braglia}}, \bibinfo {author} {\bibfnamefont {Xingang}\ \bibnamefont {Chen}},
  \ and\ \bibinfo {author} {\bibfnamefont {Dhiraj~Kumar}\ \bibnamefont
  {Hazra}},\ }\bibfield  {title} {\enquote {\bibinfo {title} {{Probing
  Primordial Features with the Stochastic Gravitational Wave Background}},}\
  }\href {\doibase 10.1088/1475-7516/2021/03/005} {\bibfield  {journal}
  {\bibinfo  {journal} {JCAP}\ }\textbf {\bibinfo {volume} {03}},\ \bibinfo
  {pages} {005} (\bibinfo {year} {2021})},\ \Eprint
  {http://arxiv.org/abs/2012.05821} {arXiv:2012.05821 [astro-ph.CO]}
  \BibitemShut {NoStop}%
\bibitem [{\citenamefont {Papanikolaou}(2022)}]{Papanikolaou:2022chm}%
  \BibitemOpen
  \bibfield  {author} {\bibinfo {author} {\bibfnamefont {Theodoros}\
  \bibnamefont {Papanikolaou}},\ }\bibfield  {title} {\enquote {\bibinfo
  {title} {{Gravitational waves induced from primordial black hole
  fluctuations: the~effect of an extended mass function}},}\ }\href {\doibase
  10.1088/1475-7516/2022/10/089} {\bibfield  {journal} {\bibinfo  {journal}
  {JCAP}\ }\textbf {\bibinfo {volume} {10}},\ \bibinfo {pages} {089} (\bibinfo
  {year} {2022})},\ \Eprint {http://arxiv.org/abs/2207.11041} {arXiv:2207.11041
  [astro-ph.CO]} \BibitemShut {NoStop}%
\bibitem [{\citenamefont {Yuan}\ and\ \citenamefont
  {Huang}(2021)}]{Yuan:2021qgz}%
  \BibitemOpen
  \bibfield  {author} {\bibinfo {author} {\bibfnamefont {Chen}\ \bibnamefont
  {Yuan}}\ and\ \bibinfo {author} {\bibfnamefont {Qing-Guo}\ \bibnamefont
  {Huang}},\ }\bibfield  {title} {\enquote {\bibinfo {title} {{A topic review
  on probing primordial black hole dark matter with scalar induced
  gravitational waves}},}\ }\href@noop {} {\  (\bibinfo {year} {2021})},\
  \Eprint {http://arxiv.org/abs/2103.04739} {arXiv:2103.04739 [astro-ph.GA]}
  \BibitemShut {NoStop}%
\bibitem [{\citenamefont {Dom\`enech}(2021)}]{Domenech:2021ztg}%
  \BibitemOpen
  \bibfield  {author} {\bibinfo {author} {\bibfnamefont {Guillem}\ \bibnamefont
  {Dom\`enech}},\ }\bibfield  {title} {\enquote {\bibinfo {title} {{Scalar
  Induced Gravitational Waves Review}},}\ }\href {\doibase
  10.3390/universe7110398} {\bibfield  {journal} {\bibinfo  {journal}
  {Universe}\ }\textbf {\bibinfo {volume} {7}},\ \bibinfo {pages} {398}
  (\bibinfo {year} {2021})},\ \Eprint {http://arxiv.org/abs/2109.01398}
  {arXiv:2109.01398 [gr-qc]} \BibitemShut {NoStop}%
\bibitem [{\citenamefont {Hu}\ and\ \citenamefont {Wu}(2017)}]{Hu:17}%
  \BibitemOpen
  \bibfield  {author} {\bibinfo {author} {\bibfnamefont {Wen-Rui}\ \bibnamefont
  {Hu}}\ and\ \bibinfo {author} {\bibfnamefont {Yue-Liang}\ \bibnamefont
  {Wu}},\ }\bibfield  {title} {\enquote {\bibinfo {title} {{The Taiji Program
  in Space for gravitational wave physics and the nature of gravity}},}\ }\href
  {\doibase 10.1093/nsr/nwx116} {\bibfield  {journal} {\bibinfo  {journal}
  {National Science Review}\ }\textbf {\bibinfo {volume} {4}},\ \bibinfo
  {pages} {685--686} (\bibinfo {year} {2017})},\ \Eprint
  {http://arxiv.org/abs/https://academic.oup.com/nsr/article-pdf/4/5/685/31566708/nwx116.pdf}
  {https://academic.oup.com/nsr/article-pdf/4/5/685/31566708/nwx116.pdf}
  \BibitemShut {NoStop}%
\bibitem [{\citenamefont {Luo}\ \emph {et~al.}(2016)\citenamefont {Luo} \emph
  {et~al.}}]{TianQin:2015yph}%
  \BibitemOpen
  \bibfield  {author} {\bibinfo {author} {\bibfnamefont {Jun}\ \bibnamefont
  {Luo}} \emph {et~al.} (\bibinfo {collaboration} {TianQin}),\ }\bibfield
  {title} {\enquote {\bibinfo {title} {{TianQin: a space-borne gravitational
  wave detector}},}\ }\href {\doibase 10.1088/0264-9381/33/3/035010} {\bibfield
   {journal} {\bibinfo  {journal} {Class. Quant. Grav.}\ }\textbf {\bibinfo
  {volume} {33}},\ \bibinfo {pages} {035010} (\bibinfo {year} {2016})},\
  \Eprint {http://arxiv.org/abs/1512.02076} {arXiv:1512.02076 [astro-ph.IM]}
  \BibitemShut {NoStop}%
\bibitem [{\citenamefont {Audley}\ \emph {et~al.}(2017)\citenamefont {Audley}
  \emph {et~al.}}]{Audley:2017drz}%
  \BibitemOpen
  \bibfield  {author} {\bibinfo {author} {\bibfnamefont {Heather}\ \bibnamefont
  {Audley}} \emph {et~al.} (\bibinfo {collaboration} {LISA}),\ }\bibfield
  {title} {\enquote {\bibinfo {title} {{Laser Interferometer Space Antenna}},}\
  }\href@noop {} {\  (\bibinfo {year} {2017})},\ \Eprint
  {http://arxiv.org/abs/1702.00786} {arXiv:1702.00786 [astro-ph.IM]}
  \BibitemShut {NoStop}%
\end{thebibliography}%

\end{document}